\documentclass[aps,reprint,physrev,superscriptaddress,nofootinbib,nobalancelastpage]{revtex4-2}
\usepackage{amsmath}
\usepackage{amsfonts}
\usepackage{amssymb}
\usepackage{graphicx}
\usepackage{mathtools}
\usepackage[dvipsnames]{xcolor}
\usepackage[colorlinks=True,citecolor=Blue,linkcolor=Blue,urlcolor=Blue]{hyperref}
\usepackage{hypcap}
\usepackage{braket}
\usepackage{yfonts}
\usepackage{bm}
\usepackage{tikz}
\usepackage{quantikz}
\usetikzlibrary{shapes, arrows.meta}
\usepackage[mathscr]{eucal}
\usepackage{marginnote}
\usepackage{soul}
\newcommand\cbox[1]{\vcenter{\hbox{#1}}}

\newcommand\eq[1]{\begin{align}#1\end{align}}

\definecolor{myBlue}{RGB}{113,73,75}
\definecolor{myOrange}{RGB}{229,131,46}
\definecolor{myGreen}{RGB}{44,160,44}
\definecolor{myRed}{RGB}{155,57,58}
\definecolor{myPurple}{RGB}{148,103,189}

\newcommand{\new}[1]{{\color{black} {#1}}}

\begin{document}
\title{Partial projected ensembles and spatiotemporal structure of information scrambling}
\author{Saptarshi Mandal}
\email{saptarshi.mandal@icts.res.in}
\affiliation{International Centre for Theoretical Sciences, Tata Institute of Fundamental Research, Bengaluru 560089, India}

\author{Pieter W. Claeys}
\email{claeys@pks.mpg.de}
\affiliation{Max Planck Institute for the Physics of Complex Systems, N\"othnitzer Str. 38, 01187 Dresden, Germany}

\author{Sthitadhi Roy}
\email{sthitadhi.roy@icts.res.in}
\affiliation{International Centre for Theoretical Sciences, Tata Institute of Fundamental Research, Bengaluru 560089, India}
\date{\today}

\begin{abstract}
Thermalisation and information scrambling in out-of-equilibrium quantum many-body systems are deeply intertwined: local subsystems dynamically approach thermal density matrices while their entropies track non-local information spreading. Projected ensembles—ensembles of pure states conditioned on measurement outcomes of complementary subsystems—provide higher-order probes of thermalisation,  converging at late times to universal maximum-entropy ensembles constrained by conservation laws.
In this work, we introduce the partial projected ensemble (PPE) as a framework to study how the spatiotemporal structure of information scrambling is imprinted on projected ensembles. The PPE consists of an ensemble of mixed states induced on a subsystem by measurements on a spatially separated part of its complement, while tracing out the remainder, naturally capturing scenarios involving discarded outcomes or noise-induced losses.
We show that the statistical fluctuations of the PPE faithfully track the causal lightcone of information spreading, thereby revealing how scrambling dynamics are encoded in the ensemble structure. In addition, we demonstrate that the probabilities of bit-string probabilities (PoPs) associated with the PPE exhibit distinct dynamical regimes and provide an experimentally accessible probe of scrambling. Both the PPE fluctuations and PoPs display exponential sensitivity to the size of the discarded region, reflecting an exponential degradation of quantum correlations under erasure or loss. We substantiate these findings using the non-integrable kicked Ising chain, combining numerics in the ergodic regime with exact results at its self-dual point, and extend our analysis to the many-body localised (MBL) regime using simulations supported by analytical results for the $\ell$-bit model. The linear and logarithmic lightcones characteristic of ergodic and MBL regimes, respectively, emerge naturally from the PPE dynamics, establishing it as a powerful tool for probing scrambling and deep thermalisation.
\end{abstract}

\maketitle
\tableofcontents
\section{Introduction}

Understanding the out-of-equilibrium dynamics of complex quantum systems is a question of enduring interest, cutting across fields such as condensed matter physics, statistical mechanics and quantum information science. 
In the context of isolated quantum many-body systems, a question of fundamental importance is if and how do such systems thermalise under unitary dynamics. 
Formalised by the eigenstate thermalisation hypothesis~\cite{deutsch1991quantum,srednicki1994chaos,deutsch2018eigenstate,rigol2008thermalisation,dalessio2016from}, the key idea is that ergodic quantum systems thermalise only locally and this thermalisation is effected by the rest of the system acting as a bath. 
More recently, it has been understood that the physics of thermalisation in quantum systems is intimately connected to that of information scrambling and growth of quantum entanglement~\cite{popescu2006entanglement,linden2009quantum,kaufman2016quantum,clalabrese2005evolution,kim2013ballistic,lakshminarayan2016entanglement,nahum2017quantum,nahum2018dynamics,huang2021extensive}.   

For a small subsystem to thermalise, it must exchange energy, particles, magnetisation etc. with the rest of the subsystem. Consequently, the information stored locally in the subsystem gets shared and scrambled non-locally across the entire system, and the entropy arising from the absence of full information about the subsystem in its reduced state is manifested in the growth of entanglement. 
Formally, thermalisation refers to the reduced density matrix of a subsystem $R$ being described by that of a thermodynamic ensemble such as the Gibbs state, $\rho_{R}\sim e^{-\beta H}$ with the temperature $\beta^{-1}$ set by the initial state. 
$\rho_R$ is obtained by tracing over (or, in other words, integrating out) the rest of the system, acting as a bath, from a pure state describing the entire system.

More interestingly, modern experiments allow for microscopic interrogation of parts of a system through projective measurements of local and non-local correlation functions and keeping track of the measurement outcomes~\cite{Bakr2009quantumgas,Haller2015singleatom,Neill2016Ergodic,kaufman2016quantum,Bernien2017probing,Ebadi2020quantum,Opremck2021High-Fidelity,Evered2023High-Fidelity,Moses2023Rce-Track,shaw2025experimental,choi2023preparing}. 
This approach allows for the construction of an `ensemble' of states on subsystem $R$ conditioned on the measurement outcomes on the rest of the system. Referred to as the {\it projected ensemble} (PE)~\cite{cotler2023emergent,choi2023preparing}, it can be formally expressed as
\eq{
    {\cal E}_{\rm PE} = \{ p(o_{\overline{R}}),~\ket{\psi_R(o_{\overline{R}})}\}_{o_{\overline{R}}}\,,
    \label{eq:PE-def}
}
where $p(o_{\overline{R}})$ is the probability of obtaining the measurement outcome $o_{\overline{R}}$, and $\ket{\psi_R(o_{\overline{R}})}$ is the corresponding state of $R$.
The PE contains manifestly more information than $\rho_R$ as the latter is just the first moment of the former. 
However, from the PE in Eq.~\ref{eq:PE-def}, higher moments of the ensemble can be computed which contain information not present in the first moment.
The statistical properties of the PE therefore open pathways to hitherto unexplored and higher probes of ergodicity and thermalisation, beyond the paradigms of the ETH. 

An important concept in this context is that of {\it deep thermalisation}~\cite{choi2023preparing,cotler2023emergent,ho2022exact,ippoliti2022solvable,lucas2023generalized,Ippoliti2023Dynamical,bhore2023deep,Chan2024projected,mark2024maximum,Varikuti2024unraveling,manna2025projected}.
The essential idea therein is that the PE \eqref{eq:PE-def} constructed from a many-body state evolved unitarily under ergodic dynamics approaches at late times a universal maximum entropy ensemble constrained only by the reduced density matrix of the subsystem $R$.
As such, not only is the first moment of the PE, i.e. the reduced density matrix $\rho_R$, described by the (generalised) Gibbs state, but all higher moments also correspond to those of the corresponding maximum entropy ensemble. 
Deep thermalisation hence puts more stringent conditions on the notion of thermalisation and constitutes a higher-order probe of ergodicity. 

Much of the work so far in this direction has focussed on the late time behaviour of the PEs and the characterisation of the projected ensembles that the PEs approach, leading to some remarkable results. 
In the absence of any conservation laws, it was shown how the PEs approach the Haar ensemble leading to the emergence of quantum designs from a single time-evolved wavefunction~\cite{cotler2023emergent,ho2022exact,ippoliti2022solvable}. 
In the presence of conservation laws, there is an interesting interplay of the conserved charges and their overlap with the measurement operators, leading to the PE approaching a Scrooge ensemble or a statistical mixture of Scrooge ensembles each corresponding to a charge sector~\cite{mark2024maximum,chang2025deep,manna2025projected}. 
Although such universal ensembles emerge for PEs of pure states, much more recently, there have also been works on PEs of mixed states which have extended the notion of the maximum entropy principle to define mixed-state deep thermalisation~\cite{yu2025mixed,sherry2025do} and laid out the conditions for the same.
An overwhelmingly large fraction of the above results are for the late time asymptotic ensembles that the PEs approach, except in certain fine-tuned, solvable settings where they do so at finite times~\cite{ho2022exact,claeys2022emergent}. 

However, much less is understood about the manifestation of the spatiotemporal structure of information scrambling on the PEs at intermediate times. 
This is particularly pertinent for locally interacting systems -- the default architecturally in most experimental settings~\cite{Bakr2009quantumgas,Haller2015singleatom,Neill2016Ergodic,kaufman2016quantum,Bernien2017probing,Ebadi2020quantum,Opremck2021High-Fidelity,Evered2023High-Fidelity,Moses2023Rce-Track,shaw2025experimental,choi2023preparing}.
In such systems there are fundamental bounds, in the form of causal lightcones, constraining how fast information can spread~\cite{lieb1972finite}.
This is exactly the regime we focus on in this work.
The central question we ask in this work is the following: {\it if and how does the dynamics of the PE on a subsystem 
\new{encode} 
the spatiotemporal structure of information scrambling and the causal lightcones?}
To address this question in a concrete setting, we study the structure of the PE on a subsystem $R$ induced by measurements on another disjoint subsystem, $S$, as a function of time as well as the separation between $R$ and $S$.
This spatiotemporal dependence, as we show in detail, holds the key to understanding the structure of information scrambling from the structure of the PE.
The question posed above can be interpreted in a different fashion as well as follows, which is more pertinent experimentally. 
Thermalisation and information scrambling inherently involves non-local, multipartite correlations. However, given that the measurements in $S$ can be chosen to be a set of local measurements and the subsystem $S$ is also local, a natural question is whether signatures of the spatiotemporal structure of information scrambling on the PE be interpreted as a local probe of a phenomenon whose underlying physics is rooted in non-local correlations?

One important point to note here is that the measured subsystem $S$ is not necessarily the same as the complement of $R$.
As such, in the construction of the PE on $R$, the information of the complement of $R\cup S$, denoted as $E = \overline{R\cup S}$ is traced out. 
This setup can be considered equivalent to performing projective measurements on all of $\overline{R} = E\cup S$ and then discarding the outcomes and averaging over them in $E$; we therefore refer to the PE on $R$ over the measurement outcomes in $S$ as the {\it partial projected ensemble (PPE)}. This process can either be a physical one, e.g. due to the inaccessibility of subsystem $E$ to microscopically resolved projective measurements or due to external noise, or can be realised in post-processing by discarding information from the full PE, presenting a way of resolving the spatiotemporal structure of the dynamics from the PE without the need for additional measurements.

In this way the presence of the traced out subsystem, $E$, is crucial to our setting. 
Throughout this work we will consider $S$ and $R$ to be spatially separated, such that any interactions between both are necessarily mediated by their local coupling to $E$.
At any given time, if the extent of $E$ is large enough that the subsystems $S$ and $R$ are not causally connected, one expects the measurements on the former to have no effect on the state of the latter.
On the other hand, if $S$ and $R$ become causally connected at a later time, measurements on the former are expected to have a non-trivial influence on the states of the later leading to a non-trivial distribution of states in the PPE on $R$. 
This suggests that the extent of $E$ intervening $S$ and $R$ may set a timescale across which the nature of the PE on $R$ changes qualitatively. 
Making this qualitative difference between the cases quantitative lies at the heart of this work.

While the main focus of this work is on the dynamics of the PPEs, the spatial extent of $E$ also plays a fundamental role in the asymptotic form of the PPE that it takes at very late times.
As we will show, the information of the asymptotic (in time) quantum correlations between $R$ and $S$ degrades exponentially with the extent of $E$ intervening the two subsystems. 
This result can be interpreted in two complementary ways.
On one hand, this result suggests that external agents such as noise-induced decoherence or lossy measurements on a part of a system can in fact degrade the quantum correlations between the remaining parts of the system exponentially with the size of the noisy subsystem. 
On the other hand, this same result can be interpreted as a signature of a higher probe of ergodicity or deep thermalisation.
However, the key point is that in either interpretation, it is the non-local, mutlipartite nature of correlations with information scrambled ergodically across the entire system that leads to the exponential degradation.

Having established the background, motivation, and the concrete setting of this work, we next present a brief overview of the paper.

\subsection{Overview}

In Sec.~\ref{sec:ppe}, we define in detail the partial projected ensemble for a one-dimensional system. In particular, in Sec.~\ref{sec:ppe-def}, we set up the precise geometry of the tripartition of the system into the three subsystems, $R$, $E$, and $S$ and define the notations. 
In Sec.~\ref{sec:spatiotemporal}, we identify the appropriate quantities, namely the fluctuations over the ensemble, and motivate how they may 
\new{show} 
the spatiotemporal structure of information spreading. 
In addition, in Sec.~\ref{sec:pop}, we also define the probabilities of (bit-string) probabilities (PoPs)~\cite{shaw2025experimental,claeys2024fock-space} over the ensemble and discuss how they may carry information of the causal lightcone; this serves a potential point of connection between our theoretical framework and experimental verification, since bit-string measurements are indeed quite the norm in modern platforms~\cite{Bakr2009quantumgas,Haller2015singleatom,Neill2016Ergodic,kaufman2016quantum,Bernien2017probing,Ebadi2020quantum,Opremck2021High-Fidelity,Evered2023High-Fidelity,Moses2023Rce-Track,shaw2025experimental,choi2023preparing}.

To demonstrate our ideas in a concrete settting, our model of choice is a non-integrable kicked Ising chain with a tunable parameters which can drive the dynamics from ergodic/chaotic to many-body localised~\cite{zhang2016floquet}. Moreover, the model for a specific choice of parameters also sits at a self-dual point which endows the model with dual-unitarity~\cite{ho2022exact,Ippoliti2023Dynamical}, which we leverage to get exact results in the thermodynamic limit. 
A description of this model and its representation as a Floquet brickwork circuit constitutes Sec.~\ref{sec:brickwork}. 
An important feature of such a brickwork geometry is the presence of a sharp lightcone velocity in the system. 
In Secs.~\ref{sec:ppe-lc} and \ref{sec:pop-lc}, we discuss how such a sharp velocity imposes certain bounds and features on the fluctuations in the partial projected ensemble and the PoPs respectively.

With the generalities set up, we turn towards explicit results for ergodic circuits in Sec.~\ref{sec:gen-erg}.
We consider a generic point in the ergodic phase of the kicked Ising chain in Sec.~\ref{sec:ftfi} and show numerical results for the fluctuations over the PPE as well as results for the PoPs. 
In Sec.~\ref{sec:sdki}, we show the corresponding results for the self-dual kicked Ising (SDKI) chain where its dual-unitarity is exploited to obtain exact numerical results for the PPE fluctuations in the thermodynamic limit as well as analytical expressions for the PoPs.
There are two key takeaways from these results. 
First, the PPE fluctuations mirror the information lightcone of these systems and hence explicitly carry the signatures of the spatiotemporal structure of information scrambling in such systems.
Second, the nature of the PoPs change qualitatively across a timescale which is again set by the causal lightcone velocities of the system.
In Sec.~\ref{sec:latetimes}, we discuss the asymptotic ensembles that the PPE approaches at late times. 
Given that the kicked Ising chain has no conserved quantities, the PPE approaches the so-called generalised Hilbert-Schmidt ensemble~\cite{sommers2004statistical,zyczkowski2011generating}.
Using this result we show that the PPE fluctuations, which encode the quantum correlations between $R$ and $S$, degrade exponentially with the size of intervening $E$.
This precise result can in turn be used as a probe for deep thermalisation.

While locally interacting, ergodic systems are archetypes for studying information scrambling, many-body localised (MBL) systems~\cite{nandkishore2015many,abanin2019colloquium,sierant2024manybody} constitute a class of system which robustly break ergodicity and yet show information scrambling, albeit logarithmically slow in space and time~\cite{bardarson2012unbounded,serbyn2013universal,nanduri2014entanglement,chen2016universal,chen2017out,huang2017out,deng2017logarithmic,pain2024entanglement,thakur2025logarithmic}. 
To show the generality of the connection between the structure of the partial projected ensembles and the information lightcone, we present results for MBL systems in Sec.~\ref{sec:mbl}. 
Using a combination of numerical results for a strongly disordered kicked Ising model~\cite{zhang2016floquet} and analytical results for a phenomenological $\ell$-bit model~\cite{serbyn2013local,huse2014phenomenology} of MBL, we show that the spatiotemporal dependence of the fluctuations of the partial projected ensemble again carries signatures of the logarithmic lightcone of information spreading in such systems.

We close with a summary and discussion of the main results and a future outlook in Sec.~\ref{sec:conclusion}.

\section{Partial projected ensemble \label{sec:ppe}}

\begin{figure}
    \begin{tikzpicture}[scale=0.8]
    \fill[fill=Cyan!10,rounded corners=5pt] (-0.1,-0.3) rectangle (1.5,1.3); 
    \draw[<->, >=Stealth] (-0.1,-0.4) -- (1.5,-0.4);
    \node at (0.8, 1.5) {\small $R$};
    \node at (0.8, -0.7) {\small $L_R$};
    \fill[fill=BrickRed!10,rounded corners=5pt] (1.5,-0.3) rectangle (4.5,1.3);
    \draw[<->, >=Stealth] (1.5,-0.4) -- (4.5,-0.4);
    \node at (3, 1.5) {\small $E$};
    \node at (3, -0.7) {\small $L_E$};
    \fill[fill=OliveGreen!10,rounded corners=5pt] (4.5,-0.3) rectangle (9.4,1.3); 
    \draw[<->, >=Stealth] (4.5,-0.4) -- (9.2,-0.4);
    \node at (6.85, 1.5) {\small $S$};
    \node at (6.85, -0.7) {\small $L_S$};
        \def\sx{0.15}
        \def\sy{0.1}
        \draw[thick, rounded corners=2pt, fill=Black!50] (-0.2+\sx,-0.15+\sy) rectangle (9.2+\sx,0.15+\sy);
        \draw[thick, rounded corners=2pt, fill=Black!10] (-0.2,-0.15) rectangle (9.2,0.15);
        \foreach \x in {0,...,9}{
        \draw[thick](\x,0.15)--(\x,0.5);
        \draw[thick](\x+\sx,0.15+\sy)--(\x+\sx,0.5+\sy);
        }
        \foreach \x in {2,...,9}{
        \draw[thick](\x,0.5) to[out=90,in=90] (\x+\sx,0.5+\sy);
        }
        \foreach \x in {5,...,9}{
        \draw[thick](\x +\sx/2,0.5 +\sy/2+0.05) -- (\x +\sx/2,0.5 +\sy/2+0.25);
        \draw[fill=black,thick, fill=white] (\x+\sx/2,0.5 +\sy/2+0.4) circle (0.15);
        \draw[->, >={Triangle[width=2pt, length=2pt]}] (\x+\sx/2-0.1,0.5 +\sy/2+0.4-0.1) -- (\x+\sx/2+0.1,0.5 +\sy/2+0.4+0.1);
        }
    \end{tikzpicture}
    \caption{Schematic representation of the partial projected ensemble. The system is tripartitioned into $R$, $E$, and $S$. The degrees of freedom in $E$ are traced out whereas those in $S$ are projectively measured in a 1-local basis. Conditioned on the outcome $o_S$, the state of $R$ is denoted as $\rho_R(o_S)$ (see Eq.~\ref{eq:rhoR-os}) and appears with probability $p({o_S})$ given in Eq.~\ref{eq:p-os}.}
    \label{fig:tripartition}
\end{figure}
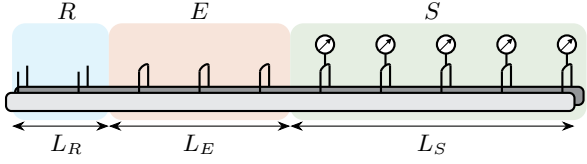

In this section, we introduce the partial projected ensemble (PPE) and quantify how the fluctuations therein may 
\new{carry imprints of} the spatiotemporal profile of information spreading in the system. 

\subsection{Defining the partial projected ensemble \label{sec:ppe-def}}

For concreteness, consider a quantum system which is tri-partitioned into three subsystems $R$, $E$, and $S$, as shown in Fig.~\ref{fig:tripartition}, and $\ket{\psi_{RSE}}$ denotes the state of the system.
The reduced density matrix of $R$,
\eq{
\rho_R = {\rm Tr}_{ES}[\rho_{RSE}]\,,
\label{eq:rhoR}
}
with $\rho_{RSE} = \ket{\psi_{RSE}}\bra{\psi_{RSE}}$, contains all the information about expectation values of any operator, $O_R$, which is supported only on $R$ as $\braket{O_R} = {\rm Tr}[O_R \rho_R ]$.
The partial trace over the complement of $R$, denoted by $\overline{R}\equiv E\cup S$, in Eq.~\ref{eq:rhoR} can be interpreted as performing projective measurements on every site of $\overline{R}$ in a local basis, and averaging over the outcomes. 

However, significantly more information, in particular about the higher moments of the observables in $R$, is contained in ensembles of states in $R$ for which $\rho_R$ is simply its first moment. These ensembles present a particular choice of unraveling of $\rho_R$.
A natural protocol for generating such an ensemble is to keep track of the measurement outcomes on a part of $\overline{R}$ as well as the state of $R$ conditioned on the measurement outcome. 
The subsystem whose measurement outcomes are kept track of is labelled as $S$ and the remaining part of $\overline{R}$, labelled as $E$, is traced out (see Fig.~\ref{fig:tripartition}). 
{\color{blue} Note that this is equivalent to first tracing out $E$ and then performing measurements over $S$.}
The PPE can then be explicitly defined as follows. Consider a set of measurements on $S$ in some local tensor-product basis, $\{\ket{o_S}\}$.
For instance, for qubits, $o_S = \{0,1\}^{\otimes L_S}$ is a bit-string. The conditional state of $R$ given an outcome $o_S$ is given by
\eq{
\rho_R(o_S) = \frac{{\rm Tr}_{ES}[\Pi_{o_S}\rho_{RSE}\Pi_{o_S}]}{p(o_S)}\,,
\label{eq:rhoR-os}
}
and the probability of the outcome is 
\eq{
p(o_S) = {\rm Tr}[\rho_{RSE}\Pi_{o_S}]\,,
\label{eq:p-os}
}
where $\Pi_{o_S} = \ket{o_S}\bra{o_S}$ is the projector onto the bit-string $o_S$. The PPE is then defined as the ensemble of states in Eq.~\ref{eq:rhoR-os} and their probabilities in Eq.~\ref{eq:p-os}, over all measurement outcomes 
\eq{
{\cal E}_{\rm PPE} \equiv \{p(o_S),~\rho_R(o_S)\}_{o_S}\,.
\label{eq:ppe-def}
}
Note that in the limit of $E$ being a null set, ${\cal E}_{\rm PPE}$ reduces to the more commonly studied projected ensemble. 

Given the PPE, as in Eq.~\ref{eq:ppe-def}, the higher moments of observables in $R$, or in general, quantities that are not linear functions of  $\rho_R$, become natural quantities of interest. Given an observable $O_R$ supported entirely in $R$, its $k^{\rm th}$ moment over the PPE is defined as 
\eq{
\braket{O_R^{(k)}} \equiv \sum_{o_S}p(o_S)\left({\rm Tr}[O_R\rho_R(o_S)]\right)^k\,,
}
which can be interpreted as the expectation value in the $k^{\rm th}$ moment of the ensemble, $\braket{O_R^{(k)}}  = {\rm Tr}[O_R^{\otimes k}\rho_R^{(k)}]$,
where
\eq{
\rho_R^{(k)} = \sum_{o_S}p(o_S)\rho_R^{\otimes k}(o_S)\,.
\label{eq:rhoR-k}
}
Note that $k=1$ in the equation above, which is the first moment of the ensemble, reduces to $\rho_R$ as given in Eq.~\ref{eq:rhoR}, whereas $k\ge 2$ carry information about higher moments which can encode non-linear functions of the state.

\subsection{Spatiotemporal structure of information spreading \label{sec:spatiotemporal}}

\begin{figure}
\begin{center}
\begin{tikzpicture}[scale=0.385]
\fill[fill=Cyan!10, rounded corners=3pt] (-0.5,-1.5) rectangle (1.5,2.75); 
\fill[fill=BrickRed!10,rounded corners=3pt] (1.5,-1.5) rectangle (5.5,2.75); 
\fill[fill=OliveGreen!10,rounded corners=3pt] (5.5,-1.5) rectangle (9.5,2.75); 
\node at (0.5,-1) {\footnotesize $R$};
\node at (3.5,-1) {\footnotesize $E$};
\node at (7.5,-1) {\footnotesize $S$};
\draw[->, >=Stealth](-0.8,-1.7)--(-0.8,1.7);
\node[rotate=90] at (-0.8,3) {\footnotesize ${\rm time}(t)$};
\draw[->, >=Stealth](-0.8,-1.7)--(0.8,-1.7);
\node at (1.8,-1.8) {\footnotesize ${\rm space}$};
\foreach \x in {0,...,9}{
\draw[thick](\x,-0.5)--(\x,2.5);
}
\draw[black, thick, fill=Black!10, rounded corners = 3pt] (-0.2,0) rectangle (9.2,2);
\shade[right color=RoyalBlue!90!white, left color=Cyan!80!white!5, opacity=0.5]
    (0,0) -- (1,0) -- (2.5,2) --(0,2) -- cycle;
\shade[left color=OliveGreen!90!white, right color=red!80!white!5, opacity=0.5] (9,0) -- (6,0) -- (6-1.5,2) --(9,2) -- cycle;
\draw[<->, >=Stealth] (0,1.8) -- (2.5,1.8);
\draw[<->, >=Stealth] (2.5,1.8) -- (4.5,1.8);
\draw[<->, >=Stealth] (4.5,1.8) -- (9,1.8);
\node at (1.25,1.25) {\footnotesize $R\cup E_1$};
\node at (3.5,1.25) {\footnotesize $E_2$};
\node at (6.75,1.25) {\footnotesize $E_3\cup S$};
\node at (4.5,6.5) {$t<t_\ast$};
\draw[rounded corners=2pt](3, 6) rectangle (6,7);
\end{tikzpicture}
\hspace{0pt}
\begin{tikzpicture}[scale=0.385]
\fill[fill=Cyan!10, rounded corners=3pt] (-0.5,-1.5) rectangle (1.5,5.75); 
\fill[fill=BrickRed!10,rounded corners=3pt] (1.5,-1.5) rectangle (5.5,5.75); 
\fill[fill=OliveGreen!10,rounded corners=3pt] (5.5,-1.5) rectangle (9.5,5.75); 
\node at (0.5,-1) {\footnotesize $R$};
\node at (3.5,-1) {\footnotesize $E$};
\node at (7.5,-1) {\footnotesize $S$};
\draw[->, >=Stealth](-0.8,-1.7)--(-0.8,1.7);
\node[rotate=90] at (-0.8,3) {\footnotesize ${\rm time}(t)$};
\draw[->, >=Stealth](-0.8,-1.7)--(0.8,-1.7);
\node at (1.8,-1.8) {\footnotesize ${\rm space}$};
\foreach \x in {0,...,9}{
\draw[ thick](\x,-0.5)--(\x,5.5);
}
\draw[black,  thick, fill=Black!10, rounded corners = 3pt] (-0.2,0) rectangle (9.2,5);
\shade[right color=RoyalBlue!90!white, left color=red!80!white!5, opacity=0.5] (0,0) -- (1,0) -- (4.75,5) --(0,5) -- cycle;
\shade[left color=OliveGreen!90!white, right color=red!80!white!5, opacity=0.5] (9,0) -- (6,0) -- (6-3.75,5) --(9,5) -- cycle;
\node at (4.5,6.5) {$t>t_\ast$};
\draw[rounded corners=2pt](3, 6) rectangle (6,7);
\end{tikzpicture}
\end{center}
\caption{Schematic plot showing the lightcones of two operators, initially supported solely on $R$ and $S$ respectively. There exists a threshold timescale proportional to the separation between $R$ and $S$, $t_\ast\propto L_E$, such that the lightcones do not overlap for $t<t_\ast$ and do so for $t>t_\ast$.}
\label{fig:lightcone-schem}
\end{figure}
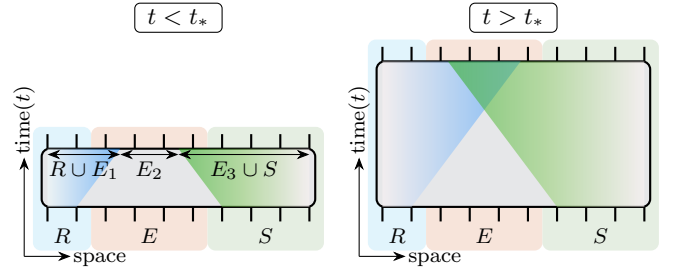

While the PPE and its moments have been defined above in all generality, let us now consider a system with spatial structure undergoing unitary dynamics governed by a strictly local circuit\footnote{By a strictly local circuit, we mean that all the gates in the circuit are of finite $O(1)$ range.}. In particular, let us consider a one-dimensional chain of qubits (spins-1/2) with the three subsystems as shown in Fig.~\ref{fig:tripartition}.
We will also denote by $D_X = 2^{L_X}$ the Hilbert-space dimension of subsystem $X$ with $X=R,E,S$.

In this setting, it is important that the traced out subsystem $E$ spatially separates the subsystem $R$ for which the PPE is defined from the projected subsystem $S$ which induces the PPE. The spatial separation between $R$ and $S$ corresponding to the length $L_E$, will be tuned, as discussed shortly, to probe the spatial structure of information spreading.

The state at any time $t$ is given by the action of a unitary operator $U(t)$ acting on an initial state which is a direct product state of each of the qubits,
\eq{
\ket{\psi_{RSE}(t)} =  U(t)\ket{\psi_{RSE}(0)}\,,
}
where 
\eq{
\ket{\psi_{RSE}(0)} = \otimes\smashoperator{\prod_{i\in R,S,E}}[A_{i\uparrow} \ket{\uparrow}_i + A_{i\downarrow}\ket{\downarrow}_i]\,,
\label{eq:psi-init}
}
with $|A_{i\uparrow}|^2+|A_{i\downarrow}|^2=1$ and $\ket{\uparrow/\downarrow}_i$ denotes the state at site $i$ polarised along the positive/negative $z$-direction. The time dependence in the state naturally endows the PPE with a time dependence. 

Within this setting, we will now argue why the moments of the PPE as a function of $L_E$ and $t$ may 
\new{contain information about} the spatiotemporal structure.
The first non-trivial moment, $k=2$, of any observable $O_R$ is given explicitly by
\eq{
\braket{O_R^{(2)}(t)}=\sum_{o_S}\frac{\braket{\psi_{RSE}(0)|\Pi_{o_S}(t)O_R(t)|\psi_{RSE}(0)}^2}{\braket{\psi_{RSE}(0)|\Pi_{o_S}(t)|\psi_{RSE}(0)}}\,.
\label{eq:OR-2}
}
The operators $\Pi_{o_S}$ and $O_R$, although initially supported solely on $S$ and $R$ respectively, grow with time. However, the strictly local nature of the circuit imposes a maximum velocity, due to the Lieb-Robinson bound, for the operator growth.
This naturally means that there exists a threshold time, $t_\ast\propto L_E$, such that for $t<t_\ast$, the support of the operators $\Pi_{o_S}(t)$ and $O_R(t)$ do not overlap whereas for $t>t_\ast$, they do. This is shown graphically in Fig.~\ref{fig:lightcone-schem}.

For $t<t_\ast$, the operator $O_R$ grows into a part of $E$ which we denote by $E_1$ and similarly $\Pi_{o_S}$ grows into a part of $E$ which label $E_3$, such that $L_{E_1}+L_{E_3}<L_E$, with the intervening region, denoted by $L_{E_2}$ of finite size, see left panel of Fig.~\ref{fig:lightcone-schem}. 
Since the initial state in Eq.~\ref{eq:psi-init} is a direct product state over all sites, it can be decomposed as $\ket{\psi_{RSE}(0)} = \ket{\psi_{R\cup E_1}}\otimes\ket{\psi_{E_2}}\otimes\ket{\psi_{E_3\cup S}}$, such that Eq.~\ref{eq:OR-2} can be written as
\eq{
\begin{split}
\braket{O_R^{(2)}(t)}=\langle\psi_{R\cup E_1}&|O_R(t)|\psi_{R\cup E_1}\rangle^2\times\\
&\sum_{o_S}\braket{\psi_{E_3\cup S}|\Pi_{o_S}(t)|\psi_{E_3\cup S}}
\end{split}\,,\nonumber\\
=\langle\psi_{R\cup E_1}&|O_R(t)|\psi_{R\cup E_1}\rangle^2\,,
\label{eq:OR-2-decomp}
}
where in the second line we used $\sum_{o_S}\Pi_{o_S} = \mathbb{I}$. Note that the result in Eq.~\ref{eq:OR-2-decomp} is simply the square of the first moment, $\braket{O_R^{(1)}(t)}^2$. 
This implies that for $t<t_\ast$, the fluctuations of the observable $O_R$ over the PPE are vanishingly small. 
On the other hand, for $t>t_\ast$, the expression in Eq.~\ref{eq:OR-2} will not factorise as in Eq.~\ref{eq:OR-2-decomp} and we will have $\braket{O_R^{(2)}}\neq \braket{O_R^{(1)}(t)}^2$. 
Since this is true for any operator $O_R$ supported in $R$, a natural measure of the fluctuations over the PPE is 
\eq{
\Delta(t,L_E) = \frac{1}{2}|| \rho_R^{(2)}(t,L_E) - \rho_R^{\otimes 2}(t,L_E)||^{\phantom\dagger}_1\,,
\label{eq:ppe-fluc}
}
where we have made the dependence on time, $t$, and the separation between $R$ and $S$, $L_E$, explicit. Much of the rest of the paper will focus on the fluctuations in the PPE as encoded by $\Delta$ in Eq.~\ref{eq:ppe-fluc}. 
In the following section, we will show rigorously for a generic class of unitary circuits that there exists a timescale $t_\ast(L_E) = v L_E$ such that $\Delta(t,L_E)=0$ for $t<t_\ast(L_E)$, where $v$ is the Lieb-Robinson velocity of the circuit. The behaviour for $t>t_\ast$ depends on the nature of the unitary circuit and reflects the lightcone of information spreading therein.

A complementary question of interest is if the PPE approaches an universal ensemble at late times, and if it does, what is the dynamics of the approach.
For ergodic systems without any conservation laws, in the limit of $t\to\infty$ and $L_S\gg L_R,L_E$ it is expected that the subsystem $R\cup E$ deep thermalises -- the projected ensemble induced on $R\cup E$ due to the measurements on $S$ is simply the Haar ensemble,
\eq{
{\cal E}_{\rm PE}\equiv \{p(o_S),~\ket{\psi_{R\cup E}(o_S)}\} = {\cal E}_{\rm Haar}\,.
\label{eq:PE-tinf-RE}
}
This result in turn implies that the PPE on $R$ is described by the so-called generalised Hilbert-Schmidt (gHS) ensemble~\cite{sommers2004statistical,zyczkowski2011generating},
\eq{
{\cal E}_{\rm PPE} \!=\! {\cal E}_{\rm{gHS}}\!\equiv\! \left\{\psi \!\sim\! {\rm Haar}(D_RD_E),{\rm Tr}_E \ket{\psi}\!\bra{\psi}\,\right\}\,.
\label{eq:gHS-ensemble}
}
The relevant quantity to study is then the distance of the PPE from the gHS ensemble as a function of time. 
This distance can be quantified by the norm of the difference of the $k^{\rm th}$ moments 
\eq{
\Delta_{\rm gHS}^{(k)}(t,L_E) = \frac{1}{2}||\rho_{R}^{(k)}(t,L_E) - \rho_{\rm gHS}^{(k)}(L_E)||_1\,.
\label{eq:Delta-gHS}
}
It was shown in Refs.~\cite{yu2025mixed,sherry2025do} that, for a certain specific system with a specific initial state and set of measurement operators, $\Delta_{\rm gHS}^{(k)}$ decays exponentially with increasing time $t$. 
However, this raises a question for generic ergodic systems, with the subsystem $R$ spatially separated from the measured subsystem $S$: how does $\Delta_{\rm gHS}^{(k)}(t,L_E)$ decay with time and how does the spatial separation $L_E$ between the two affect this decay?
We will address this question in Sec.~\ref{sec:gen-erg}. 

As an aside, while the gHS ensemble in Eq.~\ref{eq:gHS-ensemble} was constructed out of an Haar ensemble on $R\cup E$, for ergodic systems with conservation laws or for many-body localised systems, both of which satisfy the $k$-no-resonance condition, $R\cup E$ deep thermalises to a (generalised) Scrooge ensemble~\cite{mark2024maximum,chang2025deep,manna2025projected} and the corresponding gHS ensemble can be constructed by tracing out $E$ from states in the Scrooge ensemble.

\subsection{Probabilities of (bit-string) probabilities \label{sec:pop}}
Given the PPE, defined in Eq.~\ref{eq:ppe-def}, we will also study the probabilities of bit-string probabilities (PoP) over the ensemble. 
These probabilities naturally appear from the PE and provide an alternative probe of ergodicity, with direct connection to quantum information and quantum benchmarking.
The PoP can be thought of as the probability distribution of the probability of obtaining a particular bit-string denoted by $z=\{1,0,\cdots,1\}$ upon a global measurement of Pauli-$Z$ operators on all the sites of the system.
For an ensemble, such as the one defined in Eq.~\ref{eq:ppe-def}, we define an averaged PoP for the {\it conditional} probability of a bit-string in $R$, say $z_R$, as 
\eq{
{\rm PoP_{PPE}}(\tilde{p},z_R) =\sum_{o_S}p(o_S)\delta[\tilde{p}-\tilde{p}(z_R|o_S)]\,,
\label{eq:pop-ppe}
}
where the relative probability $\tilde{p}(z_R|o_S)$ is given by
\eq{
\tilde{p}(z_R|o_S) = \frac{p(z_R|o_S)}{\sum_{o_S'}p(o_S')p(z_R|o_S')}\,,
\label{eq:rel-prob}
}
with
\eq{
p(z_R|o_S) = \braket{z_R|\rho_R(o_S)|z_R}\,.
}done
Studying the PoP of the relative probability $\tilde{p}(z_R|o_S)$ ensures that the non-universal fluctuations in the mean $\tilde{p}(z_R|o_S)$ over different $z_R$ do not contaminate the results.
Note that the normalisation in Eq.~\ref{eq:rel-prob} is simply $\sum_{o_S'}p(o_S')p(z_R|o_S')=p(z_R)$, which is the probability of measuring $z_R$ in the original state.

The PoPs defined above have emerged as an important and powerful tool, both theoretically and experimentally, to characterise the ergodicity of states beyond the conventional notions of local thermalisation.
For example, they are key towards understanding the emergence of quantum designs and higher-order Hilbert space ergodicity, and to distinguish genuine (deep) thermalisation from decoherence therefore serving as resources for benchmarking of quantum devices~\cite{boixo2018characterizing,arute2019quantum,mark2023benchmarking,choi2023preparing,shaw2025experimental}.
For ensembles of random states, such as those which form a design, the PoPs show universal Porter-Thomas (PT) distributions 
\eq{
P_{\rm PT}(\tilde{p}) = \exp[-\tilde{p}]\,.
\label{eq:PT-dist}
}
On the other hand, it was recently argued that for mixed state ensembles, obtained by tracing out subsystems from design-forming pure state ensembles, the relevant universal distribution is the {\it Erlang distribution}~\cite{shaw2025experimental},
\eq{
P_{\rm Erlang}(\tilde{p},D_E)  = \frac{D_E^{D_E}}{(D_E-1)!}\exp(-D_E\tilde{p})\tilde{p}^{D_E-1}\,,
}
where $D_E$ is the Hilbert-space dimension of the traced out subsystem.

The connection with ensembles raises the questions of the fate of the PoPs in our setting and the implications of the spatiotemporal structure of information spreading in local models on them.
We will address this question and show that the PoPs are intimately related to the spatiotemporal structure of information spreading and in fact, carry universal signatures of the latter. 

While the PoPs above were defined for fixed bit-strings over an ensemble of states, one can also define a PoP over the set of bit-strings, $\{z\}$ given a fixed state, say $\rho$.
This variant of the PoP is defined as 
\eq{
{\rm PoP_{b{\text{-}}str}}(\tilde{p},\rho) = D^{-1}\sum_z \delta[\tilde{p}-D p(z,\rho)]\,,
\label{eq:pop-bit-strings}
}
where $p(z,\rho) = \braket{z|\rho|z}$ is the probability of getting the obtaining the bit-string $z$ in the state $\rho$, and the factor of $D$, the Hilbert-space dimension of $\rho$, is the normalisation of the relative probability. In the following, we will show how the PoP highlights features of the spatial structure of the information in the state.

\section{Model and generalities \label{sec:model-gen}}

\subsection{Floquet brickwork circuits \label{sec:brickwork}}
We will consider Floquet unitary circuits which can be expressed in a brickwork form. 
For the circuits we employ, the time-evolution operator over one period, i.e. the Floquet unitary, can in general be expressed as 
\eq{
U_F = \left[\prod_{i=1}^{L/2-1} u_{2i,2i+1}\right]\left[\prod_{i=1}^{L/2} u_{2i-1,2i}\right]\,,
\label{eq:UF-gen}
}
where $u_{i,j}$ is a two-site unitary gate acting on the pair of sites $(i,j)$.
Before considering specific models, we argue for a lightcone bound for the fluctuations in the PPE very generally simply by exploiting the brickwork geometry of the circuit. 
The Floquet unitary in Eq.~\ref{eq:UF-gen} can be represented graphically in the brickwork form as 
\eq{\label{eq:brickwork}
U_F=\cbox{
\begin{tikzpicture}[scale=0.8]
\def\s{0.2}
\foreach \x in {0,...,6}{
    \draw[black] (\x-0.35,-0.35)--(\x+0.35,0.35);
    \draw[black] (\x-0.35,0.35)--(\x+0.35,-0.35);
    \draw[black] (\x-0.35+0.5,-0.35+0.5)--(\x+0.35+0.5,0.35+0.5);
    \draw[black] (\x-0.35+0.5,0.35+0.5)--(\x+0.35+0.5,-0.35+0.5);
}
\def\x{7}
\draw[black] (\x-0.35,-0.35)--(\x+0.35,0.35);
\draw[black] (\x-0.35,0.35)--(\x+0.35,-0.35);
\foreach \x in {0,...,6}{
    \draw[rounded corners=2pt, fill=NavyBlue!30] (\x-\s,-\s) rectangle (\x+\s,\s); 
    \draw[black] (\x+\s-0.15,\s-0.05) -- (\x+\s-0.05,\s-0.05) -- (\x+\s-0.05,\s-0.15);
    \draw[rounded corners=2pt, fill=NavyBlue!30] (\x+0.5-\s,-\s+0.5) rectangle (\x+0.5+\s,\s+0.5); 
    \draw[black] (0.5+\x+\s-0.15,0.5+\s-0.05) -- (0.5+\x+\s-0.05,0.5+\s-0.05) -- (0.5+\x+\s-0.05,0.5+\s-0.15);
}
\def\x{7}
\draw[rounded corners=2pt, fill=NavyBlue!30] (\x-\s,-\s) rectangle (\x+\s,\s); 
\draw[black] (\x+\s-0.15,\s-0.05) -- (\x+\s-0.05,\s-0.05) -- (\x+\s-0.05,\s-0.15);
\end{tikzpicture}}\,,
}
where the blue squares denote the $u$-gates in Eq.~\ref{eq:UF-gen}.
It will be useful to recall the so-called folded picture where a unitary $u$ and its hermitian conjugate $u^\dagger$ are together represented as
\eq{\nonumber
\cbox{\begin{tikzpicture}[scale=0.8]
\def\x{0}
\def\s{0.2}
\draw[black] (\x-0.35,-0.35)--(\x+0.35,0.35);
\draw[black] (\x-0.35,0.35)--(\x+0.35,-0.35);
\draw[rounded corners=2pt, fill=NavyBlue!60] (\x-\s,-\s) rectangle (\x+\s,\s); 
\draw[black] (\x+\s-0.15,-\s+0.05) -- (\x+\s-0.05,-\s+0.05) -- (\x+\s-0.05,-\s+0.15);
\def\sx{-0.17}
\def\sy{-0.06}
\draw[black] (\x-0.35+\sx,-0.35+\sy)--(\x+0.35+\sx,0.35+\sy);
\draw[black] (\x-0.35+\sx,0.35+\sy)--(\x+0.35+\sx,-0.35+\sy);
\draw[rounded corners=2pt, fill=NavyBlue!30] (\x-\s+\sx,-\s+\sy) rectangle (\x+\s+\sx,\s+\sy); 
\draw[black] (\x+\s-0.15+\sx,\s-0.05+\sy) -- (\x+\s-0.05+\sx,\s-0.05+\sy) -- (\x+\s-0.05+\sx,\s-0.15+\sy);
\end{tikzpicture}}=
\cbox{\begin{tikzpicture}[scale=0.8]
\def\x{0}
\def\s{0.2}
\draw[black,thick] (\x-0.35,-0.35)--(\x+0.35,0.35);
\draw[black, thick] (\x-0.35,0.35)--(\x+0.35,-0.35);
\draw[rounded corners=2pt, fill=NavyBlue, thick] (\x-\s,-\s) rectangle (\x+\s,\s); 
\draw[white!5] (\x+\s-0.15,\s-0.05) -- (\x+\s-0.05,\s-0.05) -- (\x+\s-0.05,\s-0.15);
\end{tikzpicture}}\,,
}
where the thick lines in the gates in the RHS of the above equations denote that they now carry two indices. Unitarity implies $uu^\dagger=\mathbb{I}=u^\dagger u$ which is diagrammatically represented as 
\eq{
\cbox{\begin{tikzpicture}[scale=0.8]
\def\x{0}
\def\s{0.2}
\draw[black] (\x-0.35,-0.35)--(\x+0.35,0.35);
\draw[black] (\x-0.35,0.35)--(\x+0.35,-0.35);
\draw[rounded corners=2pt, fill=NavyBlue!60] (\x-\s,-\s) rectangle (\x+\s,\s); 
\draw[black] (\x+\s-0.15,-\s+0.05) -- (\x+\s-0.05,-\s+0.05) -- (\x+\s-0.05,-\s+0.15);
\def\sx{-0.17}
\def\sy{-0.06}
\draw[black] (\x-0.35+\sx,-0.35+\sy)--(\x+0.35+\sx,0.35+\sy);
\draw[black] (\x-0.35+\sx,0.35+\sy)--(\x+0.35+\sx,-0.35+\sy);
\draw[rounded corners=2pt, fill=NavyBlue!30] (\x-\s+\sx,-\s+\sy) rectangle (\x+\s+\sx,\s+\sy); 
\draw[black] (\x+\s-0.15+\sx,\s-0.05+\sy) -- (\x+\s-0.05+\sx,\s-0.05+\sy) -- (\x+\s-0.05+\sx,\s-0.15+\sy);
\draw[black] (\x-0.35,0.35) to[out=135,in=135] (\x-0.35+\sx,0.35+\sy);
\draw[black] (\x+0.35,0.35) to[out=45,in=45] (\x+0.35+\sx,0.35+\sy);
\end{tikzpicture}}=
\cbox{\begin{tikzpicture}[scale=0.8]
\def\x{0}
\def\s{0.2}
\draw[black,thick] (\x-0.35,-0.35)--(\x+0.35,0.35);
\draw[black, thick] (\x-0.35,0.35)--(\x+0.35,-0.35);
\draw[rounded corners=2pt, fill=NavyBlue, thick] (\x-\s,-\s) rectangle (\x+\s,\s); 
\draw[white!5] (\x+\s-0.15,\s-0.05) -- (\x+\s-0.05,\s-0.05) -- (\x+\s-0.05,\s-0.15);
\draw[fill=black,thick, fill=white] (\x-0.35,0.35) circle (0.07);
\draw[fill=black,thick, fill=white] (\x+0.35,0.35) circle (0.07);
\end{tikzpicture}}=
\cbox{\begin{tikzpicture}[scale=0.8]
\draw[thick,black](-0.2,-0.2)--(-0.2,0.2);
\draw[thick,black](0.2,-0.2)--(0.2,0.2);
\draw[fill=black,thick, fill=white] (-0.2,0.2) circle (0.07);
\draw[fill=black,thick, fill=white] (0.2,0.2) circle (0.07);
\end{tikzpicture}}\,;~~
\cbox{\begin{tikzpicture}[scale=0.8]
\def\x{0}
\def\s{0.2}
\draw[black] (\x-0.35,-0.35)--(\x+0.35,0.35);
\draw[black] (\x-0.35,0.35)--(\x+0.35,-0.35);
\draw[rounded corners=2pt, fill=NavyBlue!60] (\x-\s,-\s) rectangle (\x+\s,\s); 
\draw[black] (\x+\s-0.15,-\s+0.05) -- (\x+\s-0.05,-\s+0.05) -- (\x+\s-0.05,-\s+0.15);
\def\sx{-0.17}
\def\sy{-0.06}
\draw[black] (\x-0.35+\sx,-0.35+\sy)--(\x+0.35+\sx,0.35+\sy);
\draw[black] (\x-0.35+\sx,0.35+\sy)--(\x+0.35+\sx,-0.35+\sy);
\draw[rounded corners=2pt, fill=NavyBlue!30] (\x-\s+\sx,-\s+\sy) rectangle (\x+\s+\sx,\s+\sy); 
\draw[black] (\x+\s-0.15+\sx,\s-0.05+\sy) -- (\x+\s-0.05+\sx,\s-0.05+\sy) -- (\x+\s-0.05+\sx,\s-0.15+\sy);
\draw[black] (\x-0.35,-0.35) to[out=-135,in=-135] (\x-0.35+\sx,-0.35+\sy);
\draw[black] (\x+0.35,-0.35) to[out=-45,in=-45] (\x+0.35+\sx,-0.35+\sy);
\end{tikzpicture}}=
\cbox{\begin{tikzpicture}[scale=0.8]
\def\x{0}
\def\s{0.2}
\draw[black,thick] (\x-0.35,-0.35)--(\x+0.35,0.35);
\draw[black, thick] (\x-0.35,0.35)--(\x+0.35,-0.35);
\draw[rounded corners=2pt, fill=NavyBlue, thick] (\x-\s,-\s) rectangle (\x+\s,\s); 
\draw[white!5] (\x+\s-0.15,\s-0.05) -- (\x+\s-0.05,\s-0.05) -- (\x+\s-0.05,\s-0.15);
\draw[fill=black,thick, fill=white] (\x-0.35,-0.35) circle (0.07);
\draw[fill=black,thick, fill=white] (\x+0.35,-0.35) circle (0.07);
\end{tikzpicture}}
=
\cbox{\begin{tikzpicture}[scale=0.8]
\draw[thick,black](-0.2,-0.2)--(-0.2,0.2);
\draw[thick,black](0.2,-0.2)--(0.2,0.2);
\draw[fill=black,thick, fill=white] (-0.2,-0.2) circle (0.07);
\draw[fill=black,thick, fill=white] (0.2,-0.2) circle (0.07);
\end{tikzpicture}}\,.
\label{eq:diag-id}
}

\subsection{Lightcone-imposed bounds on PPE fluctuations \label{sec:ppe-lc}}

An immediate consequence of the brickwork geometry, made explicit by the diagrammatic notation above, is a bound for the $\Delta(t,L_E)$ lightcone. 
To see this consider, for example $L_R=1$, $L_E=8$ and $L_S=7$. Consistent with Fig.~\ref{fig:tripartition}, we will denote the three subsystems with blue, red, and green shades in the following.
The reduced density matrix of $R\cup S$ upon tracing out $E$, for instance for $t=2$, is given in the diagrammatic notation by
\eq{\rho_{RS}(t=2)=
\cbox{\begin{tikzpicture}[scale=0.8,thick]
\fill[fill=Cyan!10] (-0.5,-0.5) rectangle (0,1.75); 
\fill[fill=BrickRed!10] (0,-0.5) rectangle (4,1.75); 
\fill[fill=OliveGreen!10] (4,-0.5) rectangle (7.5,1.75); 
\def\s{0.2}
\foreach \y in {0,1}{
\foreach \x in {0,...,6}{
    \draw[black,thick] (\x-0.35,\y-0.35)--(\x+0.35,\y+0.35);
    \draw[black] (\x-0.35,\y+0.35)--(\x+0.35,-0.35+\y);
    \draw[black] (\x-0.35+0.5,-0.35+0.5+\y)--(\x+0.35+0.5,0.35+0.5+\y);
    \draw[black] (\x-0.35+0.5,0.35+0.5+\y)--(\x+0.35+0.5,-0.35+0.5+\y);
}
\def\x{7}
\draw[black] (\x-0.35,-0.35+\y)--(\x+0.35,0.35+\y);
\draw[black] (\x-0.35,0.35+\y)--(\x+0.35,-0.35+\y);
}
\foreach \y in {0,1}{
\foreach \x in {0,...,6}{
    \draw[rounded corners=2pt, fill=NavyBlue] (\x-\s,-\s+\y) rectangle (\x+\s,\s+\y); 
    \draw[white,thin] (\x+\s-0.15,\s-0.05+\y) -- (\x+\s-0.05,\s-0.05+\y) -- (\x+\s-0.05,\s-0.15+\y);
    \draw[rounded corners=2pt, fill=NavyBlue] (\x+0.5-\s,-\s+0.5+\y) rectangle (\x+0.5+\s,\s+0.5+\y); 
    \draw[white,thin] (0.5+\x+\s-0.15,0.5+\s-0.05+\y) -- (0.5+\x+\s-0.05,0.5+\s-0.05+\y) -- (0.5+\x+\s-0.05,0.5+\s-0.15+\y);
}
\def\x{7}
\draw[rounded corners=2pt, fill=NavyBlue] (\x-\s,-\s+\y) rectangle (\x+\s,\s+\y); 
\draw[white, thin] (\x+\s-0.15,\s-0.05+\y) -- (\x+\s-0.05,\s-0.05+\y) -- (\x+\s-0.05,\s-0.15+\y);
}
\draw[black](-0.35,0.35)--(-0.35,1-0.35);
\draw[black](7.35,0.35)--(7.35,1-0.35);
\draw[black](-0.35,1.35)--(-0.35,1.85);
\draw[black](7.35,1.35)--(7.35,1.85);
\foreach \x in {0,...,7}{
    \draw[fill=Black] (\x-0.35-0.1,-0.35-0.1) rectangle (\x-0.35,-0.35);
    \draw[fill=Black] (\x+0.35,-0.35-0.1) rectangle (\x+0.35+0.1,-0.35);
}
\foreach \x in {0,1,2,3}{
    \draw[fill=black,thick, fill=white] (\x+0.5-0.35,1.85) circle (0.07);
    \draw[fill=black,thick, fill=white] (\x+0.5+0.35,1.85) circle (0.07);
}
\end{tikzpicture}}\,.\nonumber
}
Using the unitary condition from Eq.~\ref{eq:diag-id}, this diagram can be reduced to
\eq{\rho_{RS}(t=2)=
\cbox{\begin{tikzpicture}[scale=0.8,thick]
\fill[fill=Cyan!10] (-0.5,-0.5) rectangle (0,1.75); 
\fill[fill=BrickRed!10] (0,-0.5) rectangle (4,1.75); 
\fill[fill=OliveGreen!10] (4,-0.5) rectangle (7.5,1.75); 
\def\s{0.2}
\def\y{0}
\foreach \x in {0,1,3,4,5,6,7}{
    \draw[black,thick] (\x-0.35,\y-0.35)--(\x+0.35,\y+0.35);
    \draw[black] (\x-0.35,\y+0.35)--(\x+0.35,-0.35+\y);
}
\foreach \x in {0,3,4,5,6}{
    \draw[black] (\x-0.35+0.5,-0.35+0.5+\y)--(\x+0.35+0.5,0.35+0.5+\y);
    \draw[black] (\x-0.35+0.5,0.35+0.5+\y)--(\x+0.35+0.5,-0.35+0.5+\y);
}
\def\y{1}
\foreach \x in {0,4,5,6,7}{
    \draw[black,thick] (\x-0.35,\y-0.35)--(\x+0.35,\y+0.35);
    \draw[black] (\x-0.35,\y+0.35)--(\x+0.35,-0.35+\y);
}
\foreach \x in {4,5,6}{
    \draw[black] (\x-0.35+0.5,-0.35+0.5+\y)--(\x+0.35+0.5,0.35+0.5+\y);
    \draw[black] (\x-0.35+0.5,0.35+0.5+\y)--(\x+0.35+0.5,-0.35+0.5+\y);
}
\def\y{0}
\foreach \x in {0,1,3,4,5,6,7}{
    \draw[rounded corners=2pt, fill=NavyBlue] (\x-\s,-\s+\y) rectangle (\x+\s,\s+\y); 
    \draw[white,thin] (\x+\s-0.15,\s-0.05+\y) -- (\x+\s-0.05,\s-0.05+\y) -- (\x+\s-0.05,\s-0.15+\y);
}
\foreach \x in {0,3,4,5,6}{
    \draw[rounded corners=2pt, fill=NavyBlue] (\x+0.5-\s,-\s+0.5+\y) rectangle (\x+0.5+\s,\s+0.5+\y); 
    \draw[white,thin] (0.5+\x+\s-0.15,0.5+\s-0.05+\y) -- (0.5+\x+\s-0.05,0.5+\s-0.05+\y) -- (0.5+\x+\s-0.05,0.5+\s-0.15+\y);
}
\def\y{1}
\foreach \x in {0,4,5,6,7}{
    \draw[rounded corners=2pt, fill=NavyBlue] (\x-\s,-\s+\y) rectangle (\x+\s,\s+\y); 
    \draw[white,thin] (\x+\s-0.15,\s-0.05+\y) -- (\x+\s-0.05,\s-0.05+\y) -- (\x+\s-0.05,\s-0.15+\y);
}
\foreach \x in {4,5,6}{
    \draw[rounded corners=2pt, fill=NavyBlue] (\x+0.5-\s,-\s+0.5+\y) rectangle (\x+0.5+\s,\s+0.5+\y); 
    \draw[white,thin] (0.5+\x+\s-0.15,0.5+\s-0.05+\y) -- (0.5+\x+\s-0.05,0.5+\s-0.05+\y) -- (0.5+\x+\s-0.05,0.5+\s-0.15+\y);
}
\draw[black](-0.35,0.35)--(-0.35,1-0.35);
\draw[black](7.35,0.35)--(7.35,1-0.35);
\draw[black](-0.35,1.35)--(-0.35,1.85);
\draw[black](7.35,1.35)--(7.35,1.85);
\foreach \x in {0,...,7}{
    \draw[fill=Black] (\x-0.35-0.1,-0.35-0.1) rectangle (\x-0.35,-0.35);
    \draw[fill=Black] (\x+0.35,-0.35-0.1) rectangle (\x+0.35+0.1,-0.35);
}
\draw(2-0.35,-0.35)--(2-0.2,-0.2);
\draw(2+0.35,-0.35)--(2+0.2,-0.2);
\def\y{-0.5}
\def\x{1.5}
\draw[fill=black,thick, fill=white] (\x+0.35,\y+0.35) circle (0.07);
\def\x{2.5}
\draw[fill=black,thick, fill=white] (\x-0.35,\y+0.35) circle (0.07);
\def\y{0}
\def\x{1}
\draw[fill=black,thick, fill=white] (\x+0.35,\y+0.35) circle (0.07);
\def\x{3}
\draw[fill=black,thick, fill=white] (\x-0.35,\y+0.35) circle (0.07);
\def\y{0.5}
\def\x{0.5}
\draw[fill=black,thick, fill=white] (\x+0.35,\y+0.35) circle (0.07);
\def\x{3.5}
\draw[fill=black,thick, fill=white] (\x-0.35,\y+0.35) circle (0.07);
\def\y{1}
\def\x{0}
\draw[fill=black,thick, fill=white] (\x+0.35,\y+0.35) circle (0.07);
\def\x{4}
\draw[fill=black,thick, fill=white] (\x-0.35,\y+0.35) circle (0.07);
\end{tikzpicture}}\,.\nonumber
}

The key point to notice in the above equation is that the states of $R$ and $S$ are decoupled from each other, $\rho_{RS}(t)=\rho_{R}(t)\otimes\rho_S(t)$. This decoupling directly implies $\rho_R(o_S)=\rho_R$ such that $\Delta=0$ identically.
On the other hand, for $t=3$, the states of $R$ and $S$ do not decouple 
\eq{\rho_{RS}(t=3)=
\cbox{\begin{tikzpicture}[scale=0.8,thick]
\fill[fill=Cyan!10] (-0.5,-0.5) rectangle (0,2.75); 
\fill[fill=BrickRed!10] (0,-0.5) rectangle (4,2.75); 
\fill[fill=OliveGreen!10] (4,-0.5) rectangle (7.5,2.75); 
\def\s{0.2}
\def\y{0}
\foreach \x in {0,1,2,3,4,5,6,7}{
    \draw[black,thick] (\x-0.35,\y-0.35)--(\x+0.35,\y+0.35);
    \draw[black] (\x-0.35,\y+0.35)--(\x+0.35,-0.35+\y);
}
\foreach \x in {0,1,2,3,4,5,6}{
    \draw[black] (\x-0.35+0.5,-0.35+0.5+\y)--(\x+0.35+0.5,0.35+0.5+\y);
    \draw[black] (\x-0.35+0.5,0.35+0.5+\y)--(\x+0.35+0.5,-0.35+0.5+\y);
}
\def\y{1}
\foreach \x in {0,1,3,4,5,6,7}{
    \draw[black,thick] (\x-0.35,\y-0.35)--(\x+0.35,\y+0.35);
    \draw[black] (\x-0.35,\y+0.35)--(\x+0.35,-0.35+\y);
}
\foreach \x in {0,3,4,5,6}{
    \draw[black] (\x-0.35+0.5,-0.35+0.5+\y)--(\x+0.35+0.5,0.35+0.5+\y);
    \draw[black] (\x-0.35+0.5,0.35+0.5+\y)--(\x+0.35+0.5,-0.35+0.5+\y);
}
\def\y{2}
\foreach \x in {0,4,5,6,7}{
    \draw[black,thick] (\x-0.35,\y-0.35)--(\x+0.35,\y+0.35);
    \draw[black] (\x-0.35,\y+0.35)--(\x+0.35,-0.35+\y);
}
\foreach \x in {4,5,6}{
    \draw[black] (\x-0.35+0.5,-0.35+0.5+\y)--(\x+0.35+0.5,0.35+0.5+\y);
    \draw[black] (\x-0.35+0.5,0.35+0.5+\y)--(\x+0.35+0.5,-0.35+0.5+\y);
}
\def\y{0}
\foreach \x in {0,1,2,3,4,5,6,7}{
    \draw[rounded corners=2pt, fill=NavyBlue] (\x-\s,-\s+\y) rectangle (\x+\s,\s+\y); 
    \draw[white,thin] (\x+\s-0.15,\s-0.05+\y) -- (\x+\s-0.05,\s-0.05+\y) -- (\x+\s-0.05,\s-0.15+\y);
}
\foreach \x in {0,1,2,3,4,5,6}{
    \draw[rounded corners=2pt, fill=NavyBlue] (\x+0.5-\s,-\s+0.5+\y) rectangle (\x+0.5+\s,\s+0.5+\y); 
    \draw[white,thin] (0.5+\x+\s-0.15,0.5+\s-0.05+\y) -- (0.5+\x+\s-0.05,0.5+\s-0.05+\y) -- (0.5+\x+\s-0.05,0.5+\s-0.15+\y);
}
\def\y{1}
\foreach \x in {0,1,3,4,5,6,7}{
    \draw[rounded corners=2pt, fill=NavyBlue] (\x-\s,-\s+\y) rectangle (\x+\s,\s+\y); 
    \draw[white,thin] (\x+\s-0.15,\s-0.05+\y) -- (\x+\s-0.05,\s-0.05+\y) -- (\x+\s-0.05,\s-0.15+\y);
}
\foreach \x in {0,3,4,5,6}{
    \draw[rounded corners=2pt, fill=NavyBlue] (\x+0.5-\s,-\s+0.5+\y) rectangle (\x+0.5+\s,\s+0.5+\y); 
    \draw[white,thin] (0.5+\x+\s-0.15,0.5+\s-0.05+\y) -- (0.5+\x+\s-0.05,0.5+\s-0.05+\y) -- (0.5+\x+\s-0.05,0.5+\s-0.15+\y);
}
\def\y{2}
\foreach \x in {0,4,5,6,7}{
    \draw[rounded corners=2pt, fill=NavyBlue] (\x-\s,-\s+\y) rectangle (\x+\s,\s+\y); 
    \draw[white,thin] (\x+\s-0.15,\s-0.05+\y) -- (\x+\s-0.05,\s-0.05+\y) -- (\x+\s-0.05,\s-0.15+\y);
}
\foreach \x in {4,5,6}{
    \draw[rounded corners=2pt, fill=NavyBlue] (\x+0.5-\s,-\s+0.5+\y) rectangle (\x+0.5+\s,\s+0.5+\y); 
    \draw[white,thin] (0.5+\x+\s-0.15,0.5+\s-0.05+\y) -- (0.5+\x+\s-0.05,0.5+\s-0.05+\y) -- (0.5+\x+\s-0.05,0.5+\s-0.15+\y);
}
\draw[black](-0.35,0.35)--(-0.35,1-0.35);
\draw[black](7.35,0.35)--(7.35,1-0.35);
\draw[black](-0.35,1.35)--(-0.35,2-0.35);
\draw[black](7.35,1.35)--(7.35,2-0.35);
\draw[black](-0.35,2+0.35)--(-0.35,2.85);
\draw[black](7.35,2+0.35)--(7.35,2.85);
\foreach \x in {0,...,7}{
    \draw[fill=Black] (\x-0.35-0.1,-0.35-0.1) rectangle (\x-0.35,-0.35);
    \draw[fill=Black] (\x+0.35,-0.35-0.1) rectangle (\x+0.35+0.1,-0.35);
}
\def\y{0.5}
\def\x{1.5}
\draw[fill=black,thick, fill=white] (\x+0.35,\y+0.35) circle (0.07);
\def\x{2.5}
\draw[fill=black,thick, fill=white] (\x-0.35,\y+0.35) circle (0.07);
\def\y{1}
\def\x{1}
\draw[fill=black,thick, fill=white] (\x+0.35,\y+0.35) circle (0.07);
\def\x{3}
\draw[fill=black,thick, fill=white] (\x-0.35,\y+0.35) circle (0.07);
\def\y{1.5}
\def\x{0.5}
\draw[fill=black,thick, fill=white] (\x+0.35,\y+0.35) circle (0.07);
\def\x{3.5}
\draw[fill=black,thick, fill=white] (\x-0.35,\y+0.35) circle (0.07);
\def\y{2}
\def\x{0}
\draw[fill=black,thick, fill=white] (\x+0.35,\y+0.35) circle (0.07);
\def\x{4}
\draw[fill=black,thick, fill=white] (\x-0.35,\y+0.35) circle (0.07);
\end{tikzpicture}}\,,\nonumber
}
which naturally implies that $\Delta$ does not necessarily vanish.
A geometric interpretation of the above is that a wedge of contractions with the lightcone velocity develops starting from the boundaries between $R$ and $E$, and between $E$ and $S$, from the top of the circuit and grows downwards.
At early times, the wedge goes through the entire depth of the circuit thereby decoupling $R$ and $S$ whereas at late times the wedge does not separate $R$ and $S$ leading to a finite $\Delta$.
Analysing the equations above for arbitrary $L_E$ and $t$, it can be shown that the states of $R$ and $S$ decouple if $L_E > 4t-3$ which gives us a lightcone induced bound for $\Delta$ as,
\eq{
\Delta(t,L_E) = 0\,;~\quad t<\lfloor(L_E+3)/4\rfloor\,.
\label{eq:delta-bound}
}

\subsection{Lightcone-imposed features of PoPs \label{sec:pop-lc}}

The brickwork geometry of the circuit which leads to the bound in Eq.~\ref{eq:delta-bound} also mandates some universal features of the PoPs (defined in Sec.~\ref{sec:pop}). 
In particular for $t<t_\ast$, since $\rho_{RS}(t) = \rho_R(t)\otimes\rho_S(t)$, the probability $\tilde{p}(z_R|o_S) = \braket{z_R|\rho_R|z_R}$ is independent of $o_S$ and we hence find that $p(z_R|o_S) = p(z_R)$.
This naturally implies $\tilde{p}(z_R|o_S)=\tilde{p}(z_R)= 1,~\forall z_R$.
We obtain the `universal' result that for $t<t_\ast$,
\eq{
{\rm PoP_{PPE}}(\tilde{p},z_R) = \delta(\tilde{p}-1)\,,~\forall z_R\,.
}
Therefore, the collapse of ${\rm PoP_{PPE}}$ to a Dirac-delta function or deviations directly carries information of the lightcone velocity. 

In fact, the factorisation of $\rho_{RS}$ for $t<t_\ast$ also leads to an interesting result for the PoP over the bit-strings.
Consider the PoPs over the bit-strings in the reduced states $\rho_R$, $\rho_S$ and $\rho_{RS}$, defined via Eq.~\ref{eq:pop-bit-strings}.
The last one among the three is given by
\eq{
{\rm PoP_{b{\text{-}}str}}(\tilde{p},\rho_{RS}) = \frac{\sum_{z_Rz_S}\delta(\tilde{p}-D_RD_Sp(z_Rz_S))}{D_R D_S}\,.
\label{eq:pop-bstr-rs}
}
However, for $t<t_\ast$, the factorisation of $\rho_{RS}$ into $\rho_R \otimes \rho_S$ also means that the probability $p(z_Rz_S) = \braket{z_Rz_S|\rho_{RS}|z_Rz_S}$ factorises into $p(z_R)\times p(z_S)$.
Using this factorisation in Eq.~\ref{eq:pop-bstr-rs}, we have
\eq{
{\rm PoP_{b{\text{-}}str}}&(\tilde{p},\rho_{RS}) =\nonumber\\ &\int dp'~\frac{1}{p'}{\rm PoP_{b{\text{-}}str}}(p',\rho_{R}){\rm PoP_{b{\text{-}}str}}\left(\frac{\tilde{p}}{p'},\rho_{S}\right)\,,
\label{eq:mellin-integ}
}
which shows that for $t<t_\ast$, the ${\rm PoP_{b{\text{-}}str}}$ over bit-strings in $R\cup S$ is just the Mellin convolution, 
\eq{{\rm PoP_{b{\text{-}}str}}(\tilde{p},\rho_{RS})={\rm PoP_{b{\text{-}}str}}(\tilde{p},\rho_{R})\ast_{\rm M}{\rm PoP_{b{\text{-}}str}}(\tilde{p},\rho_{S})\,,\label{eq:mellin}}
of the ${\rm PoP_{b{\text{-}}str}}$ over the bit-strings in $R$ and the ${\rm PoP_{b{\text{-}}str}}$ over the bit-strings in $S$.
Deviations of ${\rm PoP_{b{\text{-}}str}}$ over $R\cup S$ from the Mellin convolution of the PoPs in $R$ and $S$ therefore suggest that $\rho_{RS}$ does not factorise and hence that ${\rm PoP_{b{\text{-}}str}}$ can also be used to the probe the lightcone of information spreading.

\section{Ergodic circuits \label{sec:gen-erg}}
Having established the generalities, we now present explicit results for ergodic circuits in this section which put the aforementioned ideas on concrete footing.
In  particular, we consider a non-integrable kicked Ising (KI) chain described by the Floquet unitary
\eq{
\begin{split}
U_F^{\rm KI} &= U_X U_Z~~{\rm with}\\
U_X&=\exp\left[-ig\sum_{j=1}^L X_j\right]\,,\\
U_Z&=\exp\left[-i\left(\sum_{j=1}^L h_jZ_j +J\sum_{j=1}^{L-1}Z_jZ_{j+1}\right) \right]\,,\\
\end{split}
\label{eq:UF-KI}
}
where $X_j (Z_j)$ denotes the Pauli-$X(Z)$ operator acting on site $j$. 
In Sec.~\ref{sec:ftfi} we consider the model in Eq.~\ref{eq:UF-KI} with generic parameters for which it is ergodic.
In addition, we also consider the model at a fine-tuned self-dual point (in Sec.~\ref{sec:sdki}) which allows for exact computations in the limit of $L_S\to\infty$.
Since in both cases the model is ergodic and devoid of any conservation laws, the PPE is expected to approach the gHSE at late times, as discussed in Sec.~\ref{sec:spatiotemporal}; we present pertinent results in Sec.~\ref{sec:latetimes}.

\subsection{Ergodic kicked Ising chain \label{sec:ftfi}}
\begin{figure}
\includegraphics[width=\linewidth]{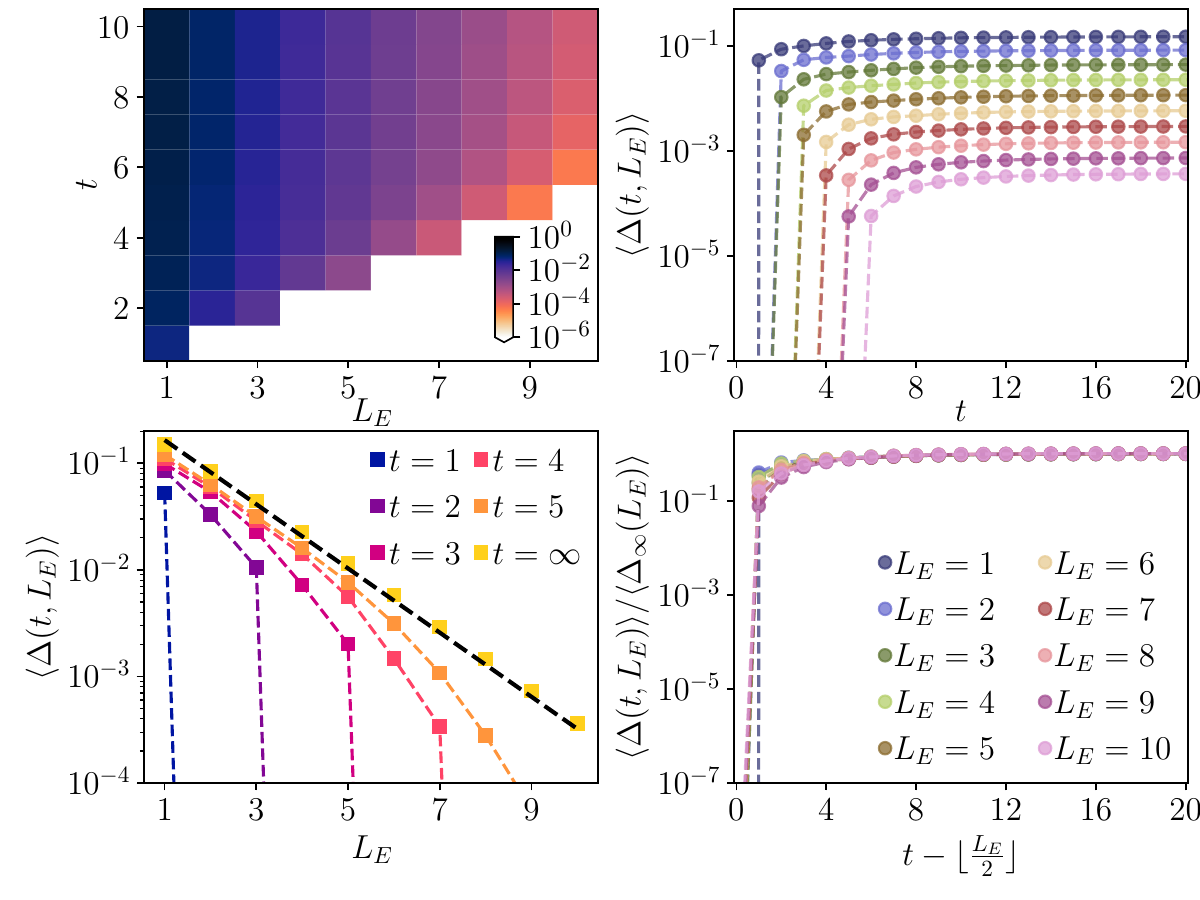}
\caption{Numerical results for $\Delta(t,L_E)$ for the kicked Ising chain in the  ergodic regime. The upper left panel shows $\Delta(t,L_E)$ as a heatmap in the $(t,L_E)$ plane where the linear lightcone is evident. The upper right panel shows $\Delta(t,L_E)$ as a function of $t$ for different $L_E$ where the onset happens at $t=\lfloor \frac{L_E}{2}\rfloor+1$ implying $t_\ast(L_E) = \lfloor \frac{L_E}{2}\rfloor$ making the linear lightcone quantitative. The lower left panel shows that $\braket{\Delta_\infty(L_E)}$ decays as $2^{-L_E}$ (indicated by the black dashed line). Using the onset timescale and the saturation value, the data for different $L_E$ can be collapsed onto each other as shown in the lower right panel. For the data $L_R=1$ and $L_S=15$, and the parameters are mentioned in Eq.~\ref{eq:params}.
}
\label{fig:tfi-erg}
\end{figure}

Consider the KI chain in Eq.~\ref{eq:UF-KI}, with the parameter values 
\eq{
J=0.8\,,\,g_i=0.5788\,,\,h_i=0.6472+0.4342\epsilon_i\,,\label{eq:params}
}
with $\epsilon_i\sim{\cal N}(0,1)$. 
These parameters place the model firmly in an ergodic phase~\cite{zhang2016floquet} and all results presented in this section correspond to these parameter values.
The results for the fluctuations in the PPE, $\Delta(t,L_E)$, as defined in Eq.~\ref{eq:ppe-fluc}, is shown in Fig.~\ref{fig:tfi-erg}.
The heatmap of $\Delta(t,L_E)$ in the $(t,L_E)$ plane, shown in the upper left panel, clearly shows the emergence of a lightcone. This lightcone is made more explicit in the upper right panel where we show $\Delta(t,L_E)$ as a function of $t$ for different values of $L_E$. The data clearly shows that for a given $L_E$, there exists a timescale $t_\ast(L_E) = \lfloor L_E/2\rfloor$ such that $\Delta(t,L_E) = 0$ for $t\leq t_\ast(L_E)$.
On the other hand, at $t=t_\ast+1$, the information lightcones emanating from the boundary between $R$ and $E$, and the boundary between $E$ and $S$ meet each other (see Fig.~\ref{fig:lightcone-schem}) resulting in $\Delta(t,L_E)$ picking up a finite value. 
The result in the lower right panel of Fig.~\ref{fig:tfi-erg} shows that the above result can be expressed in terms of the scaling form
\eq{
\Delta({t,L_E}) = \begin{cases}
0\,;~& t\leq \lfloor L_E/ 2\rfloor\,\\
{\rm constant}\times 2^{-L_E}\,;~& t\gg \lfloor L_E/2\rfloor\,
\end{cases}\,.
}

Note that the lightcones in this case are only half as fast, or equivalently, $t_\ast(L_E)$ is twice as large compared to those from the general bounds obtained from generic brickwork circuits.
This is straightforwardly understood by representing the Floquet unitary for the KI chain as a tensor network in terms of Ising phases and quantum kicks.
Introducing a diagrammatic notation as
\eq{
\cbox{
\begin{tikzpicture}[scale=0.8]
\draw(0,0.2) -- (0,.8);
\node[draw, fill=RoyalBlue,diamond, minimum size=6,inner sep=0pt] at (0,0.5) {};
\end{tikzpicture}
}=e^{-igX}\,,~
\cbox{\begin{tikzpicture}[scale=0.8]
    \draw(-0.5,0) -- (0.5,0);
    \draw(-0.5,-0.2) -- (-0.5,0.2);
    \draw(0.5,-0.2) -- (0.5,0.2);
    \draw[fill=Orange] (-0.1,-0.1) rectangle (0.1,0.1);
\end{tikzpicture}}=e^{-iJZ_1Z_2}\,,\nonumber\\
\cbox{\begin{tikzpicture}[scale=0.8,]\draw[fill=black] (0,0) circle (0.1);\draw(0,0) -- (0.5,0);\draw(0,0) -- (-0.35,-0.35);\draw(0,0) -- (-0.35,0.35);\node at (0.7,0){\small$z_1$};\node at (-0.6,0.35){\small$z_2$};\node at (-0.6,-0.35){\small$z_3$};\node at (0,-0.35){\small$h$};\end{tikzpicture}} = \delta_{z_1z_2z_3}e^{-ihz_1}\,,
\label{eq:sdki-motif}}
$U_F^{\rm KI}$ can be represented as 
\eq{U_F^{\rm KI}\!=\!
\cbox{
\begin{tikzpicture}[scale=0.8]
\def\y{0}
\draw(0,\y) -- (7,\y);
\foreach \x in {0,...,7}{
\draw[fill=black] (1*\x,1*\y) circle (0.1);
\draw(\x,-0.25) -- (\x,.75);
\node[draw, fill=RoyalBlue,diamond, minimum size=6,inner sep=0pt] at (\x,\y+0.4) {};
}
\foreach \x in {0,...,6}{
\draw[fill=Orange] (\x+0.5-0.1,\y-0.1) rectangle (\x+0.5+0.1,\y+0.1);
}
\end{tikzpicture}
}\,. 
\label{eq:UFKI-tens-net}
}
Defining a 2-site unitary gate as 
\eq{\cbox{\begin{tikzpicture}[scale=0.8]
\def\x{0}
\def\s{0.2}
\draw[black] (\x-0.35,-0.35)--(\x+0.35,0.35);
\draw[black] (\x-0.35,0.35)--(\x+0.35,-0.35);
\draw[rounded corners=2pt, fill=Salmon!80] (\x-\s,-\s) rectangle (\x+\s,\s); 
\draw[black] (\x+\s-0.15,\s-0.05) -- (\x+\s-0.05,\s-0.05) -- (\x+\s-0.05,\s-0.15);
\end{tikzpicture}}=
    \cbox{
    \begin{tikzpicture}[scale=0.8]
    \draw[fill=Salmon!30, rounded corners=3pt] (-0.2,-0.2) rectangle (1.2,1.2);
    \foreach \x in {0,1}{
    \draw(\x,0)--(\x,1);
    \node[draw, fill=RoyalBlue,diamond, minimum size=6,inner sep=0pt] at (\x,0.5) {};
    \draw(0,\x)--(1,\x);
    \draw[fill=Orange] (0.5-0.1,\x-0.1) rectangle (0.5+0.1,\x+0.1);
    }
    \foreach \x in {0,1}{
    \ifthenelse{\x=0}{\draw(0,\x)--(-0.35,\x-0.35);\draw(1,\x)--(1.35,\x-0.35);}{\draw(0,\x)--(-0.35,\x+0.35);\draw(1,\x)--(1.35,\x+0.35);}
    \draw[fill=gray] (0,\x) circle (0.1);
    \draw[fill=gray] (1,\x) circle (0.1);
    }
    \end{tikzpicture}
    }\,,
}
where the grey circles denote that the phase picked up is $h/2$ and not $h$ as it is the case for the black circles in Eq.~\ref{eq:sdki-motif}, it is straightforward to see that the bulk dynamics generated by the Floquet unitary
\eq{\tilde{U}_F^{\rm KI}=\cbox{
\begin{tikzpicture}[scale=0.8]
\def\s{0.2}
\foreach \x in {0,...,2}{
    \draw[black] (\x-0.35,-0.35)--(\x+0.35,0.35);
    \draw[black] (\x-0.35,0.35)--(\x+0.35,-0.35);
    \draw[black] (\x-0.35+0.5,-0.35+0.5)--(\x+0.35+0.5,0.35+0.5);
    \draw[black] (\x-0.35+0.5,0.35+0.5)--(\x+0.35+0.5,-0.35+0.5);
}
\def\x{3}
\draw[black] (\x-0.35,-0.35)--(\x+0.35,0.35);
\draw[black] (\x-0.35,0.35)--(\x+0.35,-0.35);
\foreach \x in {0,...,2}{
    \draw[rounded corners=2pt, fill=Salmon!80] (\x-\s,-\s) rectangle (\x+\s,\s); 
    \draw[black] (\x+\s-0.15,\s-0.05) -- (\x+\s-0.05,\s-0.05) -- (\x+\s-0.05,\s-0.15);
    \draw[rounded corners=2pt, fill=Salmon!80] (\x+0.5-\s,-\s+0.5) rectangle (\x+0.5+\s,\s+0.5); 
    \draw[black] (0.5+\x+\s-0.15,0.5+\s-0.05) -- (0.5+\x+\s-0.05,0.5+\s-0.05) -- (0.5+\x+\s-0.05,0.5+\s-0.15);
}
\def\x{3}
\draw[rounded corners=2pt, fill=Salmon!80] (\x-\s,-\s) rectangle (\x+\s,\s); 
\draw[black] (\x+\s-0.15,\s-0.05) -- (\x+\s-0.05,\s-0.05) -- (\x+\s-0.05,\s-0.15);
\end{tikzpicture}}\,,
\label{eq:UFKI-brickwork}
}
is the same as that generated by $U_F^{\rm KI}$ in Eq.~\ref{eq:UFKI-tens-net} up to additional boundary gates.
However, more importantly, each application of $\tilde{U}_{\rm F}^{\rm KI}$ corresponds to two applications of $U_F^{\rm KI}$ such that we have
\eq{
\left[U_F^{\rm KI}\right]^{2t} \propto \left[\tilde{U}_F^{\rm KI}\right]^{t}\,.
}
This naturally implies that the lightcone for the KI chain is effectively half as slow as that in generic brickwork models. 
Therefore, $R$ and $S$ to decouple at a given $t$, $L_E$ needs to be twice as large for the class of models described by Eq.~\ref{eq:UF-gen}, which in turn explains the scaling of $t_\ast(L_E) = \lfloor L_E/2\rfloor$ instead of the $L_E/4$ scaling as in Eq.~\ref{eq:delta-bound}.

Away from this decoupled limit, we observe that increasing $L_E$ for a fixed time $t$ results in an exponential suppression of $\Delta(t,L_E)$. For $t \to \infty$ this exponential decay holds for all $L_E$, and in this limit we find that $\Delta(t \to \infty,L_E) \sim 2^{-L_E}$ (dashed line in the lower left panel of Fig.~\ref{fig:tfi-erg}). This exponent can be understood by noting that in the limit $t \to \infty$ the region $R \cup E$ is expected to have deep thermalised and the ensemble approaches the gHSE. We will return to this point in Sec.~\ref{sec:latetimes}. At shorter times this exponent needs to be modified and has a non-universal dependence on $t$. 
This scaling indicates an imperfect transfer of quantum fluctuations induced by the measurements on $S$ to the subregion $R$, where increasing $L_E$ exponentially suppresses this transfer of quantum information. This argument can be made more precise at the self-dual point, which we turn to in the next section.

\begin{figure}
\includegraphics[width=\linewidth]{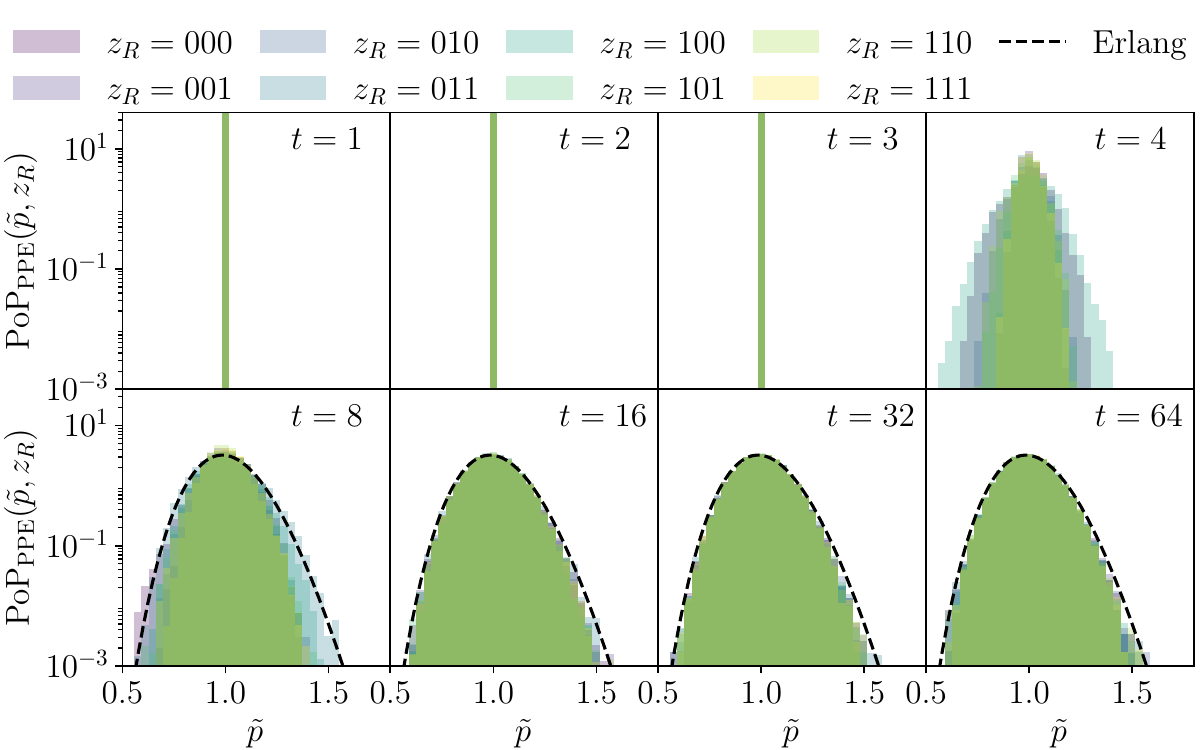}
\caption{The PoPs over the PPE (Eq.~\ref{eq:pop-ppe}) for the ergodic kicked Ising chain. Different panels correspond to different times and $L_E=6$ which implies $t_\ast(L_E)=3$. The data consistently shows the PoPs to be Dirac-delta functions at unity for $t\leq t_\ast$ whereas for $t>t_\ast$, they deviate from it. At longer times, the PoPs approach the universal Erlang distribution characterised by the Hilbert-space dimension of $L_E$ indicated by the black dashed line. For these plots, $L_R=3$ and $L_S=15$ and other parameters of the model are same as in Eq.~\ref{eq:params}.
}
\label{fig:tfi-erg-pop}
\end{figure}

We next consider the results for the PoPs for this model. The PoPs over the PPE of the individual $z_R$-bit-strings, defined in Eq.~\ref{eq:pop-ppe}, are shown in Fig.~\ref{fig:tfi-erg-pop} for the case of $L_E=6$. 
Consistent with the arguments presented in Sec.~\ref{sec:pop-lc}, the PoPs are given by a Dirac-delta function at $\tilde{p}=1$ for all the $z_R$-bit-strings for $t\leq t_\ast(L_E)$ which is a manifestation of the fact that each $\rho_R(z_S)$ in the PPE is independent of $z_S$ and, hence, the same.
On the other hand, for $t>t_\ast(L_E)$ the PPE becomes non-trivial in the sense that the different states $\rho_R(z_S)$ in the ensemble are different from each other, and hence the PoP deviates from the Dirac-delta function.
In this way, the ${\rm PoP}_{\rm PPE}$ carries signatures of the lightcone velocity via the information of $t_\ast(L_E)$.
At late times, we observe that the PPE approaches the expected Erlang distribution. 
Consistent with Ref.~\cite{shaw2025experimental}, we find that the Erlang distribution only depends on the dimension of the traced-out region $L_E$. This dimension grows exponentially quickly, and for large values of $L_E$ the Erlang distribution approaches a normal distribution with mean $1$ and variance $1/D_E$. Note that such a normal distribution would be expected from the central limit theorem if the bit-string probabilities were statistically independent variables. We therefore observe that both the PPE and the PoP indicate an exponential suppression of quantum fluctuations.

\begin{figure}
\includegraphics[width=\linewidth]{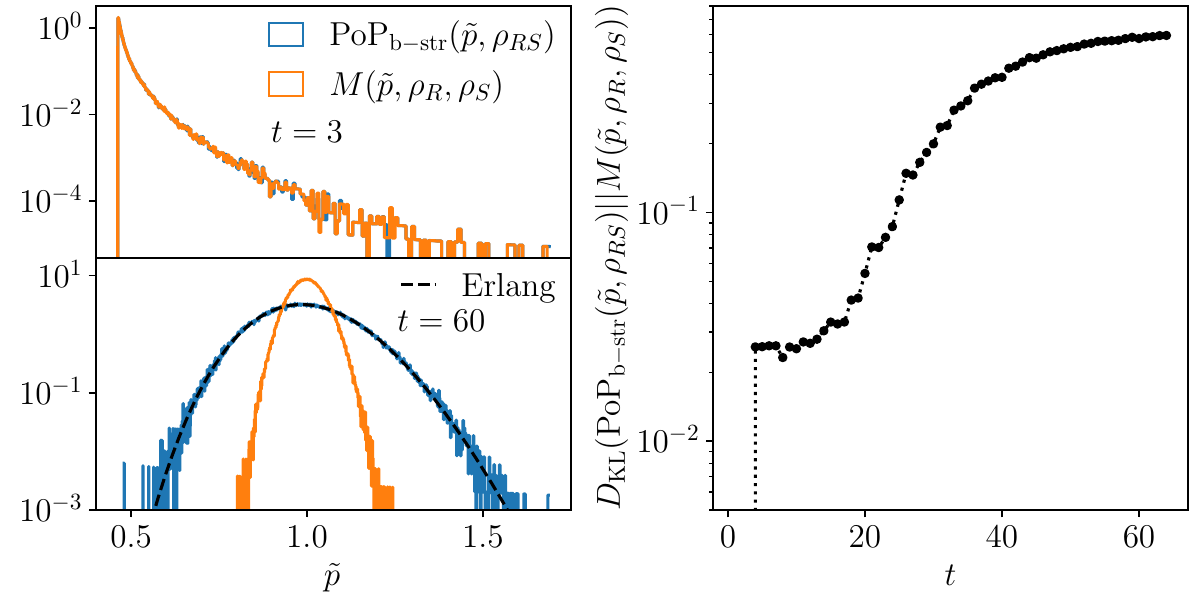}
\caption{Comparison between ${\rm PoP_{b{\text{-}}str}}(\tilde{p},\rho_{RS})$ and the Mellin convolution of ${\rm PoP_{b{\text{-}}str}}(\tilde{p},\rho_{R})$ and ${\rm PoP_{b{\text{-}}str}}(\tilde{p},\rho_{S})$, the latter denoted by $M(\tilde{p},\rho_R,\rho_S)$. The left two panels show the PoPs at $t<t_\ast(L_E)$ (top) and $t>t_\ast(L_E)$ (bottom). There is excellent agreement between the two distributions in the former case and a clear distinction in the latter case. The right panel shows the KLd between the two distributions as a function of time $t$. For these plots, $L_E=6$, and as expected the KLd is vanishingly small for $t\leq t_\ast(L_E=6)=3$ whereas it picks up a finite value for $t>t_\ast(L_E)$. Note that these results for a single time-evolved state of the kicked Ising chain with parameters in Eq.~\ref{eq:params}. Other parameters are $L_R=3$ and $L_S=15$. As an aside note that ${\rm PoP_{b{\text{-}}str}}(\tilde{p},\rho_{RS})$ at late times is well described the Erlang distribution (black dashed line) characterised by the Hilbert-space dimension of $E$. For the data we have $L_R=1$.
}
\label{fig:tfi-mellin-pop}
\end{figure}

Finally, we close the section on the ergodic kicked-Ising chain with a brief discussion of ${\rm PoP_{b{\text{-}}str}}$.
In Sec.~\ref{sec:pop}, it was discussed that for $t<t_\ast(L_E)$, the decoupling of $R$ and $S$ implies that ${\rm PoP_{b{\text{-}}str}}(\tilde{p},\rho_{RS})$ is given by the Mellin convolution of ${\rm PoP_{b{\text{-}}str}}(\tilde{p},\rho_{R})$ and ${\rm PoP_{b{\text{-}}str}}(\tilde{p},\rho_{S})$ whereas for $t>t_\ast(L_E)$, we expect ${\rm PoP_{b{\text{-}}str}}(\tilde{p},\rho_{RS})$ to deviate from the latter.
To quantify this deviation we study the Kullback-Leibler divergence (KLd) between ${\rm PoP_{b{\text{-}}str}}(\tilde{p},\rho_{RS})$ and the Mellin convolution denoted by $D_{\rm KL}({\rm PoP_{b{\text{-}}str}}(\tilde{p},\rho_{RS})||M(\tilde{p},\rho_R,\rho_S))$ where
\eq{M(\tilde{p},\rho_R,\rho_S))\equiv
{\rm PoP_{b{\text{-}}str}}(\tilde{p},\rho_{R})\ast_{\rm M}{\rm PoP_{b{\text{-}}str}}(\tilde{p},\rho_{S})\,,
}
and the KLd between two distributions $P$ and $Q$ is defined as 
\eq{
D_{\rm KL}(p||q) = \int dx~p(x)\ln\frac{p(x)}{q(x)}\,.
}
The results are shown in Fig.~\ref{fig:tfi-mellin-pop} where it is evident that ${\rm PoP_{b{\text{-}}str}}(\tilde{p},\rho_{RS})$ satisfies Eq.~\ref{eq:mellin}  for $t<t_\ast(L_E)$ signalled by the vanishing KLd whereas it deviates from the Mellin convolution for $t>t_\ast(L_E)$.
This result therefore presents a complementary signature of information propagation in the PoPs, but crucially in ${\rm PoP}_{\text{b-str}}$. This has the significance that this signature is manifested in the PoPs over bit-strings in a single time-evolved state and thus, one does not need access to individual members of the PPE.

\subsection{Self-dual kicked Ising chain \label{sec:sdki}}

Further analytical insight can be obtained by considering the kicked Ising chain at the self-dual point.
Fixing $|J| = |g_i| = \pi/4$, $\forall i$, the time-evolution operator~\eqref{eq:UF-gen} and the constituting gates~\eqref{eq:UFKI-tens-net} exhibit a space-time duality, which allows for the exact characterisation of various dynamical quantities using techniques from dual-unitarity~\cite{gutkin2020exact,bertini2019exact,gopalakrishnan2019unitary} (for a recent review, see Ref.~\cite{bertini2025exactly}). 
In the current context, the SDKI chain was the first model in which deep thermalisation could be analytically established~\cite{ho2022exact} and the full evolution of the PoP characterised~\cite{claeys2024fock-space}. 
For the former, the self-duality additionally leads to a collapse of design times: for a subsystem of size $L_R$, its projected ensemble forms an exact quantum state design after $t=L_R$ time steps~\cite{ho2022exact}. 
Conversely, generic circuits are expected to exhibit a hierarchy of design times, with higher designs realised at increasingly later times~\cite{ippoliti2022solvable}.
As a direct consequence, the PoP of the corresponding pure state rapidly approaches the Porter-Thomas distribution under unitary evolution. 
Self-duality imposes that the time scale at which the PoP approaches the PT distribution is independent of system size~\cite{claeys2024fock-space}, again to be contrasted with generic circuits where the relevant time scale is expected to scale logarithmically with system size~\cite{turkeshi2024hilbert}.

In this section, we consider PPEs and PoPs under self-dual dynamics. 
A key point is that the SDKI chain allows for the limit of $L_S\to\infty$ to be taken by replacing the average over measurement outcomes in $S$ by an average over Haar-random states contracted along the temporal lattice sites~\cite{ho2022exact}. 
This allows us to obtain exact results, either analytically or numerically in the thermodynamic limit.
Using this framework, we consider the PPE in the limit $L_S \to \infty$, where the known deep thermalisation of the self-dual kicked Ising chain directly implies that the mixed states form a gHSE at late times, as also recently established by Ref.~\cite{yu2025mixed}. We show an exponential suppression of the fluctuations in the PPE, as quantified by Eq.~\eqref{eq:ppe-fluc}, as the size of the traced-out subsystem increases. We identify three different dynamical regimes at different timescales: two universal regimes, characterized by either identically vanishing fluctuations due to the lightcone-imposed bound (early times), or fluctuations that are exponentially suppressed in the Hilbert-space dimension of the traced out region (late times), separated by a non-universal regime in which the exponential suppression is set by the efficiency of the spatial transfer of quantum information, i.e., quantum teleportation (intermediate times).
We then analytically establish the emergence of the Erlang distribution and the spatiotemporal information contained therein.

\subsubsection{PPEs and exponential suppression of fluctuations}

\begin{figure}
\includegraphics[width=\linewidth]{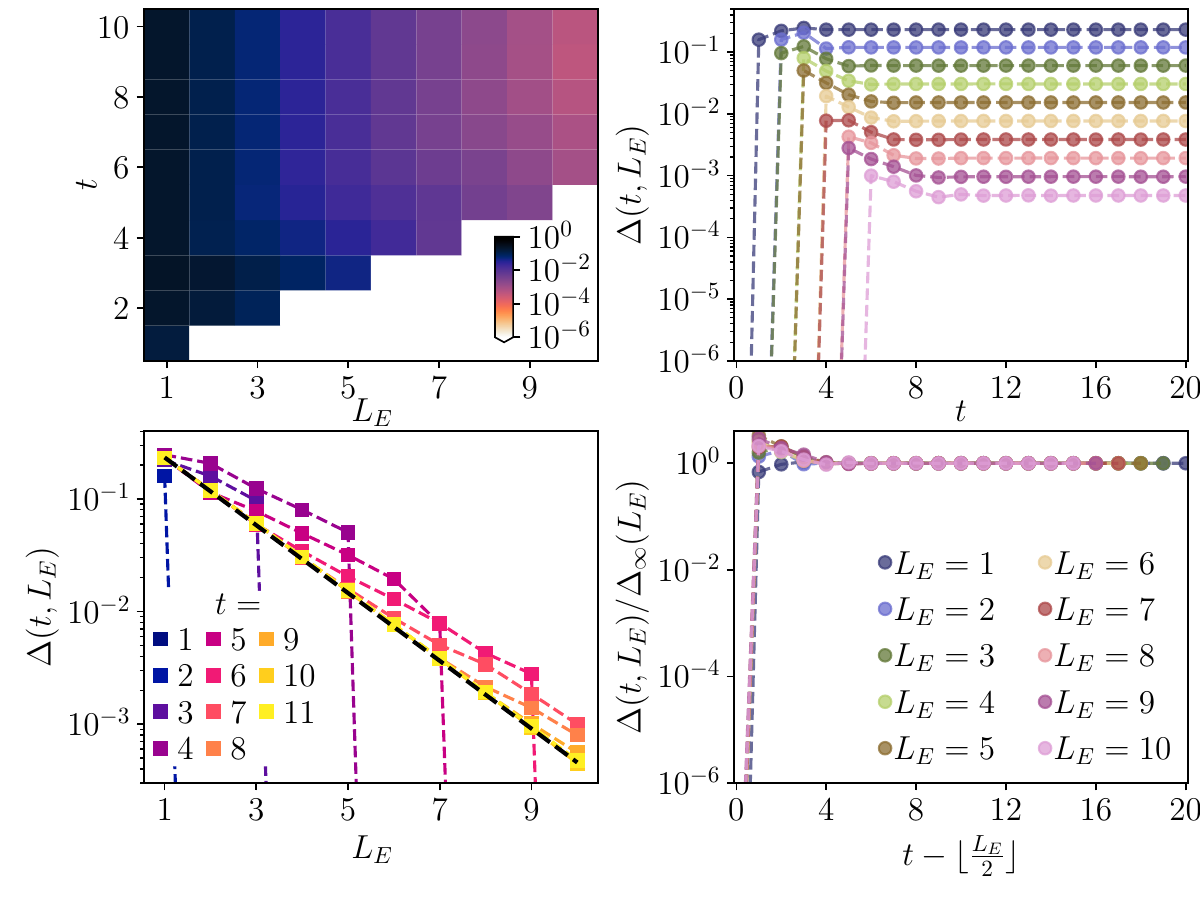}
\caption{Numerical results for $\Delta(t,L_E)$ for the kicked Ising chain at the self-dual point with $\ket{+}^{{\otimes}L}$ initial state. The upper left panel shows $\Delta(t,L_E)$ as a heatmap in the $(t,L_E)$ plane. The upper right panel shows $\Delta(t,L_E)$ as a function of $t$ for different $L_E$ where the onset happens at $t=\lfloor \frac{L_E}{2}\rfloor+1$ implying $t_\ast(L_E) = \lfloor \frac{L_E}{2}\rfloor$ making the linear lightcone quantitative. The lower left panel shows a crossover from a decay of $\braket{\Delta_\infty(L_E)}$ as $2^{-L_E/2}$ to $2^{-L_E}$ (indicated by the black dashed line) as $t$ is increased. Using the onset timescale and the saturation value, the data for different $L_E$ can be collapsed onto each other as shown in the lower right panel.}
\label{fig:SDKI-Delta}
\end{figure}

The results for the fluctuations in the PPE, as characterised by $\Delta(t,L_E)$ defined in Eq.~\ref{eq:ppe-fluc}, for the SDKI chain, are shown in Fig.~\ref{fig:SDKI-Delta}.
The lightcone structure of $\Delta$ for the self-dual case is the same as that of the generic KI chain as in Fig.~\ref{fig:tfi-erg}. Note that the results are directly in the limit of $L_S\to\infty$, which can be accessed using a special feature of the SDKI chain, which will be mentioned shortly.
We next discuss the suppression of fluctuations in the PPE, as characterised by $\Delta(t, L_E)$ in Eq.~\ref{eq:ppe-fluc} decreasing as $L_E$ is increased. 
Away from the lightcone-imposed bound, where $\Delta(t,L_E \geq t)$ vanishes identically, we argue that these fluctuations are exponentially suppressed in $L_E$.
Exact results on deep thermalization for the SDKI chain can be used to establish this exponential suppression, both at early and late times. The appearance of different time scales is directly related to the different length scales $L_R$ and $L_E$. 

Before continuing, we note that in this subsection we will consider the brickwork circuit from Eq.~\ref{eq:UFKI-brickwork} with both the initial state and the projective measurements in the $z$-basis. This dynamics is equivalent to the dynamics of an initial state in the $x$-basis under the Floquet unitary from Eq.~\ref{eq:UF-KI} after a single layer of two-site gates~\cite{bertini2025exactly}. This choice will allow us to simplify the graphical derivations.

We focus on the case where $t > L_R$. 
In this regime the subsystem $R$ has fully deep thermalised to an infinite-temperature state~\cite{ho2022exact}, such that the reduced density matrix is given by
\eq{
\rho_{R}^{(1)} = \frac{\mathbb{I}_R}{D_R}.
}
Underlying this result is the observation that in the limit $L_S \to \infty$ the environment can be replaced by a Haar-random state acting on the temporal lattice of $t$ sites~(as originally observed in Ref.~\cite{ho2022exact}). 
This is also exactly the idea that was used to obtain the results in Fig.~\ref{fig:SDKI-Delta} directly in the limit of $L_S\to\infty$.
For the second moment this returns
\eq{
\rho_R^{(2)}(t,L_E) \propto 
\cbox{
    \begin{tikzpicture}[scale=0.8]
    \fill[fill=Cyan!10] (-0.5,-1.1) rectangle (0,2.5); 
    \fill[fill=BrickRed!10] (0,-1.1) rectangle (1,2.5); 
    \fill[fill=OliveGreen!10] (1,-1.1) rectangle (1.8,2.5); 
        \def\s{0.2}
        \foreach \y in {0,1,2}{
            \draw[very thick](1.35,\y-0.35)--(1.6,\y-0.35);
            \draw[very thick](1.35,\y+0.35)--(1.6,\y+0.35);
            \draw[fill=gray,very thick]  (1.6,\y-0.35) circle (0.1);
            \draw[fill=gray,very thick]  (1.6,\y+0.35) circle (0.1);
            \def\xx{0.5}
            \draw[black,very thick] (\xx-0.35,\y-0.35-0.5)--(\xx+0.35,\y+0.35-0.5);
            \draw[black,very thick] (\xx-0.35,\y+0.35-0.5)--(\xx+0.35,\y-0.35-0.5);
            \foreach \x in {0,1}{
                \draw[black,very thick] (\x-0.35,\y-0.35)--(\x+0.35,\y+0.35);
                \draw[black,very thick] (\x-0.35,\y+0.35)--(\x+0.35,\y-0.35);
            }
        }
        \foreach \y in {0,1,2}{
            \def\xx{0.5}
            \draw[rounded corners=2pt, fill=RedViolet!70, very thick] (\xx-\s,\y-\s-0.5) rectangle (\xx+\s,\y+\s-0.5); 
            \draw[white!5] (\xx+\s-0.15,\y+\s-0.05-0.5) -- (\xx+\s-0.05,\y+\s-0.05-0.5) -- (\xx+\s-0.05,\y+\s-0.15-0.5);
            \foreach \x in {0,1}{
                \draw[rounded corners=2pt, fill=RedViolet!70, very thick] (\x-\s,\y-\s) rectangle (\x+\s,\y+\s); 
                \draw[white!5] (\x+\s-0.15,\y+\s-0.05) -- (\x+\s-0.05,\y+\s-0.05) -- (\x+\s-0.05,\y+\s-0.15);
            }
        }
        \draw[very thick](-0.35,-0.35)--(-0.35,-0.85);
        \draw[very thick](-0.35,0.35)--(-0.35,1-0.35);
        \draw[very thick](-0.35,1.35)--(-0.35,2-0.35);
        \draw[fill=gray,very thick]  (0.35,2.35) circle (0.1);
        \draw[fill=gray,very thick]  (1-0.35,2.35) circle (0.1);
        \foreach \x in {-0.35,0.15,0.85}{
        \draw[fill=gray,very thick] (\x-0.1,-0.85) -- (\x+0.1,-0.85) -- (\x,-1) -- cycle;}
    \end{tikzpicture}
}+
\cbox{
    \begin{tikzpicture}[scale=0.8]
    \fill[fill=Cyan!10] (-0.5,-1.1) rectangle (0,2.5); 
    \fill[fill=BrickRed!10] (0,-1.1) rectangle (1,2.5); 
    \fill[fill=OliveGreen!10] (1,-1.1) rectangle (1.8,2.5); 
        \def\s{0.2}
        \foreach \y in {0,1,2}{
            \draw[very thick](1.35,\y-0.35)--(1.6,\y-0.35);
            \draw[very thick](1.35,\y+0.35)--(1.6,\y+0.35);
            \draw[fill=gray, very thick] (1.5,\y-0.45) rectangle (1.7,\y-0.25);
            \draw[fill=gray, very thick] (1.5,\y-0.45+0.7) rectangle (1.7,\y-0.25+0.7);
            \def\xx{0.5}
            \draw[black,very thick] (\xx-0.35,\y-0.35-0.5)--(\xx+0.35,\y+0.35-0.5);
            \draw[black,very thick] (\xx-0.35,\y+0.35-0.5)--(\xx+0.35,\y-0.35-0.5);
            \foreach \x in {0,1}{
                \draw[black,very thick] (\x-0.35,\y-0.35)--(\x+0.35,\y+0.35);
                \draw[black,very thick] (\x-0.35,\y+0.35)--(\x+0.35,\y-0.35);
            }
        }
        \foreach \y in {0,1,2}{
            \def\xx{0.5}
            \draw[rounded corners=2pt, fill=RedViolet!70, very thick] (\xx-\s,\y-\s-0.5) rectangle (\xx+\s,\y+\s-0.5); 
            \draw[white!5] (\xx+\s-0.15,\y+\s-0.05-0.5) -- (\xx+\s-0.05,\y+\s-0.05-0.5) -- (\xx+\s-0.05,\y+\s-0.15-0.5);
            \foreach \x in {0,1}{
                \draw[rounded corners=2pt, fill=RedViolet!70, very thick] (\x-\s,\y-\s) rectangle (\x+\s,\y+\s); 
                \draw[white!5] (\x+\s-0.15,\y+\s-0.05) -- (\x+\s-0.05,\y+\s-0.05) -- (\x+\s-0.05,\y+\s-0.15);
            }
        }
        \draw[very thick](-0.35,-0.35)--(-0.35,-0.85);
        \draw[very thick](-0.35,0.35)--(-0.35,1-0.35);
        \draw[very thick](-0.35,1.35)--(-0.35,2-0.35);
        \draw[fill=gray,very thick]  (0.35,2.35) circle (0.1);
        \draw[fill=gray,very thick]  (1-0.35,2.35) circle (0.1);
        \foreach \x in {-0.35,0.15,0.85}{
        \draw[fill=gray,very thick] (\x-0.1,-0.85) -- (\x+0.1,-0.85) -- (\x,-1) -- cycle;}
    \end{tikzpicture}
}\,,\label{eq:rhoR2-brickwork-diag}
}
where the different background shades denote the three subsystems consistent with Fig.~\ref{fig:tripartition}.
In the above equation, we introduced two-times folded gates and boundary conditions corresponding to the folded identity and SWAP operators acting on the doubled Hilbert-space,
\eq{
\cbox{\begin{tikzpicture}[scale=0.8]
\def\x{0}
\def\s{0.2}
\draw[black] (\x-0.35,-0.35)--(\x+0.35,0.35);
\draw[black] (\x-0.35,0.35)--(\x+0.35,-0.35);
\draw[rounded corners=2pt, fill=Salmon!60] (\x-\s,-\s) rectangle (\x+\s,\s); 
\draw[black] (\x+\s-0.15,-\s+0.05) -- (\x+\s-0.05,-\s+0.05) -- (\x+\s-0.05,-\s+0.15);
\def\sx{-0.17}
\def\sy{-0.06}
\draw[black] (\x-0.35+\sx,-0.35+\sy)--(\x+0.35+\sx,0.35+\sy);
\draw[black] (\x-0.35+\sx,0.35+\sy)--(\x+0.35+\sx,-0.35+\sy);
\draw[rounded corners=2pt, fill=Salmon!30] (\x-\s+\sx,-\s+\sy) rectangle (\x+\s+\sx,\s+\sy); 
\draw[black] (\x+\s-0.15+\sx,\s-0.05+\sy) -- (\x+\s-0.05+\sx,\s-0.05+\sy) -- (\x+\s-0.05+\sx,\s-0.15+\sy);
\def\sx{-0.34}
\def\sy{-0.12}
\draw[black] (\x-0.35+\sx,-0.35+\sy)--(\x+0.35+\sx,0.35+\sy);
\draw[black] (\x-0.35+\sx,0.35+\sy)--(\x+0.35+\sx,-0.35+\sy);
\draw[rounded corners=2pt, fill=Salmon!60] (\x-\s+\sx,-\s+\sy) rectangle (\x+\s+\sx,\s+\sy); 
\draw[black] (\x+\s-0.15+\sx,-\s+0.05+\sy) -- (\x+\s-0.05+\sx,-\s+0.05+\sy) -- (\x+\s-0.05+\sx,-\s+0.15+\sy);
\def\sx{-0.51}
\def\sy{-0.18}
\draw[black] (\x-0.35+\sx,-0.35+\sy)--(\x+0.35+\sx,0.35+\sy);
\draw[black] (\x-0.35+\sx,0.35+\sy)--(\x+0.35+\sx,-0.35+\sy);
\draw[rounded corners=2pt, fill=Salmon!30] (\x-\s+\sx,-\s+\sy) rectangle (\x+\s+\sx,\s+\sy); 
\draw[black] (\x+\s-0.15+\sx,\s-0.05+\sy) -- (\x+\s-0.05+\sx,\s-0.05+\sy) -- (\x+\s-0.05+\sx,\s-0.15+\sy);
\end{tikzpicture}}
=
\cbox{\begin{tikzpicture}[scale=0.8]
\def\x{0}
\def\s{0.2}
\draw[black,very thick] (\x-0.35,-0.35)--(\x+0.35,0.35);
\draw[black, very thick] (\x-0.35,0.35)--(\x+0.35,-0.35);
\draw[rounded corners=2pt, fill=RedViolet!70, very thick] (\x-\s,-\s) rectangle (\x+\s,\s); 
\draw[white!5] (\x+\s-0.15,\s-0.05) -- (\x+\s-0.05,\s-0.05) -- (\x+\s-0.05,\s-0.15);
\end{tikzpicture}}\,,~
\cbox{
    \begin{tikzpicture}[scale=0.8]
    \draw[very thick](0,0)--(0,0.5);
    \draw[fill=gray,very thick] (-0.1,0.05) -- (0.1,0.05) -- (0,-0.1) -- cycle;
    \end{tikzpicture}
}
=
\left(
\cbox{
    \begin{tikzpicture}[scale=0.8]
    \draw[thick](0,0)--(0,0.4);
    \draw[black,fill=black,thick](-0.05,-0.05) rectangle (0.05,0.05);
    \end{tikzpicture}
}\right)^2\,,~
\cbox{
    \begin{tikzpicture}[scale=0.8]
    \draw[very thick](0,0)--(0,0.5);
    \draw[fill=gray,very thick] (0,0) circle (0.1);
    \end{tikzpicture}
}
=
\cbox{
    \begin{tikzpicture}[scale=0.8]
    \foreach \x in {0,0.25,0.5,0.75}
    \draw(\x,0)--(\x,0.5);
    \draw[black] (0,0) to[out=-90,in=-90] (0.25,0);
    \draw[black] (0.5,0) to[out=-90,in=-90] (0.75,0);
    \end{tikzpicture}
}\,,~
\cbox{
    \begin{tikzpicture}[scale=0.8]
    \draw[very thick](0,0)--(0,0.5);
    \draw[fill=gray, very thick] (-0.1,0.-0.1) rectangle (0.1,0.1);
    \end{tikzpicture}
}=
\cbox{
    \begin{tikzpicture}[scale=0.8]
    \foreach \x in {0,0.25,0.5,0.75}
    \draw(\x,0)--(\x,0.5);
    \draw[black] (0,0) to[out=-90,in=-90] (0.75,0);
    \draw[black] (0.25,0) to[out=-90,in=-90] (0.5,0);
    \end{tikzpicture}
}\,.
}
These operators directly appear in the second moment of the Haar-averaged state in the temporal direction.
Also, in Eq.~\ref{eq:rhoR2-brickwork-diag}, we ignore an unimportant prefactor (and therefore use the proportionality sign), which can always be determined from the normalisation. The first term factorises in the same way as $\rho_R^{\otimes 2}$, whereas the second term encodes the quantum fluctuations, such that $\Delta$ will be set by the relative weight of the second term with respect to the first term.

Unitarity and dual-unitarity of the folded gates result in the graphical identities,
\eq{
\cbox{\begin{tikzpicture}[scale=0.8]
\def\x{0}
\def\s{0.2}
\draw[very thick] (\x-0.35,-0.35)--(\x+0.35,0.35);
\draw[very thick] (\x-0.35,0.35)--(\x+0.35,-0.35);
\draw[rounded corners=2pt, fill=RedViolet!70, very thick] (\x-\s,-\s) rectangle (\x+\s,\s); 
\draw[white!5] (\x+\s-0.15,\s-0.05) -- (\x+\s-0.05,\s-0.05) -- (\x+\s-0.05,\s-0.15);
\draw[fill=gray,very thick]  (\x-0.35,0.35) circle (0.1);
\draw[fill=gray,very thick]  (\x+0.35,0.35) circle (0.1);
\end{tikzpicture}}
=
\cbox{\begin{tikzpicture}[scale=0.8]
\draw[very thick,black](-0.25,-0.25)--(-0.25,0.25);
\draw[very thick,black](0.25,-0.25)--(0.25,0.25);
\draw[very thick, fill=gray] (-0.25,0.25) circle (0.1);
\draw[very thick, fill=gray] (0.25,0.25) circle (0.1);
\end{tikzpicture}}\,,~
\cbox{\begin{tikzpicture}[scale=0.8]
\def\x{0}
\def\s{0.2}
\draw[very thick] (\x-0.35,-0.35)--(\x+0.35,0.35);
\draw[very thick] (\x-0.35,0.35)--(\x+0.35,-0.35);
\draw[rounded corners=2pt, fill=RedViolet!70, very thick] (\x-\s,-\s) rectangle (\x+\s,\s); 
\draw[white!5] (\x+\s-0.15,\s-0.05) -- (\x+\s-0.05,\s-0.05) -- (\x+\s-0.05,\s-0.15);
\draw[fill=gray,very thick]  (\x-0.35,-0.35) circle (0.1);
\draw[fill=gray,very thick]  (\x+0.35,-0.35) circle (0.1);
\end{tikzpicture}}
=
\cbox{\begin{tikzpicture}[scale=0.8]
\draw[very thick,black](-0.25,-0.25)--(-0.25,0.25);
\draw[very thick,black](0.25,-0.25)--(0.25,0.25);
\draw[very thick, fill=gray] (-0.25,-0.25) circle (0.1);
\draw[very thick, fill=gray] (0.25,-0.25) circle (0.1);
\end{tikzpicture}}\,,~
\cbox{\begin{tikzpicture}[scale=0.8]
\def\x{0}
\def\s{0.2}
\draw[very thick] (\x-0.35,-0.35)--(\x+0.35,0.35);
\draw[very thick] (\x-0.35,0.35)--(\x+0.35,-0.35);
\draw[rounded corners=2pt, fill=RedViolet!70, very thick] (\x-\s,-\s) rectangle (\x+\s,\s); 
\draw[white!5] (\x+\s-0.15,\s-0.05) -- (\x+\s-0.05,\s-0.05) -- (\x+\s-0.05,\s-0.15);
\draw[fill=gray,very thick]  (\x-0.35,-0.35) circle (0.1);
\draw[fill=gray,very thick]  (\x-0.35,0.35) circle (0.1);
\end{tikzpicture}}
=
\cbox{\begin{tikzpicture}[scale=0.8]
\draw[very thick,black](-0.25,-0.25)--(0.25,-0.25);
\draw[very thick,black](-0.25,0.25)--(0.25,0.25);
\draw[very thick, fill=gray] (-0.25,-0.25) circle (0.1);
\draw[very thick, fill=gray] (-0.25,0.25) circle (0.1);
\end{tikzpicture}}\,,~
\cbox{\begin{tikzpicture}[scale=0.8]
\def\x{0}
\def\s{0.2}
\draw[very thick] (\x-0.35,-0.35)--(\x+0.35,0.35);
\draw[very thick] (\x-0.35,0.35)--(\x+0.35,-0.35);
\draw[rounded corners=2pt, fill=RedViolet!70, very thick] (\x-\s,-\s) rectangle (\x+\s,\s); 
\draw[white!5] (\x+\s-0.15,\s-0.05) -- (\x+\s-0.05,\s-0.05) -- (\x+\s-0.05,\s-0.15);
\draw[fill=gray,very thick]  (\x+0.35,-0.35) circle (0.1);
\draw[fill=gray,very thick]  (\x+0.35,0.35) circle (0.1);
\end{tikzpicture}}
=
\cbox{\begin{tikzpicture}[scale=0.8]
\draw[very thick,black](-0.25,-0.25)--(0.25,-0.25);
\draw[very thick,black](-0.25,0.25)--(0.25,0.25);
\draw[very thick, fill=gray] (0.25,-0.25) circle (0.1);
\draw[very thick, fill=gray] (0.25,0.25) circle (0.1);
\end{tikzpicture}}\,.
\label{eq:dual_unitarity}
}
The explicit parameterization of the kicked Ising chain results in an additional identity~\cite{piroli2020exactDU,claeys2022emergent},
\eq{\label{eq:sdki-init}
\cbox{\begin{tikzpicture}[scale=0.8]
\def\x{0}
\def\s{0.2}
\draw[very thick] (\x-0.35,-0.35)--(\x+0.35,0.35);
\draw[very thick] (\x-0.35,0.35)--(\x+0.35,-0.35);
\draw[rounded corners=2pt, fill=RedViolet!70, very thick] (\x-\s,-\s) rectangle (\x+\s,\s); 
\draw[white!5] (\x+\s-0.15,\s-0.05) -- (\x+\s-0.05,\s-0.05) -- (\x+\s-0.05,\s-0.15);
\draw[fill=gray,very thick]  (\x+0.35,0.35) circle (0.1);
\def\x{-0.35}
\def\y{-0.35}
\draw[fill=gray,very thick] (\x-0.1,\y) -- (\x+0.1,\y) -- (\x,\y-0.15) -- cycle;
\def\x{0.35}
\def\y{-0.35}
\draw[fill=gray,very thick] (\x-0.1,\y) -- (\x+0.1,\y) -- (\x,\y-0.15) -- cycle;
\end{tikzpicture}
}
=
\cbox{\begin{tikzpicture}[scale=0.8]
\draw[very thick,black](-0.25,-0.25)--(0.25,-0.25);
\draw[very thick, fill=gray] (0.25,-0.25) circle (0.1);
\end{tikzpicture}}\,,~~
\cbox{\begin{tikzpicture}[scale=0.8]
\def\x{0}
\def\s{0.2}
\draw[very thick] (\x-0.35,-0.35)--(\x+0.35,0.35);
\draw[very thick] (\x-0.35,0.35)--(\x+0.35,-0.35);
\draw[rounded corners=2pt, fill=RedViolet!70, very thick] (\x-\s,-\s) rectangle (\x+\s,\s); 
\draw[white!5] (\x+\s-0.15,\s-0.05) -- (\x+\s-0.05,\s-0.05) -- (\x+\s-0.05,\s-0.15);
\draw[fill=gray,very thick]  (\x+0.35,0.35) rectangle (\x+0.55,0.55);
\def\x{-0.35}
\def\y{-0.35}
\draw[fill=gray,very thick] (\x-0.1,\y) -- (\x+0.1,\y) -- (\x,\y-0.15) -- cycle;
\def\x{0.35}
\def\y{-0.35}
\draw[fill=gray,very thick] (\x-0.1,\y) -- (\x+0.1,\y) -- (\x,\y-0.15) -- cycle;
\end{tikzpicture}
}
=
\cbox{\begin{tikzpicture}[scale=0.8]
\draw[very thick,black](-0.25,0)--(0.25,0);
\draw[fill=gray, very thick] (0.25-0.1,-0.1) rectangle (0.25+0.1,0.1);
\end{tikzpicture}}
}
provided the initial states are product states in the $z$-basis, motivating our choice of initial state and measurement basis.
Using these identities, it directly follows that the first term factorises and reproduces $\rho_{R}^{(1)} \otimes \rho_{R}^{(1)} = \mathbb{I}_R \otimes \mathbb{I}_R / D_R^2$. 
The second term encodes the fluctuations in the PPE, such that $\Delta$ from Eq.~\ref{eq:ppe-fluc} corresponds to the norm of this term.

Away from the lightcone bound, i.e. for $L_E < t$, we distinguish two different regimes. For $L_E < t-L_R$ the combined $RE$ region has deep thermalised, whereas for $ L_E > t - L_R$ only the $R$ region has deep thermalised. First consider the former region. Using unitarity and dual-unitarity (Eq.~\ref{eq:dual_unitarity}), it is possible to fully propagate the boundary conditions through the bulk, such that we find 
\eq{
\rho_R^{(2)}(t,L_E) &\propto 
\cbox{
    \begin{tikzpicture}[scale=0.8]
    \fill[fill=Cyan!10] (-0.5,-1.1) rectangle (0,3.5); 
    \fill[fill=BrickRed!10] (0,-1.1) rectangle (1,3.5); 
    \fill[fill=OliveGreen!10] (1,-1.1) rectangle (1.8,3.5); 
    \draw[line width=3pt, red!50] (1.35,3.35)--(1.2,3.2)--(0.8,3.2)--(1-0.35,3.35);
    \draw[line width=3pt, red!50] (1.35,3-.35)--(1.2,3-.2)--(0.8,3-.2)--(0.5+0.2,2.5+0.2)--(0.5-0.2,2.5+0.2)--(0.2,3-0.2)--(0.2,3.2)--(0.35,3.35);
    \draw[line width=3pt, red!50] (1.35,2.35)--(1.2,2.2)--(0.8,2.2)--(0.7,2.3)--(0.3,2.3)--(0.2,2.2)--(-0.2,2.2)--(-0.35,2.35)--(-0.35,3-0.35)--(-0.2,2.8)--(-0.2,3.2)--(-0.35,3.35);
    \draw[line width=3pt, red!50] (1.35,1.35)--(1.2,1.2)--(0.8,1.2)--(0.7,1.3)--(0.3,1.3)--(0.2,1.2)--(-0.2,1.2)--(-0.35,1.35)--(-0.35,2-0.35)--(-0.2,1.8)--(0.2,1.8)--(0.3,1.7)--(0.7,1.7)--(0.8,1.8)--(1.2,1.8)--(1.35,2-0.35);
    \draw[line width=3pt, red!50] (1.35,0.35)--(1.2,0.2)--(0.8,0.2)--(0.7,0.3)--(0.3,0.3)--(0.2,0.2)--(-0.2,0.2)--(-0.35,0.35)--(-0.35,1-0.35)--(-0.2,0.8)--(0.2,0.8)--(0.3,0.7)--(0.7,0.7)--(0.8,0.8)--(1.2,0.8)--(1.35,1-0.35);
    \draw[line width=3pt, red!50] (1.35,-0.35)--(1.2,-0.2)--(0.8,-0.2)--(0.7,-0.3)--(0.7,-0.7)--(0.85,-0.85);
    \draw[line width=3pt, red!50] (-0.35,-0.85)--(-0.35,-0.35)--(-0.2,-0.2)--(0.2,-0.2)--(0.3,-0.3)--(0.3,-0.7)--(0.5-0.35,-0.85);
        \def\s{0.2}
        \foreach \y in {0,1,2,3}{
            \draw[very thick](1.35,\y-0.35)--(1.6,\y-0.35);
            \draw[very thick](1.35,\y+0.35)--(1.6,\y+0.35);
            \draw[fill=gray,very thick]  (1.6,\y-0.35) circle (0.1);
            \draw[fill=gray,very thick]  (1.6,\y+0.35) circle (0.1);
            \def\xx{0.5}
            \draw[black,very thick] (\xx-0.35,\y-0.35-0.5)--(\xx+0.35,\y+0.35-0.5);
            \draw[black,very thick] (\xx-0.35,\y+0.35-0.5)--(\xx+0.35,\y-0.35-0.5);
            \foreach \x in {0,1}{
                \draw[black,very thick] (\x-0.35,\y-0.35)--(\x+0.35,\y+0.35);
                \draw[black,very thick] (\x-0.35,\y+0.35)--(\x+0.35,\y-0.35);
            }
        }
        \foreach \y in {0,1,2,3}{
            \def\xx{0.5}
            \draw[rounded corners=2pt, fill=RedViolet!70, very thick] (\xx-\s,\y-\s-0.5) rectangle (\xx+\s,\y+\s-0.5); 
            \draw[white!5] (\xx+\s-0.15,\y+\s-0.05-0.5) -- (\xx+\s-0.05,\y+\s-0.05-0.5) -- (\xx+\s-0.05,\y+\s-0.15-0.5);
            \foreach \x in {0,1}{
                \draw[rounded corners=2pt, fill=RedViolet!70, very thick] (\x-\s,\y-\s) rectangle (\x+\s,\y+\s); 
                \draw[white!5] (\x+\s-0.15,\y+\s-0.05) -- (\x+\s-0.05,\y+\s-0.05) -- (\x+\s-0.05,\y+\s-0.15);
            }
        }
        \draw[very thick](-0.35,-0.35)--(-0.35,-0.85);
        \draw[very thick](-0.35,0.35)--(-0.35,1-0.35);
        \draw[very thick](-0.35,1.35)--(-0.35,2-0.35);
        \draw[very thick](-0.35,2.35)--(-0.35,3-0.35);
        \draw[fill=gray,very thick]  (0.35,3.35) circle (0.1);
        \draw[fill=gray,very thick]  (1-0.35,3.35) circle (0.1);
        \foreach \x in {-0.35,0.15,0.85}{
        \draw[fill=gray,very thick] (\x-0.1,-0.85) -- (\x+0.1,-0.85) -- (\x,-1) -- cycle;}
    \end{tikzpicture}
}+
\cbox{
    \begin{tikzpicture}[scale=0.8]
    \fill[fill=Cyan!10] (-0.5,-1.1) rectangle (0,3.5); 
    \fill[fill=BrickRed!10] (0,-1.1) rectangle (1,3.5); 
    \fill[fill=OliveGreen!10] (1,-1.1) rectangle (1.8,3.5); 
    \draw[line width=3pt, red!50] (1.35,3.35)--(1.2,3.2)--(0.8,3.2)--(1-0.35,3.35);
    \draw[line width=3pt, red!50] (1.35,3-.35)--(1.2,3-.2)--(0.8,3-.2)--(0.5+0.2,2.5+0.2)--(0.5-0.2,2.5+0.2)--(0.2,3-0.2)--(0.2,3.2)--(0.35,3.35);
    \draw[line width=3pt, red!50] (1.35,2.35)--(1.2,2.2)--(0.8,2.2)--(0.7,2.3)--(0.3,2.3)--(0.2,2.2)--(-0.2,2.2)--(-0.35,2.35)--(-0.35,3-0.35)--(-0.2,2.8)--(-0.2,3.2)--(-0.35,3.35);
    \draw[line width=3pt, red!50] (1.35,1.35)--(1.2,1.2)--(0.8,1.2)--(0.7,1.3)--(0.3,1.3)--(0.2,1.2)--(-0.2,1.2)--(-0.35,1.35)--(-0.35,2-0.35)--(-0.2,1.8)--(0.2,1.8)--(0.3,1.7)--(0.7,1.7)--(0.8,1.8)--(1.2,1.8)--(1.35,2-0.35);
    \draw[line width=3pt, red!50] (1.35,0.35)--(1.2,0.2)--(0.8,0.2)--(0.7,0.3)--(0.3,0.3)--(0.2,0.2)--(-0.2,0.2)--(-0.35,0.35)--(-0.35,1-0.35)--(-0.2,0.8)--(0.2,0.8)--(0.3,0.7)--(0.7,0.7)--(0.8,0.8)--(1.2,0.8)--(1.35,1-0.35);
    \draw[line width=3pt, red!50] (1.35,-0.35)--(1.2,-0.2)--(0.8,-0.2)--(0.7,-0.3)--(0.7,-0.7)--(0.85,-0.85);
    \draw[line width=3pt, red!50] (-0.35,-0.85)--(-0.35,-0.35)--(-0.2,-0.2)--(0.2,-0.2)--(0.3,-0.3)--(0.3,-0.7)--(0.5-0.35,-0.85);
        \def\s{0.2}
        \foreach \y in {0,1,2,3}{
            \draw[very thick](1.35,\y-0.35)--(1.6,\y-0.35);
            \draw[very thick](1.35,\y+0.35)--(1.6,\y+0.35);
            \draw[fill=gray, very thick] (1.5,\y-0.45) rectangle (1.7,\y-0.25);
            \draw[fill=gray, very thick] (1.5,\y-0.45+0.7) rectangle (1.7,\y-0.25+0.7);
            \def\xx{0.5}
            \draw[black,very thick] (\xx-0.35,\y-0.35-0.5)--(\xx+0.35,\y+0.35-0.5);
            \draw[black,very thick] (\xx-0.35,\y+0.35-0.5)--(\xx+0.35,\y-0.35-0.5);
            \foreach \x in {0,1}{
                \draw[black,very thick] (\x-0.35,\y-0.35)--(\x+0.35,\y+0.35);
                \draw[black,very thick] (\x-0.35,\y+0.35)--(\x+0.35,\y-0.35);
            }
        }
        \foreach \y in {0,1,2,3}{
            \def\xx{0.5}
            \draw[rounded corners=2pt, fill=RedViolet!70, very thick] (\xx-\s,\y-\s-0.5) rectangle (\xx+\s,\y+\s-0.5); 
            \draw[white!5] (\xx+\s-0.15,\y+\s-0.05-0.5) -- (\xx+\s-0.05,\y+\s-0.05-0.5) -- (\xx+\s-0.05,\y+\s-0.15-0.5);
            \foreach \x in {0,1}{
                \draw[rounded corners=2pt, fill=RedViolet!70, very thick] (\x-\s,\y-\s) rectangle (\x+\s,\y+\s); 
                \draw[white!5] (\x+\s-0.15,\y+\s-0.05) -- (\x+\s-0.05,\y+\s-0.05) -- (\x+\s-0.05,\y+\s-0.15);
            }
        }
        \draw[very thick](-0.35,-0.35)--(-0.35,-0.85);
        \draw[very thick](-0.35,0.35)--(-0.35,1-0.35);
        \draw[very thick](-0.35,1.35)--(-0.35,2-0.35);
        \draw[very thick](-0.35,2.35)--(-0.35,3-0.35);
        \draw[fill=gray,very thick]  (0.35,3.35) circle (0.1);
        \draw[fill=gray,very thick]  (1-0.35,3.35) circle (0.1);
        \foreach \x in {-0.35,0.15,0.85}{
        \draw[fill=gray,very thick] (\x-0.1,-0.85) -- (\x+0.1,-0.85) -- (\x,-1) -- cycle;}
    \end{tikzpicture}
}\,,\nonumber\\
&=\cbox{
    \begin{tikzpicture}[scale=0.8]
    \draw[very thick](0,0)--(0,0.5);
    \draw[fill=gray, very thick] (0,0) circle (0.1);
    \end{tikzpicture}}^{\otimes L_R}
    \left(\cbox{\begin{tikzpicture}[scale=0.8]\draw[very thick](0,0)--(0,0.5);
    \draw[fill=gray, very thick] (0,0) circle (0.1);
    \draw[fill=gray, very thick] (0,0.5) circle (0.1);
    \end{tikzpicture}}\right)^{t-L_R}
+\cbox{
    \begin{tikzpicture}[scale=0.8]
    \draw[very thick](0,0)--(0,0.5);
    \draw[fill=gray, very thick] (-0.1,-0.1) rectangle (0.1,0.1);
    \end{tikzpicture}}^{\otimes L_R}
    \left(\cbox{\begin{tikzpicture}[scale=0.8]\draw[very thick](0,0)--(0,0.5);
    \draw[fill=gray, very thick] (-0.1,0.-0.1) rectangle (0.1,0.1);
    \draw[fill=gray, very thick] (0,0.5) circle (0.1);
    \end{tikzpicture}}\right)^{L_E}
    \left(\cbox{\begin{tikzpicture}[scale=0.8]\draw[very thick](0,0)--(0,0.5);
    \draw[fill=gray, very thick] (-0.1,0.-0.1) rectangle (0.1,0.1);
    \draw[fill=gray, very thick] (-0.1,0.4) rectangle (0.1,0.6);
    \end{tikzpicture}}\right)^{t-L_R-L_E}\,,\nonumber\\
& \propto \frac{1}{D_R^2}\left[{\mathbb{I}}_{R^{\otimes 2}}+\frac{1}{D_E}\mathbb{S}_{R^{\otimes 2}}\right]\,,
\label{eq:rhoR2-SDKI-DT}
}
where in the circuit diagram the thick red lines denote the paths along which the boundary contractions can be propagated.
It directly follows that the fluctuations scale as $1/D_E$. In this regime the exact deep thermalisation of $R \cup E$ implies that the PPE forms an exact gHSE, which in turn directly implies the exponential scaling of $\Delta$. We will return to this result in Sec.~\ref{sec:latetimes}.

In the opposite regime, $L_E > t - L_R$, the region $R \cup E$ is not deep thermalised. While no exact result for $\Delta$ can be obtained, it is possible to obtain a bound on the fluctuations which returns the numerically observed scaling. Graphically,
\eq{
\rho_R^{(2)}(t,L_E) \!&\propto\! 
\cbox{
    \begin{tikzpicture}[scale=0.8]
    \fill[fill=Cyan!10] (-0.5,-1.1) rectangle (0,1.5); 
    \fill[fill=BrickRed!10] (0,-1.1) rectangle (2,1.5); 
    \fill[fill=OliveGreen!10] (2,-1.1) rectangle (2.8,1.5); 
        \def\s{0.2}
        \foreach \y in {0,1}{
            \draw[very thick](2.35,\y-0.35)--(2.6,\y-0.35);
            \draw[very thick](2.35,\y+0.35)--(2.6,\y+0.35);
            \draw[fill=gray,very thick]  (2.6,\y-0.35) circle (0.1);
            \draw[fill=gray,very thick]  (2.6,\y+0.35) circle (0.1);
            \foreach \xx in {0.5,1.5}{
            \draw[black,very thick] (\xx-0.35,\y-0.35-0.5)--(\xx+0.35,\y+0.35-0.5);
            \draw[black,very thick] (\xx-0.35,\y+0.35-0.5)--(\xx+0.35,\y-0.35-0.5);}
            \foreach \x in {0,1,2}{
                \draw[black,very thick] (\x-0.35,\y-0.35)--(\x+0.35,\y+0.35);
                \draw[black,very thick] (\x-0.35,\y+0.35)--(\x+0.35,\y-0.35);
            }
        }
        \foreach \y in {0,1}{
            \foreach \xx in {0.5,1.5}{
            \draw[rounded corners=2pt, fill=RedViolet!70, very thick] (\xx-\s,\y-\s-0.5) rectangle (\xx+\s,\y+\s-0.5); 
            \draw[white!5] (\xx+\s-0.15,\y+\s-0.05-0.5) -- (\xx+\s-0.05,\y+\s-0.05-0.5) -- (\xx+\s-0.05,\y+\s-0.15-0.5);}
            \foreach \x in {0,1,2}{
                \draw[rounded corners=2pt, fill=RedViolet!70, very thick] (\x-\s,\y-\s) rectangle (\x+\s,\y+\s); 
                \draw[white!5] (\x+\s-0.15,\y+\s-0.05) -- (\x+\s-0.05,\y+\s-0.05) -- (\x+\s-0.05,\y+\s-0.15);
            }
        }
        \draw[very thick](-0.35,-0.35)--(-0.35,-0.85);
        \draw[very thick](-0.35,0.35)--(-0.35,1-0.35);
        \foreach \x in {0.35,1-0.35,1+.35,2-.35}{
        \draw[fill=gray,very thick]  (\x,1.35) circle (0.1);
        \draw[fill=gray,very thick]  (\x,1.35) circle (0.1);}
        \foreach \x in {-0.35,0.15,0.85,1.15,1.85}{
        \draw[fill=gray,very thick] (\x-0.1,-0.85) -- (\x+0.1,-0.85) -- (\x,-1) -- cycle;}
    \end{tikzpicture}
}\!+\!
\cbox{
    \begin{tikzpicture}[scale=0.8]
    \fill[fill=Cyan!10] (-0.5,-1.1) rectangle (0,1.5); 
    \fill[fill=BrickRed!10] (0,-1.1) rectangle (2,1.5); 
    \fill[fill=OliveGreen!10] (2,-1.1) rectangle (2.8,1.5); 
        \def\s{0.2}
        \foreach \y in {0,1}{
            \draw[very thick](2.35,\y-0.35)--(2.6,\y-0.35);
            \draw[very thick](2.35,\y+0.35)--(2.6,\y+0.35);
            \draw[fill=gray,very thick]  (2.6-0.1,\y-0.35-0.1) rectangle (2.6+0.1,\y-0.35+0.1);
            \draw[fill=gray,very thick]  (2.6-0.1,\y+0.35-0.1) rectangle (2.6+0.1,\y+0.35+0.1);
            \foreach \xx in {0.5,1.5}{
            \draw[black,very thick] (\xx-0.35,\y-0.35-0.5)--(\xx+0.35,\y+0.35-0.5);
            \draw[black,very thick] (\xx-0.35,\y+0.35-0.5)--(\xx+0.35,\y-0.35-0.5);}
            \foreach \x in {0,1,2}{
                \draw[black,very thick] (\x-0.35,\y-0.35)--(\x+0.35,\y+0.35);
                \draw[black,very thick] (\x-0.35,\y+0.35)--(\x+0.35,\y-0.35);
            }
        }
        \foreach \y in {0,1}{
            \foreach \xx in {0.5,1.5}{
            \draw[rounded corners=2pt, fill=RedViolet!70, very thick] (\xx-\s,\y-\s-0.5) rectangle (\xx+\s,\y+\s-0.5); 
            \draw[white!5] (\xx+\s-0.15,\y+\s-0.05-0.5) -- (\xx+\s-0.05,\y+\s-0.05-0.5) -- (\xx+\s-0.05,\y+\s-0.15-0.5);}
            \foreach \x in {0,1,2}{
                \draw[rounded corners=2pt, fill=RedViolet!70, very thick] (\x-\s,\y-\s) rectangle (\x+\s,\y+\s); 
                \draw[white!5] (\x+\s-0.15,\y+\s-0.05) -- (\x+\s-0.05,\y+\s-0.05) -- (\x+\s-0.05,\y+\s-0.15);
            }
        }
        \draw[very thick](-0.35,-0.35)--(-0.35,-0.85);
        \draw[very thick](-0.35,0.35)--(-0.35,1-0.35);
        \foreach \x in {0.35,1-0.35,1+.35,2-.35}{
        \draw[fill=gray,very thick]  (\x,1.35) circle (0.1);
        \draw[fill=gray,very thick]  (\x,1.35) circle (0.1);}
        \foreach \x in {-0.35,0.15,0.85,1.15,1.85}{
        \draw[fill=gray,very thick] (\x-0.1,-0.85) -- (\x+0.1,-0.85) -- (\x,-1) -- cycle;}
    \end{tikzpicture}
}.\nonumber
}
Performing the contractions where possible, this diagram can be simplified to
\eq{\nonumber\rho_R^{(2)}(t,L_E)\propto
\cbox{
	\begin{tikzpicture}[scale=0.8]
    \def\s{0.2}
	\draw[black,very thick] (0-0.35,-0.35)--(0-0.35,-0.85);
	\draw[black,very thick] (0-0.35,-0.35)--(0+0.35,+0.35);
    \draw[black,very thick] (0-0.35,+0.35)--(0+0.35,-0.35);
    \draw[black,very thick] (0.5-0.35,0.5-0.35)--(0.5+0.35,0.5+0.35);
    \draw[black,very thick] (0.5-0.35,0.5+0.35)--(0.5+0.35,0.5-0.35);
    \draw[black,very thick] (0-0.35,1-0.35)--(0+0.35,1+0.35);
    \draw[black,very thick] (0-0.35,1+0.35)--(0+0.35,1-0.35);
    \draw[black,very thick] (-0.35,0.35)--(-0.35,1-0.35);
	\draw[rounded corners=2pt, fill=RedViolet!70, very thick] (0-\s,0-\s) rectangle (0+\s,0+\s);
	\draw[rounded corners=2pt, fill=RedViolet!70, very thick] (0.5-\s,0.5-\s) rectangle (0.5+\s,0.5+\s);
	\draw[rounded corners=2pt, fill=RedViolet!70, very thick] (0-\s,1-\s) rectangle (0+\s,1+\s);
    \draw[white!5] (0+\s-0.15,0+\s-0.05) -- (0+\s-0.05,0+\s-0.05) -- (0+\s-0.05,0+\s-0.15);
    \draw[white!5] (0.5+\s-0.15,0.5+\s-0.05) -- (0.5+\s-0.05,0.5+\s-0.05) -- (0.5+\s-0.05,0.5+\s-0.15);
    \draw[white!5] (0+\s-0.15,1+\s-0.05) -- (0+\s-0.05,1+\s-0.05) -- (0+\s-0.05,1+\s-0.15);
    \def\x{-0.35}
    \draw[fill=gray,very thick] (\x-0.1,-0.85) -- (\x+0.1,-0.85) -- (\x,-1) -- cycle;
    \draw[fill=gray,very thick]  (0.35,1.35) circle (0.1);
    \draw[fill=gray,very thick]  (0.35+0.5,1.35-0.5) circle (0.1);
    \draw[fill=gray,very thick]  (0.35,-0.35) circle (0.1);
    \draw[fill=gray,very thick]  (0.35+0.5,-0.35+0.5) circle (0.1);
	\end{tikzpicture}
}\times
\left(\cbox{\begin{tikzpicture}[scale=0.8]\draw[very thick](0,0)--(0,0.5);
    \draw[fill=gray, very thick] (0,0) circle (0.1);
    \draw[fill=gray, very thick] (0,0.5) circle (0.1);
    \end{tikzpicture}}\right)^{L_E/2}
+
\cbox{
	\begin{tikzpicture}[scale=0.8]
    \def\s{0.2}
	\draw[black,very thick] (0-0.35,-0.35)--(0-0.35,-0.85);
	\draw[black,very thick] (0-0.35,-0.35)--(0+0.35,+0.35);
    \draw[black,very thick] (0-0.35,+0.35)--(0+0.35,-0.35);
    \draw[black,very thick] (0.5-0.35,0.5-0.35)--(0.5+0.35,0.5+0.35);
    \draw[black,very thick] (0.5-0.35,0.5+0.35)--(0.5+0.35,0.5-0.35);
    \draw[black,very thick] (0-0.35,1-0.35)--(0+0.35,1+0.35);
    \draw[black,very thick] (0-0.35,1+0.35)--(0+0.35,1-0.35);
    \draw[black,very thick] (-0.35,0.35)--(-0.35,1-0.35);
	\draw[rounded corners=2pt, fill=RedViolet!70, very thick] (0-\s,0-\s) rectangle (0+\s,0+\s);
	\draw[rounded corners=2pt, fill=RedViolet!70, very thick] (0.5-\s,0.5-\s) rectangle (0.5+\s,0.5+\s);
	\draw[rounded corners=2pt, fill=RedViolet!70, very thick] (0-\s,1-\s) rectangle (0+\s,1+\s);
    \draw[white!5] (0+\s-0.15,0+\s-0.05) -- (0+\s-0.05,0+\s-0.05) -- (0+\s-0.05,0+\s-0.15);
    \draw[white!5] (0.5+\s-0.15,0.5+\s-0.05) -- (0.5+\s-0.05,0.5+\s-0.05) -- (0.5+\s-0.05,0.5+\s-0.15);
    \draw[white!5] (0+\s-0.15,1+\s-0.05) -- (0+\s-0.05,1+\s-0.05) -- (0+\s-0.05,1+\s-0.15);
    \def\x{-0.35}
    \draw[fill=gray,very thick] (\x-0.1,-0.85) -- (\x+0.1,-0.85) -- (\x,-1) -- cycle;
    \draw[fill=gray,very thick]  (0.35,1.35) circle (0.1);
    \draw[fill=gray,very thick]  (0.35+0.5,1.35-0.5) circle (0.1);
    \draw[fill=gray,very thick]  (0.35,-0.55) rectangle (0.55,-0.35);
    \draw[fill=gray,very thick]  (0.35+0.5,-0.55+0.5) rectangle (0.55+0.5,-0.35+0.5);
	\end{tikzpicture}
}\times
\left(\cbox{\begin{tikzpicture}[scale=0.8]\draw[very thick](0,0)--(0,0.5);
    \draw[fill=gray, very thick] (-0.1,0.-0.1) rectangle (0.1,0.1);
    \draw[fill=gray, very thick] (0,0.5) circle (0.1);
    \end{tikzpicture}}\right)^{L_E/2}\,.
}
Every contraction between the identity and the SWAP returns a factor which is half of that returned by the contraction between two identities (or more generally a factor set by the inverse of the local Hilbert space dimension), and the total number of such contractions here scales as $L_E/2$. These overlaps already guarantee a suppression of $\Delta$ by a factor $2^{L_E/2} = D_E^{1/2}$. This scaling agrees with the numerical results of Fig.~\ref{fig:SDKI-Delta}. This reduced suppression follows from the inability of the spatial region $L_E$ to fully transfer the SWAPs to the spatial region $R$, a signature of imperfect transfer of quantum information or quantum teleportation. In the fully deep thermalised regime, by contrast, the dynamics effectively encodes an isometric transformation from the temporal lattice of $t$ sites to the spatial lattice of $L_R$ sites, indicating perfect transfer of quantum information.

\subsubsection{PoPs and Erlang distribution}

We next turn to results for the PoPs for the SDKI chain. 
The results for ${\rm PoP}_{\rm PPE}$ are qualitatively the same as that of the kicked Ising chain at generic parameters (see Fig.~\ref{fig:tfi-erg-pop}), with the unerlying physics also being identical. 
However, the self-duality induces some additional features, which contain spatiotemporal information, in ${\rm PoP_{b{\text{-}}str}}$ which can be computed analytically.
In particular, for $t<L_E/2$, which corresponds to the decoupled regime, we will show in the following that ${\rm PoP_{b{\text{-}}str}}(\tilde{p},\rho_{RS})$ is a Dirac-delta function at $\tilde{p}=1$.
More generally, we will also show that the ${\rm PoP_{b{\text{-}}str}}$ of any contiguous subsystem with one of its ends at the boundary of the system shows qualitatively different behaviour across different timescales set by the lengths of traced out regions, thereby 
\new{showing} again the spatiotemporal structure of the dynamics. 

As a first step, let us first compute ${\rm PoP_{b{\text{-}}str}}(\tilde{p},\rho_{S})$.
The Porter-Thomas distribution arises as the PoP when performing projective measurements on the full system for a pure state, i.e. fixing $L_R = L_E = 0$. Note that these probabilities are absolute probabilities rather than the relative probabilities of Sec.~\ref{sec:pop}. We now consider bit-string probabilities for measurement outcomes $\{\ket{o_S}\}$ on $S$ when $R$ and $E$ are not measured. These probabilities correspond exactly to the probabilities appearing in the definition of the PPE (see Eq.~\ref{eq:p-os}). In this definition $R$ and $E$ are treated on an equal footing, since both are traced over. The probabilities $p(o_S)$ can be interpreted as bit-string probabilities for the full system, marginalised with respect to the measurement outcomes on $R$ and $E$ as
\eq{
p(o_S) = \sum_{o_{RE}} p(o_{RES}), \quad \ket{o_{RES}} = \ket{o_{RE}} \otimes \ket{o_S}.
\label{eq:p-o_s-marginalized}
}
The corresponding PoP, however, does not follow directly from the PoP for the bit-string probabilities of the full system. The different bit-string probabilities that are summed over are highly correlated due to the spatiotemporal structure of the unitary dynamics, such that the mixed-state PoP cannot be directly obtained through the repeated convolution of the pure-state PoP with itself (as would be the case if the different terms in Eq.~\ref{eq:p-o_s-marginalized} were statistically independent).

\begin{figure}[!t]
\includegraphics[width=\linewidth]{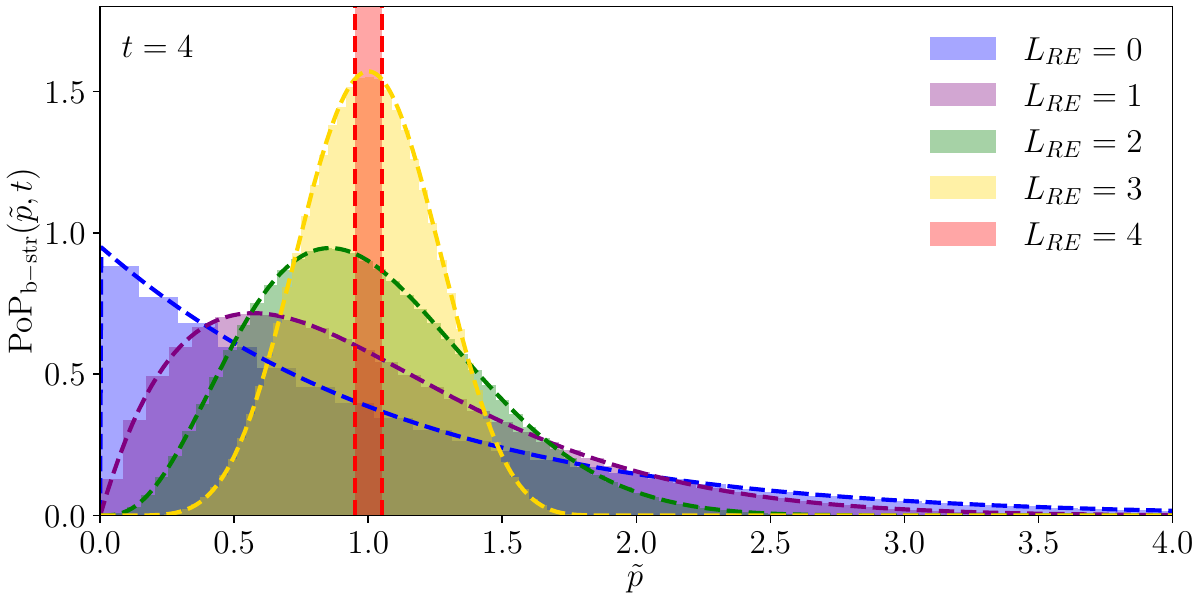}
\caption{PoP for $p(o_S)$ in the self-dual kicked Ising chain (with $\ket{+}^{\otimes L}$ initial state) for fixed total system size $L=24$, $t=4$ time steps, and varying sizes $L_{RE}$. Numerically obtained PoPs (histograms) are compared with the analytical prediction from Eq.~\ref{eq:SDKI-PoP} (dashed lines), showing excellent agreement. The beta distribution smoothly interpolates between the Porter-Thomas distribution ($L_{RE}=0$) and the Erlang distribution $0 < L_{RE} < t$, before collapsing to a delta function ($L_{RE} \geq t$).
}
\label{fig:SDKI-PoP}
\end{figure}

This PoP can be analytically calculated for the SDKI chain using the same techniques as originating in the calculation of deep thermalisation in this model~\cite{ho2022exact}. The self-duality allows for an alternative interpretation of the dynamics as unitary evolution in space, where sampling over different measurement outcomes corresponds to sampling different spatial unitaries (see also Ref.~\cite{stephen2024universal}). For ergodic Ising dynamics these spatial unitaries constitute a universal gate set, such that the spatial dynamics quickly gives rise to Haar-random states in the spatial direction. The full derivation of the resulting PoP is presented in App.~\ref{app:PoP_SDKI} and we here only quote the final result. In the thermodynamic limit $L_S \to \infty$, the PoP takes the form of a beta distribution depending only on the total traced out region $L_{RE} = L_R+L_E$ for $t > L_{RE}$ time steps:
\eq{
{\rm PoP_{b{\text{-}}str}}(\tilde{p};t) \propto  &\left(\frac{\tilde{p}}{2^{t-{L_{RE}}}}\right)^{\alpha-1} \left(1-\frac{\tilde{p}}{2^{t-L_{RE}}}\right)^{\beta-1}\nonumber\\
&\qquad \times\, \theta(2^{t-L_{RE}}-\tilde{p}),
\label{eq:SDKI-PoP}
}
with $\alpha = 2^{L_{RE}} = D_{RE}$, $\beta = 2^{t}-2^{L_{RE}} = D_t - D_{RE}$, and where $\theta$ is the step function. The probabilities are normalized as $\tilde{p}(o_S) = 2^{L_S} p(o_S)$ such that $\tilde{p}$ has average value 1.
For $t \leq L_{RE}$ the tracing out of the $R \cup E$ region results in a collapse of the PoP to the delta function. This latter collapse is particular to the setup considered and hence nonuniversal, such that we will focus on the late-time appearance of beta distribution.
This distribution is illustrated in Fig.~\ref{fig:SDKI-PoP} for a fixed number of time steps $t=4$ and different values of $L_{RE}$. Numerical results for the PoP obtained for a fixed system size of $L = 14$ are compared with the analytical expression from Eq.~\ref{eq:SDKI-PoP}, showing excellent agreement even for finite system sizes.

\begin{figure}[!b]
\includegraphics[width=\linewidth]{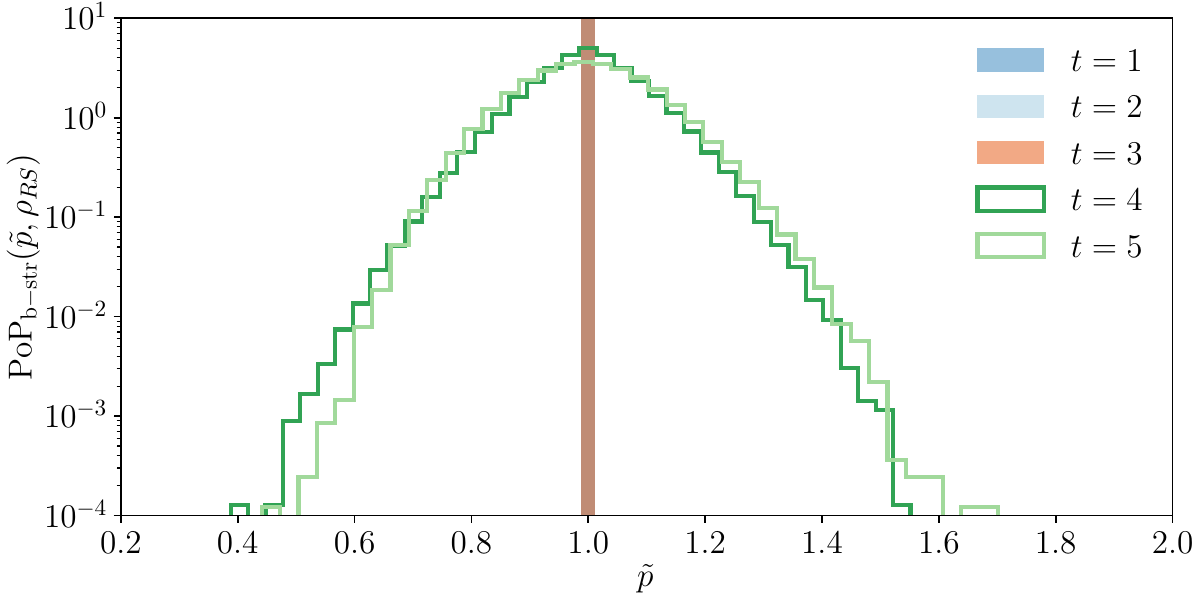}
\caption{PoP over the bit-strings of $R\cup S$ for the self-dual kicked Ising chain at different times $t$. The results are for $L_E=6$ which corresponds to $t_\ast=3$. For $t\leq t_\ast$, {\it i.e.} for $R$ and $S$ decoupled, the PoPs collapse to the Dirac-delta function whereas the the PoP deviates from the Dirac-delta function once the lightcones intersect. For the data we have $L_R=3$ and $L_S=15$.
}
\label{fig:SDKI-PoPRS}
\end{figure}

Different limits can be clearly identified. For $L_{RE}=0$ the beta distribution with $\alpha=0$ reduces to a rescaled Porter-Thomas distribution, recovering the results of Ref.~\cite{claeys2024fock-space}. This PoP approaches the exponential distribution exponentially quickly as $t$ increases:
\eq{
{\rm PoP_{b{\text{-}}str}}(\tilde{p};t) = \left(1-\frac{\tilde{p}}{2^{t}}\right)^{2^{t}-2} \approx e^{-\tilde{p}} \quad \textrm{for} \quad t \gg 1,
}
where we have dropped the step function since it is always satisfied in this limit.
In the limit $t \gg L_{RE}$ a similar approximation can be performed for the mixed-state PoP to return the Erlang distribution:
\eq{
{\rm PoP_{b{\text{-}}str}}(\tilde{p};t) \propto \tilde{p}^{D_{RE}-1} e^{- D_{RE} \tilde{p}} \quad \textrm{for} \quad t \gg L_{RE}.
}
We therefore analytically recover the expected universal distribution of the mixed-state PoP. For large $D_{RE}$, this distribution in turn approaches a Gaussian distribution with unit mean and variance $1/D_{RE}$. This crossover from a Porter-Thomas to a Gaussian distribution can be seen as a quantum-to-classical crossover due to the loss of information about the $RE$ region~\cite{shaw2025experimental}.

Let us now relate this result to the spatiotemporal lightcone. Just as for $t\leq L_{RE}$ we have ${\rm PoP_{b{\text{-}}str}}(\tilde{p},\rho_{S})=\delta(\tilde{p}-1)$, by the same token, for $t\leq L_{SE}$ we have ${\rm PoP_{b{\text{-}}str}}(\tilde{p},\rho_{R})=\delta(\tilde{p}-1)$.
This naturally implies that for the timescales $t\leq t_\ast = L_E/2$, both ${\rm PoP_{b{\text{-}}str}}(\tilde{p},\rho_{S})$ and ${\rm PoP_{b{\text{-}}str}}(\tilde{p},\rho_{R})$ collapse to the Dirac-delta functions. 
This timescale also corresponds to the one throughout which $R$ and $S$ are decoupled from each other.
Therefore, using the results in Eq.~\ref{eq:mellin-integ}-\ref{eq:mellin}, we have
\eq{
{\rm PoP_{b{\text{-}}str}}(\tilde{p},\rho_{RS}) =\delta(\tilde{p}-1)\,;~t\leq t_\ast\,,
}
which is simply due to the fact that the Mellin convolution of two Dirac-delta functions is also a Dirac-delta function. 
The PoP of $R\cup S$ over the bit-strings therefore 
\new{reflects} the lightcone of information spreading in the system. 
The results in Fig.~\ref{fig:SDKI-PoPRS} provide a numerical corroboration of the result.

\subsection{Late times and deep thermalisation \label{sec:latetimes}}

Finally we discuss the behaviour of the infinite-time profile of $\Delta_\infty(L_E)\equiv\Delta(L_E,t\to\infty)$ in the limit of $L_S\to\infty$. For the SDKI chain, the $t\to\infty$ result sets in already at $t=L_R+L_E$ due to the formation of an exact quantum design within that timescale~\cite{ho2022exact}.
In particular, we show that the exponential decay $\Delta_\infty(L_E)\sim 2^{-L_E}$ is a straightforward outcome of the fact the state of $R\cup E$ deep thermalises to the Haar ensemble (see Eq.~\ref{eq:PE-tinf-RE}).
This directly implies the second moment of the PE on $R\cup E$ is given by
\eq{
\rho_{RE}^{(2)}(\infty) = \frac{1}{D_R D_E(D_RD_E+1)}[\mathbb{I}_{RE^{\otimes2}}+\mathbb{S}_{RE^{\otimes2}}]\,,
}
${\mathbb{I}}_{RE^{\otimes2}}$ and ${\mathbb{S}}_{RE^{\otimes2}}$ are the identity and SWAP operators respectively acting on the  doubled Hilbert space of $R\cup E$. From the above result, the second moment of PPE on $R$ can be derived as
\eq{
\rho_R^{(2)}(\infty) &= {\rm Tr}_E^{\otimes 2}\left[\rho_{RE}^{(2)}(\infty)\right]\nonumber\\
&=\frac{[D_E^2\mathbb{I}_{R^{\otimes2}}+D_E\mathbb{S}_{R^{\otimes2}}]}{D_R D_E(D_RD_E+1)}\,.
\label{eq:rhoR2-tinf}
}
Note that the prefactor of $D_E^{-1}$ infront of $\mathbb{S}_{R^{\otimes 2}}$ relative to $\mathbb{I}_{R^{\otimes 2}}$ was obtained exactly in Eq.~\ref{eq:rhoR2-SDKI-DT} for the SDKI at $t>L_R+L_E$.
At the same time, thermalisation (to infinite temperature, in this case) implies 
\eq{
\rho_R^{\otimes 2}(\infty)=\frac{1}{D_R^2}\mathbb{I}_{R^{\otimes2}}\,.
\label{eq:rhoR-i2}
}
From Eq.~\ref{eq:rhoR2-tinf} and Eq.~\ref{eq:rhoR-i2} we have
\eq{
\rho_R^{(2)}(\infty)-\rho_R^{\otimes 2}(\infty) \approx \frac{1}{D_R(D_RD_E+1)}\mathbb{S}_{R^{\otimes2}}\,,
\label{eq:rho-diff-inf}
}
from which it trivially follows that $\Delta_\infty(L_E)\overset{D_E\gg 1}{\approx}D_E^{-1}$ which concludes the analytical derivation of the result for $\Delta_\infty(L_E)$.

\begin{figure}
\includegraphics[width=\linewidth]{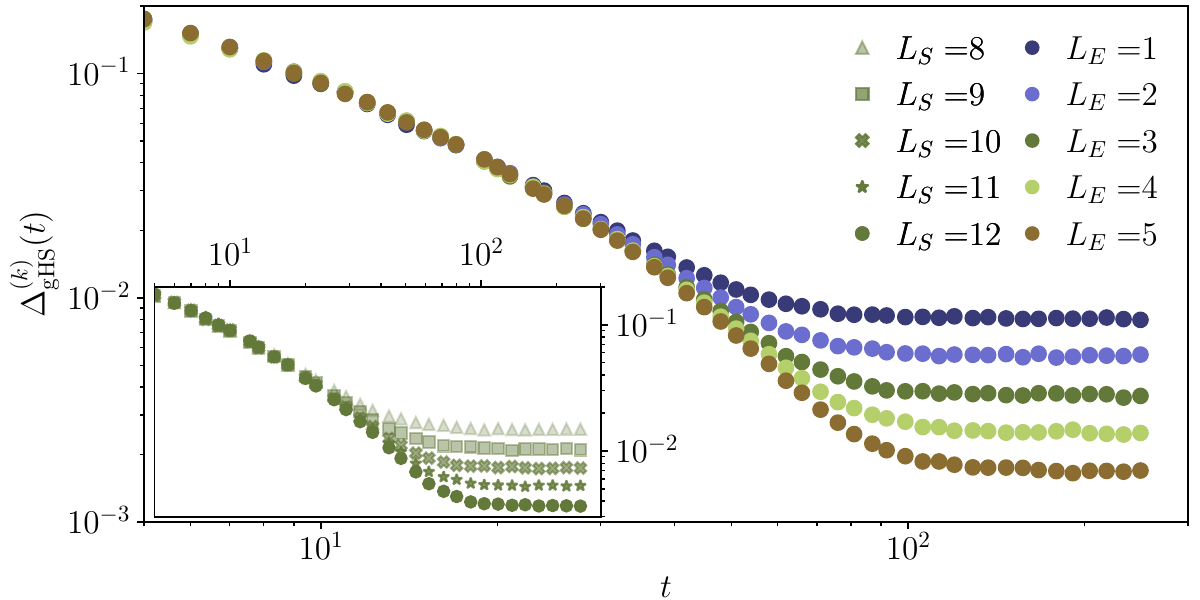}
\caption{The distance of the PPE from the corresponding gHSE, $\Delta_{\text{gHS}}^{(k=2)}(t)$, (defined in Eq.~\ref{eq:Delta-gHS}) as a function of time $t$ for the kicked Ising chain for $L_R=1$ with generic parameters. The main panel shows $\Delta_{\text{gHS}}^{(k)}(t)$ for different values of $L_E$ for $L_S=12$. The curves appear to follow a $L_E$-independent master curve except for the saturation. The inset shows that the saturation is due to finite $L_S$ and in the limit of $L_S\to\infty$, the saturation value goes to zero. The two panels together therefore show that the dynamics of $\Delta_{\text{gHS}}^{(k)}(t)$ is independent of $L_E$ in the thermodynamic limit.}
\label{fig:Delta_GHSE}
\end{figure}

Before closing this section, let us briefly comment on the temporal approach of the PPE to the gHS ensemble, in particular for the generic kicked Ising chain. 
The results for $\Delta_{\text{gHS}}^{(k)}(t)$ (defined in Eq.~\ref{eq:Delta-gHS}) shown in Fig.~\ref{fig:Delta_GHSE} show that the distance of the PPE from the appropriate gHS ensemble -- obtained by tracing out the corresponding $E$ from a Haar ensemble on $R\cup E$ -- at any particular $t$ is independent of $D_E$.
The interesting point about this result is that at any given $t$, the projected ensemble on $R\cup E$ is further away from its asymptotic ensemble, the Haar-ensemble, for larger $E$. Nevertheless, on tracing out $E$, the distance of the resultant PPE from its corresponding asymptotic gHS ensemble, becomes independent of $E$. 
However, a detailed understanding of this work is neither the focus of this work, nor within its scope and we leave it as a question for the future.

\section{Many-body localised circuits \label{sec:mbl}}
In this section, we will focus on circuits whose dynamics are in the many-body localised (MBL) regime. The MBL regime is rather unusual and interesting as the system does not thermalise under dynamics and there is no quantum chaos, and yet information spreads through the system albeit in an ultraslow logarithmic fashion~\cite{bardarson2012unbounded,serbyn2013universal,chen2016universal,chen2017out,huang2017out,deng2017logarithmic,pain2024entanglement,thakur2025logarithmic}. 
We find that the fluctuations in the PPE, $\Delta(t,L_E)$ defined in Eq.~\ref{eq:ppe-fluc}, show a logarithmic lightcone in the $(t,L_E)$ plane with the emergence of timescales which are exponentially large in $L_E$.
We show numerical results for a parameter regime of the model in Eq.~\ref{eq:UF-gen} such that it lies in the MBL regime and develop a (semi)analytical understanding of the results through the phenomenological  $\ell$-bit picture~\cite{serbyn2013local,huse2014phenomenology,ros2015integrals}.

\subsection{Many-body localised kicked Ising model {\label{sec:tfi-mbl}}}

The model in Eq.~\ref{eq:UF-KI} with parameter values 
\eq{
J = 0.8,~g_i=0.7236\Gamma, h_i = 0.6472 + 0.7236\sqrt{1-\Gamma^2}\epsilon_i
\label{eq:MBL-params}
}
with $\epsilon_i\sim {\cal N}(0,1)$ exhibits MBL dynamics for $\Gamma\lessapprox 0.3$~\cite{zhang2016floquet}. To probe the MBL regime, we therefore use $\Gamma=0.15$ along with the said parameter values throughout this section. 
We will consider $L_R=1$ and study $\Delta(t,L_E)$ for increasing values of $L_S$, where we are interested in the eventual limit of $L_S\to\infty$.
Note that the disorder in the model which effects the MBL regime couples to the $Z$-component of the spins and the eigenstates are therefore close to $z$-configurations. To avoid any pathology arising from the measurement basis being too close to the eigenbasis, we therefore construct the PPE using measurements in the $X$-basis. However, we explicitly show in the next section that the results remain qualitatively the same even if the measurement basis overlaps perfectly with the eigenbasis.

\begin{figure}
\includegraphics[width=\linewidth]{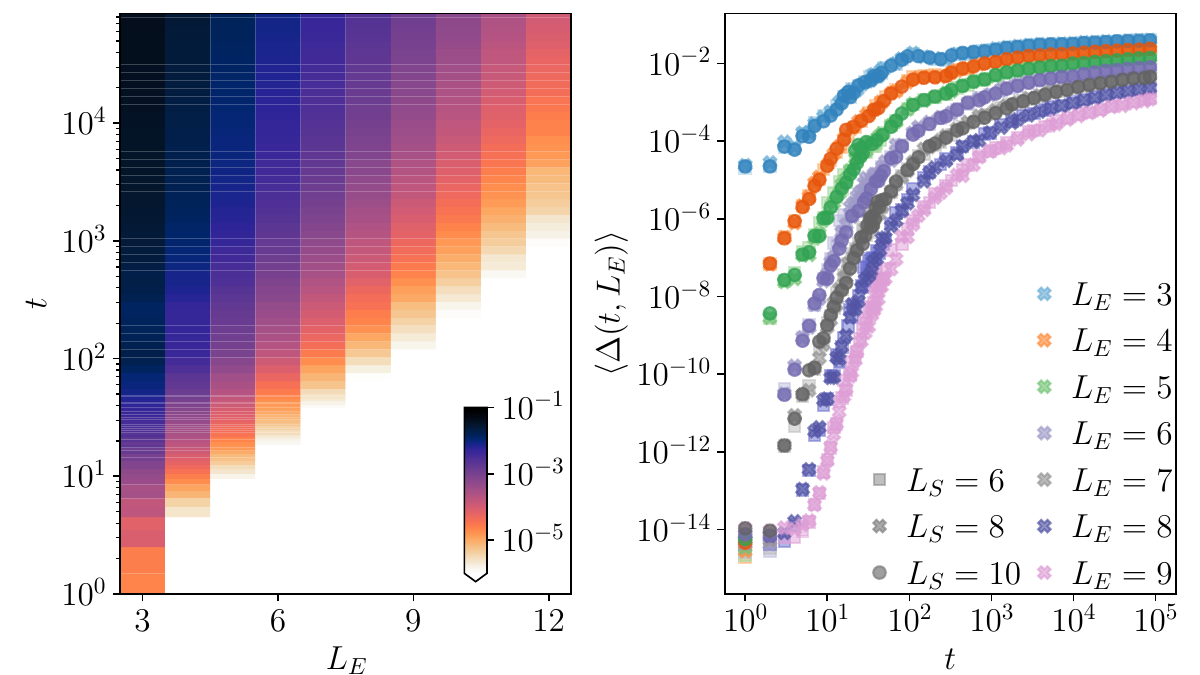}
\caption{The left panel show the heatmap of the average $\langle\Delta(t,L_E)\rangle$ for the kicked Ising chain (Eq.~\ref{eq:UF-gen}) with the parameters tuned to the MBL regime. Note that the time axis is on logarithmic scales which makes the logarithmic lightcone clear. The right panel shows $\langle\Delta(t,L_E)\rangle$ for different values of $L_E$ (different colours) and $L_S$ (different markers) as a function of $t$. The data has negligible dependence on $L_S$ suggesting the $L_S\to\infty$ limit is well approximated in the numerical results presented. All results were averaged over 1000 disorder realisation of the model and 10 initial states (Eq.~\ref{eq:psi-init} for each realisation.
}
\label{fig:mbl-tfi-lc}
\end{figure}

\begin{figure}
\includegraphics[width=\linewidth]{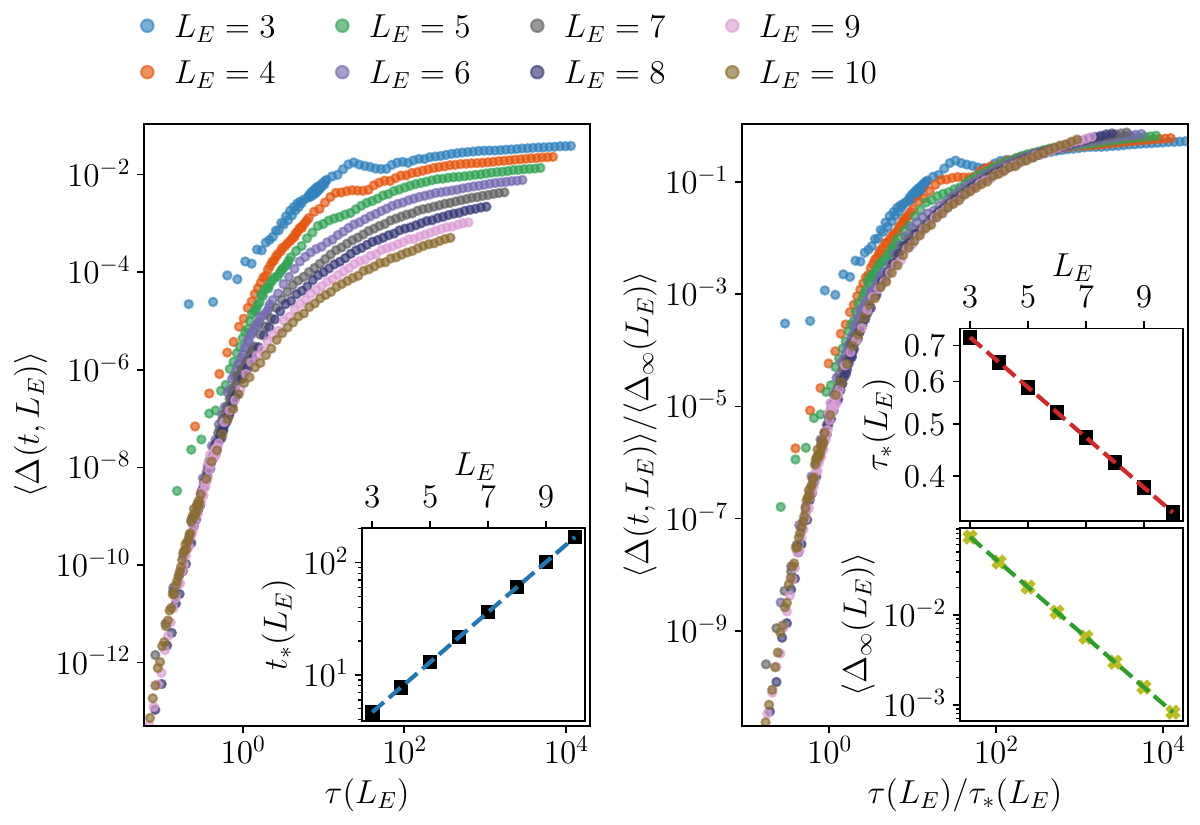}
\caption{Scaling of $\Delta(t,L_E)$ for the Floquet Ising model in the MBL regime. In the left panel $\langle\Delta(t,L_E)\rangle$ when plotted versus $\tau(L_E)\equiv t/t_\ast(L_E)$ collapses the onset of $\Delta(t,L_E)$ onto a common curve for different $L_E$. The inset shows that $t_\ast(L_E)\sim e^{L_E/\xi_t}$ with $\xi_t\approx 1.95$.
The right panel shows that rescaling $\langle\Delta(t,L_E)\rangle$ by its infinite-time saturation, $\langle\Delta_\infty(L_E)\rangle$, and rescaling $\tau(L_E)$ by $\tau_\ast(L_E)$ collapses the entire curves for different $L_E$ onto each other. The insets show that (i) the saturation value decays exponentially with $L_E$, $\braket{\Delta_\infty(L_E)}\sim 0.527^{L_E}$ and (ii) $\tau_\ast(L_E)$ decays exponentially with $L_E$, $\tau_\ast(L_E)\sim e^{-L_E/\xi_\tau}$ with $\xi_\tau\approx 9.32$. However the exponential decay is slower than the exponential growth of $t_\ast(L_E)$ such that the saturation timescale $t_\ast(L_E)\times \tau_\ast(L_E)$ still grows exponentially with $L_E$.}
\label{fig:mbl-tfi-scaling}
\end{figure}

To establish the phenomenology, we again study $\Delta(t,L_E)$ as a heatmap in the $(t,L_E)$ plane as shown in Fig.~\ref{fig:mbl-tfi-lc} (left panel). The data clearly shows that the front of $\Delta(t,L_E)$ spreads in space logarithmically in $t$. This lightcone is made quantitative in the right panel where $\Delta_{t,L_E}$ is shown as a function of $t$ for different $L_E$ and also $L_S$. 
Note that the data has a negligible dependence on $L_S$ for the values we used, indicating that it approximates the $L_S\to\infty$ limit well. 
The data suggests that for any arbitrary threshold, $\Delta_{\rm th}$, the time $t_{\rm th}(L_E)$ such that $\Delta(t_{\rm th},L_E)=\Delta_{\rm th}$ grows exponentially with $L_E$.
This motivates a rescaling of the time axis,
\eq{
\tau(L_E) = t/t_\ast(L_E)\,;\quad t_\ast(L_E) = t_0e^{L_E/\xi_t}\,,
\label{eq:tau-le}
}
such that plotting $\Delta(t,L_E)$ as a function of $\tau(L_E)$ collapses the onset of $\Delta(t,L_E)$ for different $L_E$ on top of each other.
This rescaling is shown in the left panel of Fig.~\ref{fig:mbl-tfi-scaling}. The inset confirms that $t_\ast(L_E)$ does indeed grow exponentially with $L_E$, as indicated in Eq.~\ref{eq:tau-le}.
The emergence of such a timescale which grows exponentially with $L_E$ makes the logarithmic front of $\Delta(t,L_E)$ quantitative.

The next point to note from the data is that the infinite-time saturation value, $\Delta_\infty(L_E)$, decays with $L_E$. In fact, rescaling $\Delta(t,L_E)$ by $\Delta_\infty(L_E)$, and $\tau(L_E)$ by a scale $\tau_\ast(L_E)$ collapses the entire curves (and not just the onset) of $\Delta(t,L_E)$ for different $L_E$ on top of each other; this collapse is evident in the right panel of Fig.~\ref{fig:mbl-tfi-scaling}.
The scaling collapse implies a scaling form,
\eq{
\begin{split}
\Delta(t,L_E) &= \Delta_\infty(L_E) \times f\left(\frac{\tau(L_E)}{\tau_\ast(L_E)}\right)\\
              &= \Delta_\infty(L_E) \times f\left(\frac{t}{t_\ast(L_E)\tau_\ast(L_E)}\right)
\end{split}\,,
\label{eq:lbit-full-scaling}
}
where the scaling function has the asymptotic behaviour,
\eq{
f(x) = \begin{cases}
x^a\,;&x\ll 1\\
1\,; &x\to\infty
\end{cases}\,.
\label{eq:lbit-scaling-asymp}
}
The insets to the right panel in Fig.~\ref{fig:mbl-tfi-scaling} show that $\Delta_\infty(L_E)$ decays exponentially with $L_E$; we shall provide an analytical understanding of the same in the next subsection. 
The insets also show that the timescale $\tau_\ast(L_E)$ decays exponentially with $L_E$, $\tau_\ast(L_E)\sim e^{-L_E/\xi_\tau}$. This can be understood as follows; for any given $L_E$ the onset of $\Delta(t,L_E)$ occurs at timescale which exponentially large on $L_E$.
At the same time, the saturation value is exponentially small in $L_E$. This naturally means that the dynamical range over which $\Delta(t,L_E)$ shows dynamics, namely the window of time after the onset but before the saturation, is exponentially small in $L_E$.


{\color{blue}
This discussion clarifies the physical interpretation of the two characteristic timescales, $t_\ast(L_E)$ and $\tau_\ast(L_E)$. The first, $t_\ast(L_E)$, sets the time at which the front of the light cone associated with $\Delta(t,L_E)$ reaches a distance $L_E$; it therefore captures how rapidly correlations propagate through the system. The second, $\tau_\ast(L_E)$, controls the subsequent relaxation dynamics of $\Delta(t,L_E)$ once the front has arrived, determining how quickly it approaches its stationary value at that distance. 

The overall saturation time,
\begin{equation}
    t_{\mathrm{sat}}(L_E) \equiv t_\ast(L_E)\,\tau_\ast(L_E),
\end{equation}
combines these two processes. While $t_\ast(L_E)$ increases exponentially with $L_E$, $\tau_\ast(L_E)$ decreases exponentially, reflecting the growing delay in front propagation and the simultaneously faster local equilibration near the front. Crucially, the rate of the exponential decrease of $\tau_\ast(L_E)$ is smaller than the rate of the exponential increase of $t_\ast(L_E)$. As a result, their product $t_{\mathrm{sat}}(L_E)$ still grows exponentially with $L_E$, leading to saturation times that are extremely large for distant regions. 

This hierarchy of timescales directly underlies the logarithmic light-cone structure observed in $\Delta(t,L_E)$ (see Fig.~\ref{fig:mbl-tfi-lc}). In the next subsection, we provide a semi-analytical framework that explains the emergence of this slow, logarithmic spreading in terms of the competing exponential dependences described above.

}

\subsection{Analytical results for the $\ell$-bit model}
With the phenomenology of $\Delta(t,L_E)$ in the MBL regime at hand, we now turn towards a (semi)analytical understanding of the logarithmic lightcone, based on the $\ell$-bit picture~\cite{serbyn2013local,huse2014phenomenology,ros2015integrals}. 
Within this picture, an MBL system is approximately described via extensive set of mutually commuting local integrals of motion, the $\ell$-bits, each of which also commutes with the Hamiltonian. 
Denoting the $\ell$-bit at site $i$ via $Z_i$, the $\ell$-bit Hamiltonian is given by
\begin{equation}
\label{eq:lbithamiltonian}
H_{\ell{\text {-bit}}}=\sum_{i}J_{i}Z_{i}+\sum_{i>j}J_{ij}Z_{i}Z_{j}+\sum_{i>j>k}J_{ijk}Z_{i}Z_{j}Z_{k}+\cdots
\end{equation}
where the random couplings decay exponentially in space with the form
\eq{
J_{i_1i_2\cdots i_k} = r_{i_1i_2\cdots i_k}e^{-(i_1-i_k)/\xi}\,,
}
and $r_{i_1i_2\cdots i_k}\sim{\cal N}(0,1)$ are independent Gaussian random numbers. Since the Hamiltonian in Eq.~\ref{eq:lbithamiltonian} is composed only of $Z$-operators, its eigenstates are just spin configurations polarised along the $z$-direction which we denote as $\ket{z}$ and the corresponding eigenvalues are denoted as $E_{z}$. 
The structure of the $\ell$-bit Hamiltonian implies that the $\ell$-bit at site $i$ feels an effectively magnetic field along the $z$-direction given by
\eq{
\label{eq:eff-lbit}
H_i(z^{(i)})= J_i + H_i^{(1)}(z^{(i)})+H_i^{(2)}(z^{(i)})+\cdots\,,
}
where $z^{(i)}$ is the spin-configuration barring site $i$ and $H_i^{(l)}$ denotes the magnetic field at site $i$ arising from the interactions of the spin with those within distance $l$ from it and naturally $H_i^{(l)}\sim e^{-l/\xi}$.

As before we start with an initial state of the form in Eq.~\ref{eq:psi-init}, such that the state at time $t$ is given by
\eq{
\ket{\psi_{RSE}(t)}=\sum_{z}\left(\prod_{i=1}^L A_{i z_i}\right) e^{-i E_{z} t}\ket{z}\,,
\label{eq:lbitevolvedstate}
}
As a matter of notation for later convenience, the eigenstate $\ket{z}$ can be expressed as a direct product of spin configurations on $R$, $S$, and $E$ as $\ket{z}\equiv  \ket{z_R}\otimes\ket{z_E}\otimes\ket{z_S}$, and similarly the overlap of the initial state on an eigenstate also factorises as $\prod_{i}A_{iz_i} = A_{z_R} A_{z_E} A_{z_S}$ where $A_{z_X} = \prod_{i\in X}A_{iz_i}$.  

One important point to note for the $\ell$-bit model is that since the conserved quantities are 1-local, and so are the measurements on $S$, the projected ensemble, and therefore naturally the PPE, is qualitatively different for measurements in the $Z$-basis and for those where the basis is tilted away from $Z$~\cite{manna2025projected}. 
This is intimately connected to the idea that initial states with conserved charge fluctuations or charge non-revealing measurements lead to the projected ensembles approaching Scrooge ensembles whereas for symmetric initial states with charge revealing measurements the projected ensemble is described by a generalised Scrooge ensemble~\cite{mark2024maximum,chang2025deep}. 
However, it was shown for systems with locally conserved charges that the for all measurements except the ones that correspond exactly to the charges, the projected ensemble is described by the Scrooge ensemble~\cite{manna2025projected}.
Since for the $\ell$-bit model the set $\{Z_i\}$ constitutes exactly the locally conserved charges, we treat the cases of $Z$-measurements and $X$-measurements separately, starting with the former.

\subsubsection{$Z$-measurements}

Since the $Z_i$'s are integrals of motion, in this case the probability of obtaining a specific bit-string $z_S$ after the measurements 
\eq{
p(z_S) =|A_{z_S}|^2\,,
\label{eq:pzs-lbit}
}
is independent of $t$ and set by the initial state. The conditional state of $R$ is given by
\eq{
\rho_R(t,z_S) = \sum_{z_R,z_R^\prime}\ket{z_R}\bra{z_R^\prime}\varrho_{z_Rz_R^\prime}(t,z_S)\,,
\label{eq:rho-R-lbit-gen}
}
where 
\eq{
\varrho_{z_Rz_R^\prime}(t,z_S)=A_{z_R}A_{z_R^\prime}^\ast\sum_{z_E}|A_{z_E}|^2 e^{-it(E_{z_Rz_Ez_S}-E_{z^\prime_Rz_Ez_S})}\,.
}
Specialising to the case of $L_R=1$ for simplicity, the conditional state in Eq.~\ref{eq:rho-R-lbit-gen} can be written explicitly as 
\eq{
\rho_R(t,z_S) = 
\begin{bmatrix}
|A_{\uparrow_R}|^{2} & \varrho_{\uparrow\downarrow}(t,z_{S}) \\
\varrho_{\downarrow\uparrow}(t,z_{S} )& |A_{\downarrow_R}|^{2}
\end{bmatrix}\,,
\label{eq:rho-R-lbit-Z}
}
where the off-diagonal matrix element is given by
\eq{
\label{eq:offdiagonal}
\varrho_{\uparrow\downarrow}(t,z_{S})=A_{\uparrow_R}A_{\downarrow_R}^{\ast}\sum_{z_{E}}|A_{z_{E}}|^2e^{-2iH_{R}(z_{E}, z_{S})t}
}
and $\varrho_{\downarrow\uparrow}(t,z_{S})=\varrho^{\ast}_{\uparrow\downarrow}(t,z_{S})$ with the effective field, $H_R(z_E,z_S)$, seen by the spin in $R$ defined in Eq.~\ref{eq:eff-lbit}.
With the conditional states $\rho_R(t,z_S)$ in Eq.~\ref{eq:rho-R-lbit-gen} and their probabilities $p(z_S)$ in Eq.~\ref{eq:pzs-lbit} at hand, arbitrary moments of the PPE can be constructed explicitly.
As such, the fluctuations in the PPE (defined via Eq.~\ref{eq:ppe-fluc}) turns out to be
\eq{
\Delta(t,L_E) = |X(t)| + |Y(t)|\,, 
\label{eq:Delta-lbit-XY}
}
with
\eq{
\begin{split}
X =& \sum_{z_S}p(z_S)\varrho_{\uparrow\downarrow}^2(z_S)-\left(\sum_{z_S}p(z_S)\varrho_{\uparrow\downarrow}(z_S)\right)^2\,,\\
Y =& \sum_{z_S}p(z_S)|\varrho_{\uparrow\downarrow}(z_S)|^2-|\sum_{z_S}p(z_S)\varrho_{\uparrow\downarrow}(z_S)|^2\,,
\end{split}
\label{eq:Delta-lbit-Z}
}
where we have made the dependence on $t$ implicit for brevity. In the following we show how the logarithmic lightcone $\Delta(t,L_E)$ falls out directly from the structure of the effective field in Eq.~\ref{eq:eff-lbit} when used in Eq.~\ref{eq:Delta-lbit-Z}.

Deep in the MBL regime, one expects the lengthscale $\xi\ll1$ such that there is separation in scales of the effective fields due to interactions within range $l$ and range $l+1$ with $|H_i^{(l)}|\gg |H_i^{(l+1)}|$. 
This implies that at any time $t$, the off-diagonal matrix element in Eq.~\ref{eq:offdiagonal} for the spin in $R$ in sensitive to the interactions of the spin with those within a distance $r(t)\sim \xi\ln t$.
Given that the distance between $R$ and $S$ is $L_E$, there exists a natural timescale $t_\ast(L_E)\sim e^{cL_E/\xi}$ such that for $t<t_\ast(L_E)$, $\rho_R(t,z_S)$ is in fact independent of $z_S$ and for $t>t_\ast(L_E)$, the state of $R$ depends non-trivially on $z_S$ through Eq.~\ref{eq:offdiagonal}.
In the former case, $\varrho_{\uparrow\downarrow}(z_S) = \varrho_{\uparrow\downarrow}$ such that $X=0=Y$ in Eq.~\ref{eq:Delta-lbit-Z}. On the other hand, for $t>t_\ast(L_E)$, $\varrho_{\uparrow\downarrow}(z_S)$ in Eq.~\ref{eq:offdiagonal} depends non-trivially on $z_S$ such that $X,Y\neq 0$ in general, which in turn implies $\Delta(t,L_E)\neq 0$.
The emergence of a timescale $t_\ast(L_E)\sim e^{cL_E/\xi}$ such that
$\Delta(t,L_E) \approx 0$ for $t<t_\ast(L_E)$ and $\Delta(t,L_E) > 0$ for $t>t_\ast(L_E)$
makes the logarithmic lightcone evident. 

\begin{figure}
\includegraphics[width=\linewidth]{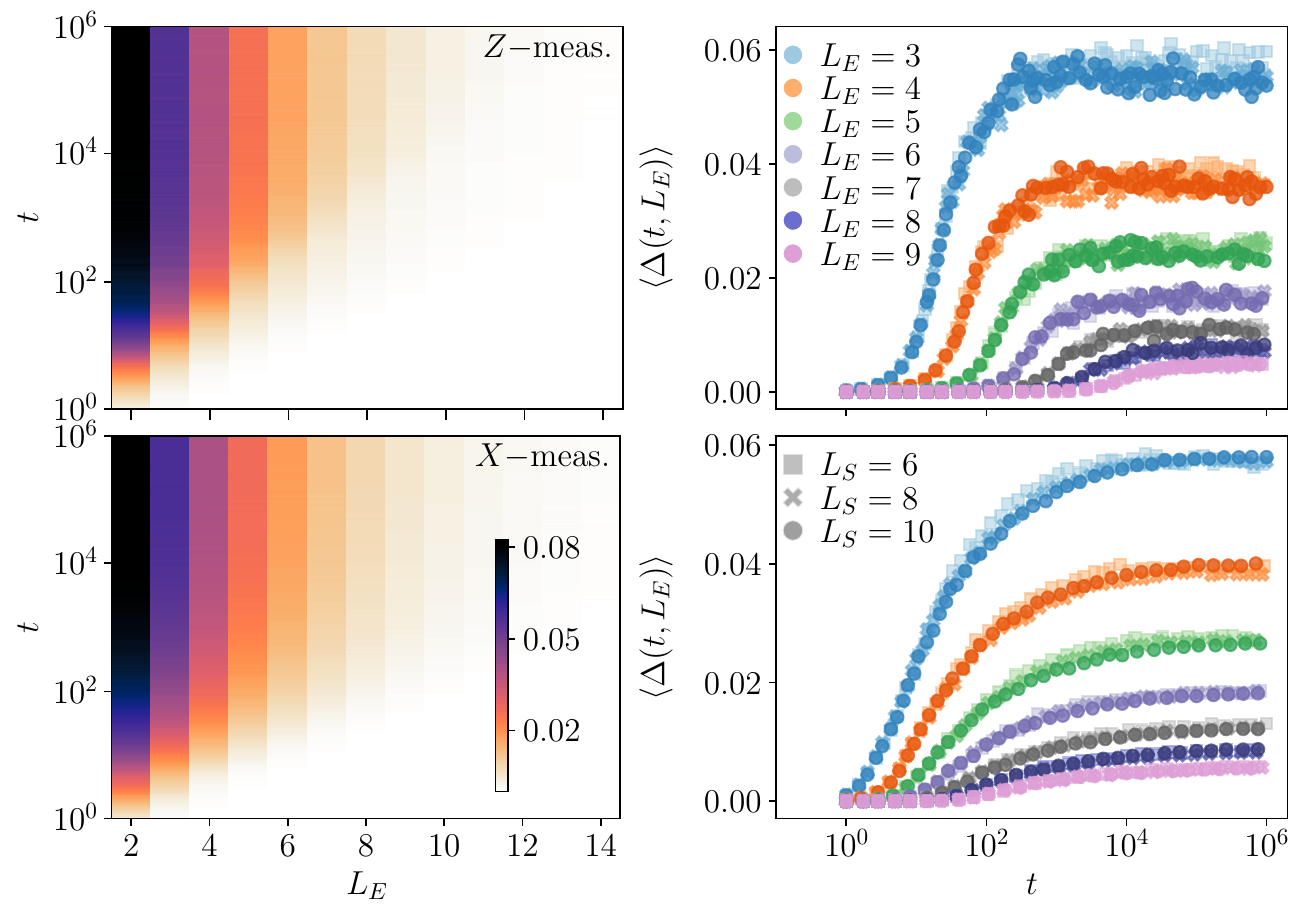}
\caption{$\Delta(t,L_E)$ for the $\ell$-bit model (defined in Eq.~\ref{eq:lbithamiltonian}) for both $Z$ (top row) and $X$ measurements (bottom row). The left panels show the data for $\Delta(t,L_E)$ as a heatmap in the $(L_E,t)$ plane which shows the logarithmic lightcone. The right panels show $\Delta(t,L_E)$ as a function of $t$ for different values of $L_E$ denoted by different colours. The data is already suggestive of a timescale for the onset of $\Delta(t,L_E)$ which grows exponentially with $L_E$. Different shades denote different values of $L_S$ which shows the negligible $L_S$ dependence in our results.}
\label{fig:lightcone-lbit}
\end{figure}

\begin{figure}
\includegraphics[width=\linewidth]{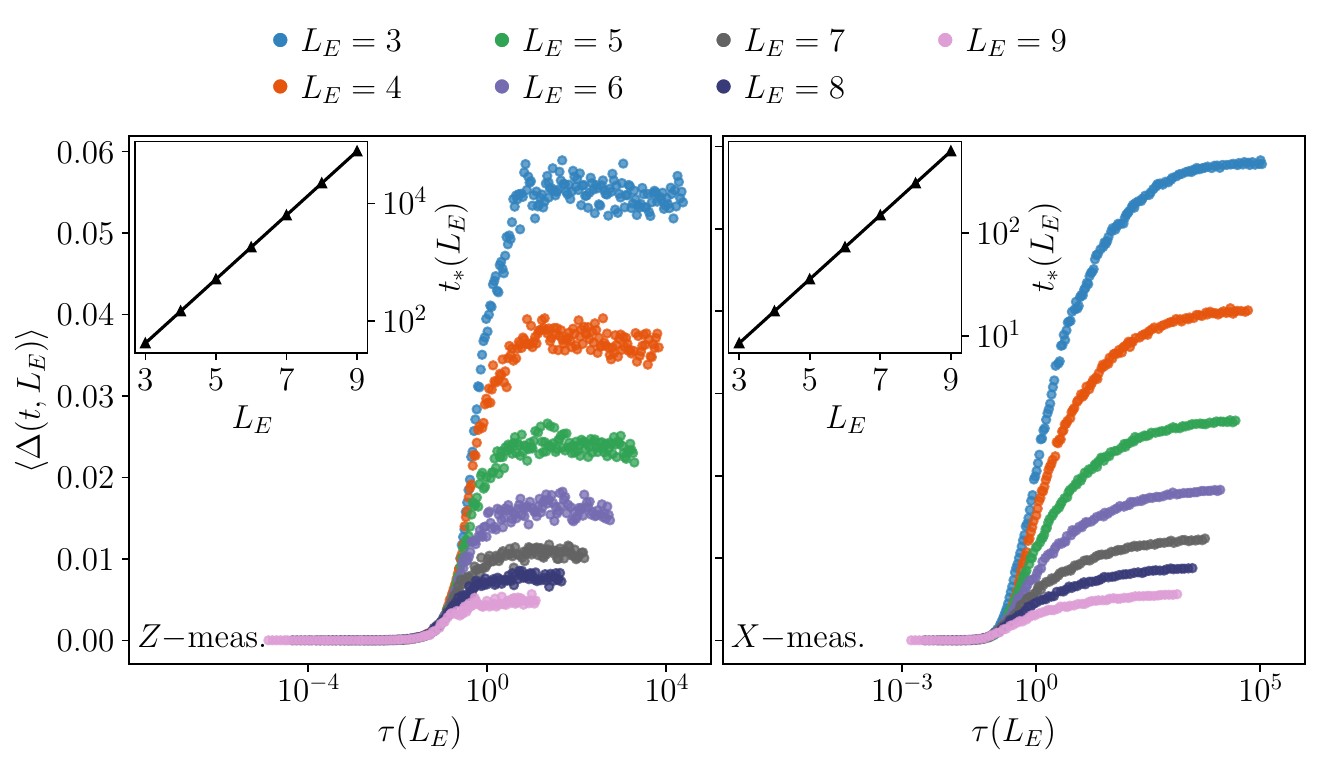}
\caption{$\Delta(t,L_E)$ as a function of $\tau(L_E)$ which collapses the data for the onset where $\tau(L_E)$ is defined in Eq.~\ref{eq:tau-le}. The left and right panels correspond to $Z$ and $X$ measurements respectively and $L_S=10$ is fixed. The collapse yields a timescale $t_{\ast}(L_E)$, which grows exponentially with $L_E$, $t_\ast\sim e^{L_E/\xi_t}$ (see insets), for the effect of the measurements in $S$ to reach $R$ and therefore for the onset of $\Delta(t,L_E)$. From the data in the insets we obtain $\xi_{t}=0.8$ for $Z$-measurements and $\xi_{t}=1.4$ for $X$-measurements.}
\label{fig:lbit-onset_scaling}
\end{figure}

We also compute $\Delta(t,L_E)$ numerically for the $\ell$-bit model with the results shown in Fig.~\ref{fig:lightcone-lbit} where the logarithmic lightcone is clearly visible. The data in the right panels show that the onset of a finite $\Delta(t,L_E)$ for a fixed $L_E$ happens at a timescale which seems to grow exponentially with $L_E$. 
To make this quantitatively concrete, as we did for the disordered Floquet Ising model (Sec.~\ref{sec:tfi-mbl}), we show in Fig.~\ref{fig:lbit-onset_scaling}, $\Delta(t,L_E)$ as a function of $\tau\equiv t/t_\ast(L_E)$ such that the data for different $L_E$ collapse on top of each other for timescales which correspond to the onset of $\Delta(t,L_E)$. The insets show that $t_\ast(L_E)$ grows exponentially with $L_E$ as predicted by the separation of scales in the effective magnetic field above, $t_\ast(L_E)\sim e^{L_E/\xi_t}$ with $\xi_t\propto\xi$.
This shows that the front of the lightcone does indeed spread logarithmically in spacetime.

\begin{figure}[!b]
\includegraphics[width=\linewidth]{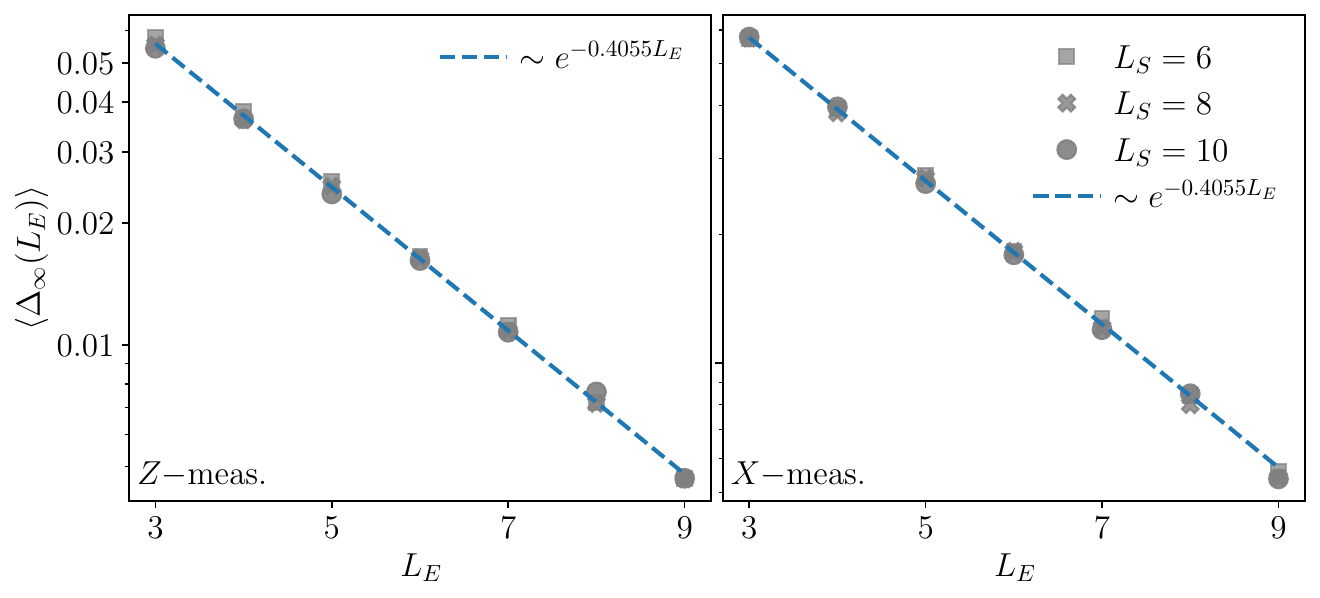}
\caption{The infinite-time result, $\Delta_\infty(L_E)$ as a function of $L_E$ which shows an exponential decay with $L_E$. The dashed lines show the analytic result of $\braket{\Delta_\infty(L_E)}\sim (2/3)^{L_E}$. The left and right panels correspond to $Z$ and $X$ measurements respectively and different shades correspond to different $L_S$.}
\label{fig:lbit-sat}
\end{figure}

The $\ell$-bit model also captures the decay of $\Delta_\infty(L_E)$ with $L_E$, as can be seen in the right panels of Fig.~\ref{fig:lightcone-lbit}. To understand this decay analytically, the limit of $t\to\infty$ can be taken in Eq.~\ref{eq:Delta-lbit-Z} which in the limit of $L_S\gg 1$ gives
\eq{
\Delta_\infty(L_E) \approx |A_{\uparrow_R} A_{\downarrow_R}|^2 \sum_{z_E}|A_{z_E}|^4\,,
}
which can be straightforwardly concluded to be exponentially small in $L_E$ from the form of $A_{z_E}$ from the initial state in Eq.~\ref{eq:psi-init}. In particular, the result averaged over random product state as initial states turns out to be $\braket{\Delta_\infty(L_E)}  \approx  (2/3)^{L_E}/3\sim e^{-0.405 L_E}$. The numerical results shown in Fig.~\ref{fig:lbit-sat} confirm this.

\begin{figure}
\includegraphics[width=\linewidth]{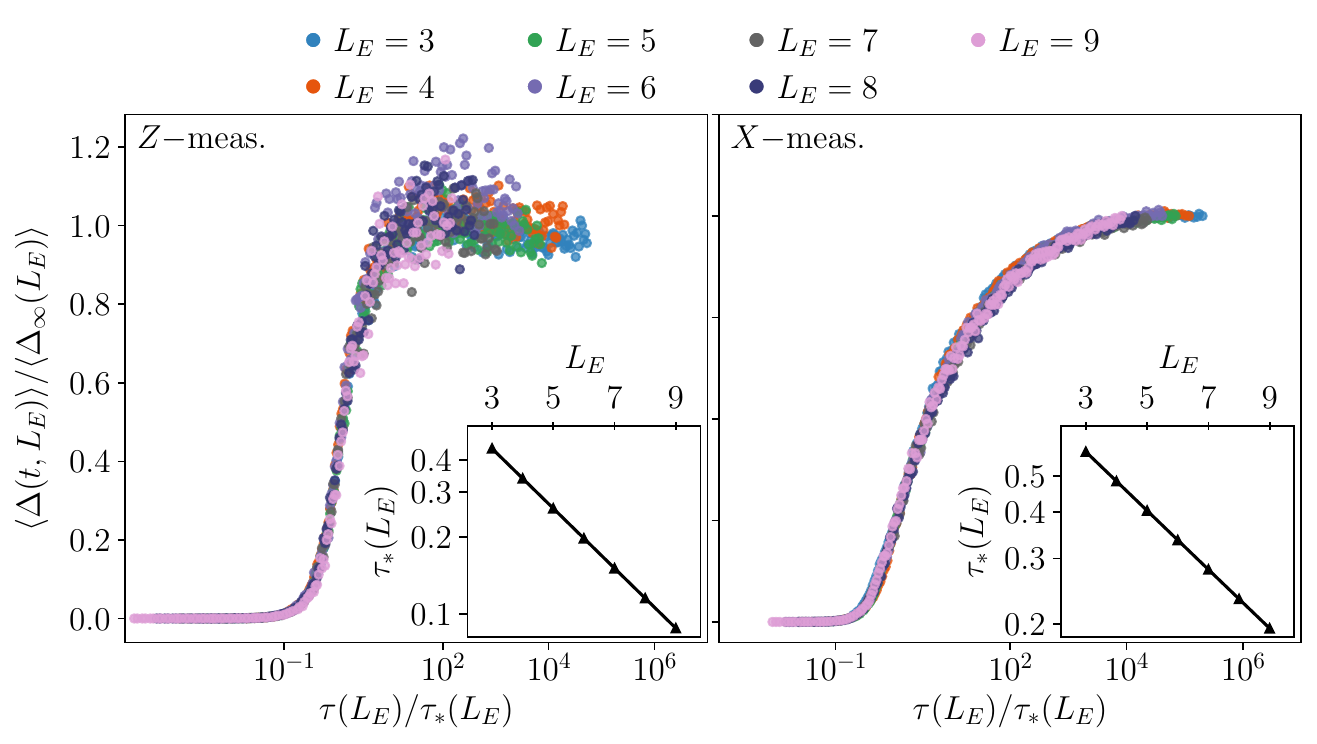}
\caption{Full scaling collapse of $\Delta(t,L_E)$ which shows evidence for the scaling form in Eq.~\ref{eq:lbit-full-scaling}. The left and right panels correspond to $Z$ and $X$ measurements respectively and $L_S=10$ is fixed. The insets show $\tau_\ast(L_E)$ which decays exponentially with $L_E$ as argued in the text, $\tau_\ast(L_E)\sim e^{-L_E/\xi_\tau}$, with $\xi_{\tau}=3.7$ for $Z$-measurements and $\xi_{\tau}=5.5$ for $X$-measurements.}
\label{fig:lbit-full-scaling}
\end{figure}

Armed with the infinite-time results, the entire data for $\Delta(t,L_E)$ for different $L_E$ can be collapsed on top of each other using the scaling form in Eq.~\ref{eq:lbit-full-scaling}. The results are shown in Fig.~\ref{fig:lbit-full-scaling}. The phenomenology is identical to that of the Floquet disordered Ising model with $\tau_\ast(L_E)$ decaying exponentially with $L_E$, $\tau_\ast(L_E)\sim e^{-L_E/\xi_\tau}$. The values of $\xi_t$ and $\xi_\tau$ mentioned in the captions to Fig.~\ref{fig:lbit-onset_scaling} and Fig.~\ref{fig:lbit-full-scaling} again ensure that the exponential growth of $t_\ast(L_E)$ with $L_E$ is faster than the exponential decay of $\tau_\ast(L_E)$ such that the saturation time scale $t_{\rm sat}(L_E)\equiv t_\ast(L_E)\times\tau_\ast(L_E)$ grows exponentially with $L_E$.
This concludes our analysis of the $Z$-measurement case with the $\ell$-bit model which offers an analytically tractable scenario in the MBL regime with a logarithmic lightcone of $\Delta(t,L_E)$.

\subsubsection{$X$-measurements}
We now turn towards the case of $X$-measurements where the measurement basis is completely orthogonal to the eigenbasis of the $\ell$-bit Hamiltonian.
The numerical results in Fig.~\ref{fig:lightcone-lbit} through Fig.~\ref{fig:lbit-full-scaling} show that the $X$-measurement case is qualitatively identical to the $Z$-measurement case.
However, in this case the analytical expressions are more complicated than the previous case due to the measurement operators not commuting with the Hamiltonian.
The probability of the outcome of an $x$-bit-string, which now depends explicitly on time is given by
\eq{
\begin{split}
p(t,x_S) = \sum_{\substack{z_R,z_E,\\z_S,z_S^\prime}}\big[|A_{z_R}|^2&|A_{z_E}|^2 B_{z_S z_S^\prime}^{x_S}\times \\&e^{-it\left(E_{z_Rz_Ez_S}-E_{z_Rz_Ez_S^\prime}\right)}\big]\,,
\end{split}
\label{eq:pxs-lbit}
}
where $B_{z_Sz_S^\prime}^{x_S} = A_{z_S^\prime}^\ast A_{z_S}\braket{z_S^\prime|x_S}\braket{x_S|z_S}$, and the corresponding conditional state of $R$ is given by
\eq{
\rho_R(t,x_S) = \sum_{z_R,z_R^\prime}\ket{z_R}\bra{z_R^\prime}\varrho_{z_Rz_R^\prime}(t,x_S)\,,
\label{eq:rho-R-lbit-gen-X}
}
where 
\eq{
\begin{split}
\varrho_{z_Rz_R^\prime}(t,x_S)=\frac{A_{z_R}A_{z_R^\prime}^\ast}{p(t,x_S)}&\smashoperator{\sum_{z_E,z_S,z_S^\prime}}\big[|A_{z_E}|^2 B_{z_Sz_S^\prime}^{x_S}\times\\&e^{-it\left(E_{z_Rz_Ez_S}-E_{z^\prime_Rz_Ez_S^\prime}\right)}\big]\,.
\end{split}
}
The PPE can then again be explicitly constructed using the states in Eq.~\ref{eq:rho-R-lbit-gen-X} and their probabilities in Eq.~\ref{eq:pxs-lbit}.
However, the expressions are unwieldy enough that we do not attempt to derive an exact expression for $\Delta(t,L_E)$ as we had done for the $Z$-measurement case.
Instead, we argue for the logarithmic lightcone by pointing out that the matrix elements of $\rho_R(t,x_S)$ develop a dependence on the measurement outcome $x_S$ only when $t>t_\ast$ where $t_\ast$ again grows exponentially with $L_E$.

To see this, consider a specific matrix element $\varrho_{\uparrow\uparrow}(t,x_S)$ which can be explicitly written as
\eq{
\varrho_{\uparrow\uparrow}(t,x_S)=\frac{|A_{\uparrow_R}|^2}{|A_{\uparrow_R}|^2+{\cal G}(t,x_S)|A_{\downarrow_R}|^2}\,,
\label{eq:lbit-X-rhoupup}
}
where 
\eq{
{\cal G} = \frac{\sum\limits_{z_E,z_S,z^\prime_S}|A_{z_E}|^2B^{x_S}_{z_S,z^\prime_S}e^{-it\delta_{z_Sz_S^\prime}^{\downarrow_R z_E}}}{\sum\limits_{z_E,z_S,z^\prime_S}|A_{z_E}|^2B^{x_S}_{z_S,z^\prime_S}e^{-it\delta_{z_Sz_S^\prime}^{\uparrow_R z_E}}}\,,
\label{eq:lbit-X-G}
}
with $\delta_{z_Sz_S^\prime}^{\uparrow_R z_E} \equiv E_{\uparrow_R z_Ez_S}-E_{\uparrow_R z_Ez_S^\prime}$ and similarly for $\delta_{z_Sz_S^\prime}^{\downarrow_R z_E}$,
and ${\cal G}(0,x_S)=1$.
The form of the matrix element in Eq.~\ref{eq:lbit-X-rhoupup} makes it clear that $\varrho_{\uparrow\uparrow}(t,x_S)$ develops a dependence on $x_S$, which in turn leads to a non-zero $\Delta(t,L_E)$ only when ${\cal G}(t,L_E)$ deviates from unity. To estimate the timescale, note that in Eq.~\ref{eq:lbit-X-G}, the phases contain the difference in the effective fields seen by the spin at $R$ for two different configurations, $z_S$ and $z^\prime_S$, in $S$ but the same configuration in $E$. While the phases in the numerator contain the corresponding energy differences for the spin at $R$ pointing down, it is those with the spin at $R$ pointing up that contribute to the denominator. 
The structure of the $\ell$-bit Hamiltonian (Eq.~\ref{eq:lbithamiltonian}) suggests that  $\delta_{z_Sz_S^\prime}^{\uparrow_R z_E}$ and $\delta_{z_Sz_S^\prime}^{\downarrow_R z_E}$ are both exponentially suppressed in the distance between $R$ and $S$ which is nothing but $L_E$.
Invoking the separation of scales, $|H_i^{(l)}|\gg |H_i^{(l+1)}|$,
again implies the presence of a timescale $t_\ast(L_E)$ which grows exponentially with $L_E$, $t_\ast(L_E)\sim e^{L_E/\xi^X}$.  For $t<t_\ast(L_E)$, we have $e^{-it\delta_{z_Sz_S^\prime}^{\uparrow_R z_E}},e^{-it\delta_{z_Sz_S^\prime}^{\downarrow_R z_E}}\approx 1$ which in turn means ${\cal G}(t,L_E)\approx 1$ and therefore $\varrho_{\uparrow\uparrow}(t,x_S)\approx|A_{\uparrow_R}|^2$, independent of $x_S$.
On the other hand, for $t>t_\ast(L_E)$, both $\delta_{z_Sz_S^\prime}^{\uparrow_R z_E}$ and $\delta_{z_Sz_S^\prime}^{\downarrow_R z_E}$ deviate from unity, and more importantly, they deviate from each other in fashion which depends on $x_S$ through $B_{z_Sz^\prime_S}^{x_S}$. This leads to ${\cal G}(t,L_E)$ deviating from unity which in turn automatically means $\varrho_{\uparrow\uparrow}(t,x_S)$ also develops a non-trivial dependence on $x_S$.
While we focussed explicitly on the matrix element ${\varrho_{\uparrow\uparrow}}$ arguments along exactly similar lines can be developed for the other matrix elements. The upshot of all this is $\rho_R(t,x_S)$ develops a non-trivial dependence on $x_S$ leading to a onset of $\Delta(t,L_E)$ only for $t>t_\ast(L_E)$ where $t_\ast(L_E)\sim e^{L_E/\xi_t^X}$. This explains the logarithmic lightcone for the $X$-measurements.

We finally turn towards the infinite-time behaviour $\Delta_{\infty}(L_E)$. We use the fact that for the $\ell$-bit model with $X$-measurements, the ensemble of pure states induced on $R\cup E$ due to the measurements on $S$ is the Scrooge ensemble corresponding to $\rho_{RE}(\infty)$, the infinite-time reduced density matrix of $R\cup E$~\cite{manna2025projected}. In this case $\rho_{RE}(\infty)$ is given explicitly as
\eq{
\rho_{RE}(\infty) = \sum_{z_R,z_E}|A_{z_R}A_{z_E}|^2\ket{z_Rz_E}\bra{z_Rz_E}\,,\label{eq:rhoRE-inf}
}
which the purity of $\rho_{RE}(\infty)$ is exponentially small in $L_E+L_R$. As such $\rho_{RE}(\infty)$ is sufficiently mixed that the second moment of the projected ensemble is well approximated by~\cite{mark2024maximum,chang2025deep}
\eq{
\rho_{RE}^{(2)}(\infty)\approx(\mathbb{I}+\mathbb{S})[\rho_{RE}(\infty)]^{\otimes 2}\,,
}
where $\mathbb{S}$ is the SWAP operator acting on the doubled Hilbert space of $R$ and $E$. The second moment of the PPE is then given by $\rho_R^{(2)}(\infty) = {\rm Tr}_E^{\otimes 2}[\rho_{RE}^{(2)}(\infty)]$, which evaluates to 
\eq{
\rho_R^{(2)}(\infty)=&[\rho_R(\infty)]^{\otimes 2} + \nonumber\\&\sum_{z_E}|A_{z_E}|^4\sum_{z_R\neq z_R^\prime}|A_{z_R}A_{z^\prime_R}|^2\ket{z_Rz_R^\prime}\bra{z_R^\prime z_R}\,,
}
which directly yields $\Delta_\infty(L_E)\sim \sum_{z_E}|A_{z_E}|^4$ which upon averaging over initial states leads to $\braket{\Delta_\infty(L_E)}\sim (2/3)^{L_E}$; this is exactly the result which emerges numerically as shown in Fig.~\ref{fig:lbit-sat} (right panel).

\section{Summary and Outlook \label{sec:conclusion}}

Let us briefly summarise the main results of this work.
We introduced the partial projected ensemble (PPE), defined by conditioning a subsystem $R$ on projective measurements performed on a spatially separated region $S$ while tracing out the intervening environment $E$. This construction was motivated by the need to probe how the spatiotemporal structure of information scrambling is imprinted on projected ensembles, particularly in realistic settings where parts of the system are inaccessible or subject to loss. 
By analysing the statistical fluctuations of the PPE, we showed that they remain trivial when $R$ and $S$ are outside each other’s causal lightcones and become non-trivial once they are causally connected, thereby directly mirroring the lightcone of information spreading.
We further demonstrated that the associated bit-string outcome probabilities (PoPs) exhibit sharp dynamical crossovers that track the same causal structure, providing an experimentally accessible probe of scrambling. 
Both the PPE fluctuations and PoPs display exponential sensitivity to the size of the discarded region $E$, signalling the exponential degradation of quantum correlations under erasure or loss -- a feature rooted in the multipartite structure of scrambling. 

These results were substantiated in the non-integrable kicked Ising chain using numerical simulations deep in the ergodic  and MBL phases, with analytical insights obtained in the ergodic phase at the self-dual point and in the MBL phase by considering the phenomenological $\ell$-bit model. 
The emergence of linear lightcones in the ergodic regime and logarithmic lightcones in the MBL regime underscores the ability of the PPE to faithfully capture distinct dynamical mechanisms of scrambling. Together, these findings establish the PPE as a versatile framework for diagnosing information scrambling, quantifying the fragility of quantum correlations, and characterising deep thermalisation in out-of-equilibrium quantum systems.

The notions of projected ensembles and deep thermalisation are still somewhat in their nascent stages. 
Work on PPEs and their connection to quantum information dynamics in systems with spatial structure is even more fledgling. 
It is therefore natural that our current work opens several new directions and raises many questions for future research. 
A question of immediate interest is a comprehensive understanding of the dynamics of the PPE at intermediate times; in particular, it will be important to understand why does the distance of the PPE from the corresponding gHS ensemble cease to depend on the size of the traced out region, as found in Sec.~\ref{sec:latetimes}.
While the gHS ensemble approached by the PPE is a derivative of a maximum-entropy ensemble, a fundamental question would be are their scenarios where the PPE at intermediate times can be understood as a derivative of a maximum-entropy Scrooge ensemble. 

In our current work, we exclusively considered one-dimensional systems with local interactions, with the traced out subsystem intervening the subsystems which were measured and on which the PPE was induced. 
As such the information between the latter two subsystems were transferred necessarily through the traced out subsystem.
The fate of the PPE can be drastically different in higher dimensions or with long-ranged interactions where the information can find multiple pathways to travel, some bypassing the traced out region. 
This remains a work for the future.
It will also be interesting to analyse the fate of the PPE in dynamics with conservation laws. Not only will the late time PPE be a derivative of a (generalised) Scrooge ensemble~\cite{cotler2023emergent,chang2025deep} but the dynamics will also be constrained by the conservation laws. In particular, it will be pertinent to ask if the dynamics of the PPE hosts manifestations of the diffusive operator hydrodynamics in such settings~\cite{khemani2018operator,rakovszky2018diffusive}.

\new{It will also be interesting to study the behaviour of the PPE in eigenstates across eigenstate transitions such as the MBL transition. Preliminary results (see Appendix~\ref{app:eigenstates}) already indicates that the PPE is drastically different in ergodic volume-law, entangled eigenstates and area-law, MBL eigenstates. In particular, it will be significant if deep thermalisation measures can shed new insights onto the nature of these transitions.}

The PoPs also present several promising avenues. They have emerged, both theoretically and experimentally, as useful tools to distinguish genuine thermalisation from decoherence as well as computationally efficient experimental noise learning~\cite{shaw2025experimental}. Building on this one can also investigate if the different dynamical regimes in the PoPs observed in this work can be reverse-engineered to classify and discriminate different kinds of noise models in experiments.

Finally, it is of pressing interest to investigate the nature of mixed-state ensembles where the mixedness could be due to the external noise, bulk measurements or classical randomness. For pure state ensembles augmented with such classical randomness, it was shown that the latter enhances the randomness in the projected ensemble~\cite{mok2025optimal}. 
It remains a question for the future the analyse the corresponding effect in mixed-state ensembles. Relevant for systems with spatial structure, it will also be useful to the study the effect on the PPE due to such classical randomness restricted to a far away subsystem.

\begin{acknowledgments}
We thank M. Dalmonte for useful discussions, and S. Manna, A. Sherry and G. J. Sreejith for useful discussions as well as collaborations on related work. SM and SR acknowledge support from the Department of Atomic Energy, Government of India under Project No. RTI4001. SR acknowledges support from SERB-DST, Government of India under Grant No. SRG/2023/000858 and from a Max Planck Partner Group grant between ICTS-TIFR, Bengaluru and MPIPKS, Dresden. PWC acknowledges support from the Max Planck Society.
\end{acknowledgments}

\appendix
\section{PoP in the self-dual kicked Ising chain}
\label{app:PoP_SDKI}
In this Appendix we present an analytic derivation for ${\rm PoP_{b{\text{-}}str}}(\tilde{p};t)$ in the kicked Ising chain at the self-dual point. This derivation follows the approach of Ref.~\cite{claeys2024fock-space}. In order to determine the full distribution, we first characterize the moments of the distribution for the unnormalized probabilities $p(o_s)$. Specifically, we consider
\eq{\label{eq:Iq-diag}
I_q = \sum_{o_s} p(o_s)^{q} = D_{S} \int_{0}^1 dp\, {\rm PoP_{b{\text{-}}str}}(p) p^q,
}
where we have made the dependence on $t$ implicit for ease of notation. The quantities $I_q, q \in \mathbb{N}$ can also be interpreted as inverse participation ratios for $\rho_S$. Graphically, $I_q$ can be represented by introducing a $q$-folded version of the gates (denoted by the grey gates),
\eq{
I_q = \frac{D_{RE}^{q/2}}{D_{RES}^q}\times
\cbox{\begin{tikzpicture}[scale=0.8]
\def\s{0.2}
\foreach \y in {0,...,2}{
\draw[very thick](-0.35,\y+.35) -- (-0.35,\y+1-0.35);
\draw[very thick](3.35,\y+.35) -- (3.35,\y+1-0.35);
\ifthenelse{\y=2}{\draw[very thick](-0.35,\y+.35) --(-0.35,\y+0.5+0.35);\draw[very thick](3.35,\y+.35) --(3.35,\y+0.5+0.35);}{}
\def\x{3}
\draw[black, very thick] (\x-0.35,\y-0.35)--(\x+0.35,\y+0.35);
\draw[black, very thick] (\x-0.35,\y+0.35)--(\x+0.35,\y-0.35);
\foreach \x in {0,...,2}{
    \draw[black, very thick] (\x-0.35,\y-0.35)--(\x+0.35,\y+0.35);
    \draw[black, very thick] (\x-0.35,\y+0.35)--(\x+0.35,\y-0.35);
    \draw[black, very thick] (\x-0.35+0.5,\y-0.35+0.5)--(\x+0.35+0.5,\y+0.35+0.5);
    \draw[black, very thick] (\x-0.35+0.5,\y+0.35+0.5)--(\x+0.35+0.5,\y-0.35+0.5);
}}
\foreach \y in {0,...,2}{
\def\x{3}
\draw[rounded corners=2pt, fill=gray, very thick] (\x-\s,\y-\s) rectangle (\x+\s,\y+\s); 
\draw[white] (\x+\s-0.15,\y+\s-0.05) -- (\x+\s-0.05,\y+\s-0.05) -- (\x+\s-0.05,\y+\s-0.15);
\foreach \x in {0,...,2}{
    \draw[rounded corners=2pt, fill=gray, very thick] (\x-\s,\y-\s) rectangle (\x+\s,\y+\s); 
    \draw[white] (\x+\s-0.15,\y+\s-0.05) -- (\x+\s-0.05,\y+\s-0.05) -- (\x+\s-0.05,\y+\s-0.15);
    \draw[rounded corners=2pt, fill=gray, very thick] (\x+0.5-\s,\y-\s+0.5) rectangle (\x+0.5+\s,\y+\s+0.5); 
    \draw[white] (0.5+\x+\s-0.15,\y+0.5+\s-0.05) -- (0.5+\x+\s-0.05,\y+0.5+\s-0.05) -- (0.5+\x+\s-0.05,\y+0.5+\s-0.15);
}}
\foreach \xx in {0,...,3}{
\def\y{-0.35}
\def\x{\xx-0.35}
\draw[fill=red,very thick] (\x-0.1,\y) -- (\x+0.1,\y) -- (\x,\y-0.15) -- cycle;
\def\x{\xx+0.35}
\draw[fill=red,very thick] (\x-0.1,\y) -- (\x+0.1,\y) -- (\x,\y-0.15) -- cycle;
}
\foreach \x in {1.5-0.35,1.5+0.35,2.5-0.35,2.5+0.35,3.35}{
\def\y{2.85}
\draw[fill=red,very thick] (\x-0.1,\y) -- (\x+0.1,\y) -- (\x,\y+0.15) -- cycle;
}
\foreach \x in {1.5-0.35,1.5+0.35,2.5-0.35,2.5+0.35,3.35}{
\def\y{2.85}
\draw[fill=red,very thick] (\x-0.1,\y) -- (\x+0.1,\y) -- (\x,\y+0.15) -- cycle;
}
\foreach \x in {-0.35,0.5-0.35,0.85}{
\def\y{2.85}
\draw[fill=YellowOrange,very thick] (\x,\y+0.2) circle (0.2);
}
\node at (1.25,3.2) {\footnotesize{$o_{1}$}};
\node at (3.35,3.2) {\footnotesize{$o_{L_S}$}};
\draw[dotted](1.5,3.2) --(3,3.2);
\end{tikzpicture}}\,,
}
where the red triangles at the bottom denote copies of $z$-product states, the red triangles at the top denote measurements with outcomes $o_1,\cdots,o_{L_S}$ and the orange circles denote the trace in the $q$-folded picture. 
The normalisation of the states denoted by the triangles is not the usual one as will be clarified shortly.
The dual-unitarity of the ergodic kicked Ising chain can now be used to replace the averaging over measurement outcomes in region $S$ by averaging over a Haar-random state in the spatial direction, {\it i.e.} acting on the temporal lattice of $t$ sites (see Refs.~\cite{ho2022exact}). For $q$ replicas and a state acting on a $D_t$-dimensional Hilbert space, the relevant average is given by
\eq{
&\mathbb{E}_{\phi_t \sim \textrm{Haar}(D_t)}\left[\left(\ket{\phi_t}\!\bra{\phi_t}\right)^{\otimes q}\right] \nonumber\\
&\qquad = \sum_{\pi \in S_q} \frac{P(\pi)}{D_t(D_t+1)\dots (D_t+q-1)}.
}
Here $P(\pi)$ is a permutation operator, permuting the $q$ different replicas according to a permutation $\pi \in S_q$ as
\eq{
P(\pi) \ket{o_1,o_2, \dots o_q} = \ket{o_{\pi(1)},o_{\pi(2)}, \dots o_{\pi(q)}}.
}
Graphically, the folded permutation operators can be represented as
\eq{
P(\pi) = 
\cbox{
\begin{tikzpicture}[scale=0.8]
\draw[very thick](-0.6,0)--(0,0);
\draw[fill=YellowOrange,very thick] (0,0) circle (0.2);
\node at (0,0) {\footnotesize$\pi$};
\end{tikzpicture}
}\,,
}
with the orange circle without $\pi$ denoting the identity permutation $P(\mathbb{I}) = \cbox{
\begin{tikzpicture}[scale=0.8]
\draw[very thick](-0.6,0)--(0,0);
\draw[fill=YellowOrange,very thick] (0,0) circle (0.2);
\end{tikzpicture}}
$.
The unitarity and the dual-unitarity of the $q$-folded gates imply that leave these permutation states invariant when contracted along the spatial and temporal directions respectively,
\eq{
\cbox{\begin{tikzpicture}[scale=0.8]
\def\s{0.2}
\draw[very thick](-0.35,-.35) -- (+0.35,+0.35);
\draw[very thick](-0.35,.35) -- (+0.35,-0.35);
\draw[rounded corners=2pt, fill=gray, very thick] (-\s,-\s) rectangle (+\s,+\s); 
\draw[white] (+\s-0.15,+\s-0.05) -- (+\s-0.05,+\s-0.05) -- (+\s-0.05,+\s-0.15);
\draw[fill=YellowOrange,very thick] (-.4,.4) circle (0.2);\node at (-.4,.4) {\footnotesize$\pi$};
\draw[fill=YellowOrange,very thick] (.4,.4) circle (0.2);\node at (.4,.4) {\footnotesize$\pi$};
\end{tikzpicture}}=
\cbox{
\begin{tikzpicture}[scale=0.8]
\foreach \y in {-0.25,0.25}{
\draw[very thick](\y,-0.6)--(\y,0);
\draw[fill=YellowOrange,very thick] (\y,0) circle (0.2);
\node at (\y,0) {\footnotesize$\pi$};}
\end{tikzpicture}}
\,;~~~
\cbox{\begin{tikzpicture}[scale=0.8]
\def\s{0.2}
\draw[very thick](-0.35,-.35) -- (+0.35,+0.35);
\draw[very thick](-0.35,.35) -- (+0.35,-0.35);
\draw[rounded corners=2pt, fill=gray, very thick] (-\s,-\s) rectangle (+\s,+\s); 
\draw[white] (+\s-0.15,+\s-0.05) -- (+\s-0.05,+\s-0.05) -- (+\s-0.05,+\s-0.15);
\draw[fill=YellowOrange,very thick] (.4,.4) circle (0.2);\node at (.4,.4) {\footnotesize$\pi$};
\end{tikzpicture}}=\cbox{
\begin{tikzpicture}[scale=0.8]
\foreach \y in {-0.25,0.25}{
\draw[very thick](-0.6,\y)--(0,\y);
\draw[fill=YellowOrange,very thick] (0,\y) circle (0.2);
\node at (0,\y) {\footnotesize$\pi$};}
\end{tikzpicture}}\,.
}
In addition, the normalisation of the states chosen in Eq.~\ref{eq:Iq-diag} is such that,
\eq{
\cbox{\begin{tikzpicture}[scale=0.8]
\def\s{0.2}
\draw[very thick](-0.35,-.35) -- (+0.35,+0.35);
\draw[very thick](-0.35,.35) -- (+0.35,-0.35);
\draw[rounded corners=2pt, fill=gray, very thick] (-\s,-\s) rectangle (+\s,+\s); 
\draw[white] (+\s-0.15,+\s-0.05) -- (+\s-0.05,+\s-0.05) -- (+\s-0.05,+\s-0.15);
\draw[fill=YellowOrange,very thick] (.4,.4) circle (0.2);\node at (.4,.4) {\footnotesize$\pi$};
\foreach \xx in {0}{
\def\y{-0.35}
\def\x{\xx-0.35}
\draw[fill=red,very thick] (\x-0.1,\y) -- (\x+0.1,\y) -- (\x,\y-0.15) -- cycle;
\def\x{\xx+0.35}
\draw[fill=red,very thick] (\x-0.1,\y) -- (\x+0.1,\y) -- (\x,\y-0.15) -- cycle;
}
\end{tikzpicture}}=\cbox{
\begin{tikzpicture}[scale=0.8]
\foreach \y in {0.25}{
\draw[very thick](-0.6,\y)--(0,\y);
\draw[fill=YellowOrange,very thick] (0,\y) circle (0.2);
\node at (0,\y) {\footnotesize$\pi$};}
\node at (0,-0.25) {};
\end{tikzpicture}}
}
Replacing the averaging over measurement outcomes by a Haar-random unitarity in the spatial direction, $I_q$ can be represented as
\eq{
I_q = \frac{D_t^{q/2} D_{RE}^{q/2}}{D_{RES}^{q}} \sum_{\pi \in S_q}\frac{1}{\prod\limits_{a=0}^{q-1}(D_t+a)}\!\times\!\!\!\!\!\cbox{
    \begin{tikzpicture}[scale=0.8]
        \def\s{0.2}
        \foreach \y in {0,1,2}{
            \draw[very thick](1.35,\y-0.35)--(1.6,\y-0.35);
            \draw[very thick](1.35,\y+0.35)--(1.6,\y+0.35);
            \draw[fill=YellowOrange,very thick]  (1.6,\y-0.35+0.1) circle (0.2);\node at (1.6,\y-0.35+0.1) {\footnotesize$\pi$};
            \draw[fill=YellowOrange,very thick]  (1.6,\y+0.35-0.1) circle (0.2);\node at (1.6,\y+0.35-0.1) {\footnotesize{$\pi$}};
            \def\xx{0.5}
            \draw[black,very thick] (\xx-0.35,\y-0.35+0.5)--(\xx+0.35,\y+0.35+0.5);
            \draw[black,very thick] (\xx-0.35,\y+0.35+0.5)--(\xx+0.35,\y-0.35+0.5);
            \foreach \x in {0,1}{
                \draw[black,very thick] (\x-0.35,\y-0.35)--(\x+0.35,\y+0.35);
                \draw[black,very thick] (\x-0.35,\y+0.35)--(\x+0.35,\y-0.35);
            }
        }
        \foreach \y in {0,1,2}{
            \def\xx{0.5}
            \draw[rounded corners=2pt, fill=gray, very thick] (\xx-\s,\y-\s+0.5) rectangle (\xx+\s,\y+\s+0.5); 
            \draw[white!5] (\xx+\s-0.15,\y+\s-0.05+0.5) -- (\xx+\s-0.05,\y+\s-0.05+0.5) -- (\xx+\s-0.05,\y+\s-0.15+0.5);
            \foreach \x in {0,1}{
                \draw[rounded corners=2pt, fill=gray, very thick] (\x-\s,\y-\s) rectangle (\x+\s,\y+\s); 
                \draw[white!5] (\x+\s-0.15,\y+\s-0.05) -- (\x+\s-0.05,\y+\s-0.05) -- (\x+\s-0.05,\y+\s-0.15);
            }
        }
        \draw[very thick](-0.35,2.35)--(-0.35,2.85);
        \draw[very thick](-0.35,0.35)--(-0.35,1-0.35);
        \draw[very thick](-0.35,1.35)--(-0.35,2-0.35);
        \foreach \x in {-0.35,0.5-0.35,0.85}{
        \draw[fill=YellowOrange,very thick]  (-0.35-0.1,3) circle (0.2);}
        \draw[fill=YellowOrange,very thick]  (0.5-0.35-0.1,3) circle (0.2);
        \draw[fill=YellowOrange,very thick]  (0.5+0.35+0.1,3) circle (0.2);
        \foreach \x in {-0.35,0.35,0.65}{
        \draw[fill=red,very thick] (\x-0.1,-0.35) -- (\x+0.1,-0.35) -- (\x,-.5) -- cycle;}
    \end{tikzpicture}
}\!.
}
The prefactor originates from the normalization of the spatial boundary and the product states.

We first consider the case of $t \geq L_{RE}$.
Using dual-unitarity and unitarity, all permutations operators can be propagated to return
\eq{
I_q =  \frac{D_t^{q/2} D_{RE}^{q/2}}{D_{RES}^{q}} \sum_{\pi \in S_q}\frac{1}{\prod\limits_{a=0}^{q-1}(D_t+a)} 
\left(\!\!\cbox{
\begin{tikzpicture}[scale=0.8]
\foreach \y in {0.25}{
\draw[very thick](\y,-0.6)--(\y,0);
\draw[fill=YellowOrange,very thick] (\y,0.1) circle (0.2);
\draw[fill=YellowOrange,very thick] (\y,-0.5) circle (0.2);
\node at (\y,0.1) {\footnotesize$\pi$};}
\end{tikzpicture}}\right)^{L_{RE}}
\left(\!\!\cbox{
\begin{tikzpicture}[scale=0.8]
\foreach \y in {0.25}{
\draw[very thick](\y,-0.6)--(\y,0);
\draw[fill=YellowOrange,very thick] (\y,0.1) circle (0.2);
\draw[fill=YellowOrange,very thick] (\y,-0.5) circle (0.2);
\node at (\y,0.1) {\footnotesize$\pi$};
\node at (\y,-0.5) {\footnotesize$\pi$};}
\end{tikzpicture}}\right)^{\frac{t-L_{RE}}{2}}\,.
}
Each overlap between a permutation state $\pi$ and the identity permutation, 
$\cbox{
\begin{tikzpicture}[scale=0.8]
\draw[very thick](-0.6,0)--(0,0);
\draw[fill=YellowOrange,very thick] (-0.6,0) circle (0.2);
\node at (-0.6,0) {\footnotesize $\pi$};
\draw[fill=YellowOrange,very thick] (0,0) circle (0.2);
\end{tikzpicture}}
$, returns a factor $2^{q|\pi|}$, with $|\pi|$ the number of cycles in $\pi$. The total number of such overlaps is given by $L_{RE}$, returning a global factor $2^{q |\pi| L_{RE} } = D_{RE}^{q|\pi|}$. The remaining $(t - L_{RE})/2$ overlaps return a factor $(D_t/D_{RE})^{q/2}$, which can be absorbed in the overall prefactor, to return
\eq{
I_q = \frac{D_t^{q}}{D_{RES}^{q}} \sum_{\pi \in S_q}\frac{D_{RE}^{q |\pi|}}{D_t(D_t+1)\dots (D_t+q-1)} 
}
The number of permutations of $q$ elements with $r$ cycles is given by the Stirling numbers of the first kind $\begin{bmatrix} q \\ r \end{bmatrix}$. These additionally satisfy 
\eq{
	\sum_{r=1}^q \begin{bmatrix}
	q \\ r
	\end{bmatrix} x^r = x (x+1) \dots (x+q-1).
}
Using this identity, $I_q$ admits a closed-form expression as
\eq{
I_q =  \frac{D_t^{q}}{D_{RES}^{q}} \frac{D_{RE} (D_{RE}+1) \dots (D_{RE}+q-1)}{D_t(D_t+1)\dots (D_t+q-1)} .
}
These corresponds to the moments of a rescaled beta distribution, $\textrm{Beta}(\alpha,\beta)$, with $\alpha = D_{RE}$ and $\beta = D_t - D_{RE}$. Specifically, we find that
\eq{
&{\rm PoP_{b{\text{-}}str}}(p) \propto  \left(\frac{D_{RES}}{D_t}p\right)^{D_{RE}-1} \left(1-\frac{D_{RES} }{D_t}p\right)^{D_t-D_{RE}-1} \nonumber\\
&\qquad \qquad\times\theta\left(1- \frac{D_{RES} }{D_t}p\right).
}
In terms of normalized probabilities $\tilde{p} = D_S \, p$ with average value $1$, the resulting distribution reads
\eq{
{\rm PoP_{b{\text{-}}str}}(\tilde{p}) =  &\frac{\Gamma(D_t)}{\Gamma(D_{RE})\Gamma(D_t-D_{RE})} \theta\left(1- \frac{D_{RE} }{D_t}\tilde{p}\right) \nonumber\\
&\,\times\left(\frac{D_{RE}}{D_t}\tilde{p}\right)^{D_{RE}-1} \left(1-\frac{D_{RE} }{D_t}\tilde{p}\right)^{D_t-D_{RE}-1} ,
}
with $\Gamma$ the gamma function fixing the normalization. It can be directly observed that the only system-size dependence of this distribution is through the Hilbert-space dimension $D_{RE}$ of the traced-out region.

For $t < L_{RE}$ the above derivation needs to be modified, since the total number of overlaps between the permutation operators and the identity permutations is set by $t$, resulting in a remaining $(L_{RE}-t)/2$ contractions between identity permutations. The resulting $I_q$ reads
\eq{
I_q &= \frac{D_{RE}^{q}}{D_{RES}^{q}} \sum_{\pi \in S_q}\frac{D_{t}^{q |\pi|}}{D_t(D_t+1)\dots (D_t+q-1)} \\
&=\frac{D_{RE}^{q}}{D_{RES}^{q}},
}
consistent with probabilities that are delta-distributed, ${\rm PoP_{b{\text{-}}str}}(p) = \delta(p - \frac{D_{RE}}{D_{RES}}) = \delta(p - \frac{1}{D_S})$.

\new{
\section{PPE fluctuations in eigenstates} 
\label{app:eigenstates}

While the central focus of the work was on the dynamics of the PPE and its connection to the information scrambling, one can also study the PPE fluctuations, specifically their dependence on $L_E$, in eigenstates.
In particular, one pertinent question is if and how are they different across an MBL transition where eigenstates transition from being volume-law entangled to area-law entangled.
While a detailed analysis is outside the scope of this work, in this Appendix we present some preliminary results which show that the statistics of the PPE fluctuations in the eigenstates are indeed qualitatively different in the ergodic and MBL phases.

To do this, we again consider the KI chain in Eq.~\ref{eq:UF-KI} with parameters in both the ergodic phase \eqref{eq:params} as well as in the MBL phase \eqref{eq:MBL-params}.
In each case we obtain the eigenstates via exact diagonalisation of the Floquet unitary. 
For each eigenstate we construct the PPE, and compute its fluctuation, $\Delta(L_E)$, exactly as in Eq.~\ref{eq:ppe-fluc} except now there is no time dependence. 

The results are shown in Fig.~\ref{fig:ppe-eig}. 
In the ergodic phase, the distribution of $\Delta(L_E)$ over eigenstates and disorder realisations is quite sharply peaked around is mean value of $\sim D_E^{-1}$. 
This is simply because the eigenstates of an ergodic system indeed show deep thermalisation~\cite{cotler2023emergent} and therefore the PPE on $R$ is well described by a gHSE. As such it follows trivially from the the result in Eq.~\ref{eq:rho-diff-inf} that $\Delta(L_E)\sim D_E^{-1}$.

The situation in the MBL phase is rather different. The MBL eigenstates are area-law entangled and therefore the quantum correlations between the qubits in $R$ with those in $S$ are typically, extremely small in the first place. This is reflected in the fact that the typical values (defined as the geometric mean $\equiv \exp [\braket{\ln \Delta(L_E)}]$) are several orders of magnitude smaller than the corresponding ones in the ergodic phase.
However, a feature of the note in this case is that the distributions of $\Delta(L_E)$ are very broad with heavy power-law tails. This is also reflected in the mean $\Delta(L_E)$ being significantly larger than the typical. 
We nominally attribute these tails to the presence of rare long-ranged resonances~\cite{morningstar2021avalanches,garratt2021resonances,garratt2022resonant,crowley2022constructive} and the multiscale structure of entanglement~\cite{hervious2019multiscale} in the MBL phase. 

To summarise, the results shown in Appendix, hint towards two interesting directions for the future. 
First the PPE fluctuations may be used as sensitive probe of entanglement phase transitions such as the MBL transition.
Second, these fluctuations also seemingly probe the rare long-ranged resonances in the MBL phase; it will therefore be interesting to develop experimentally feasible probes for the latter based on the former.

\begin{figure}[!t]
\includegraphics[width=\linewidth]{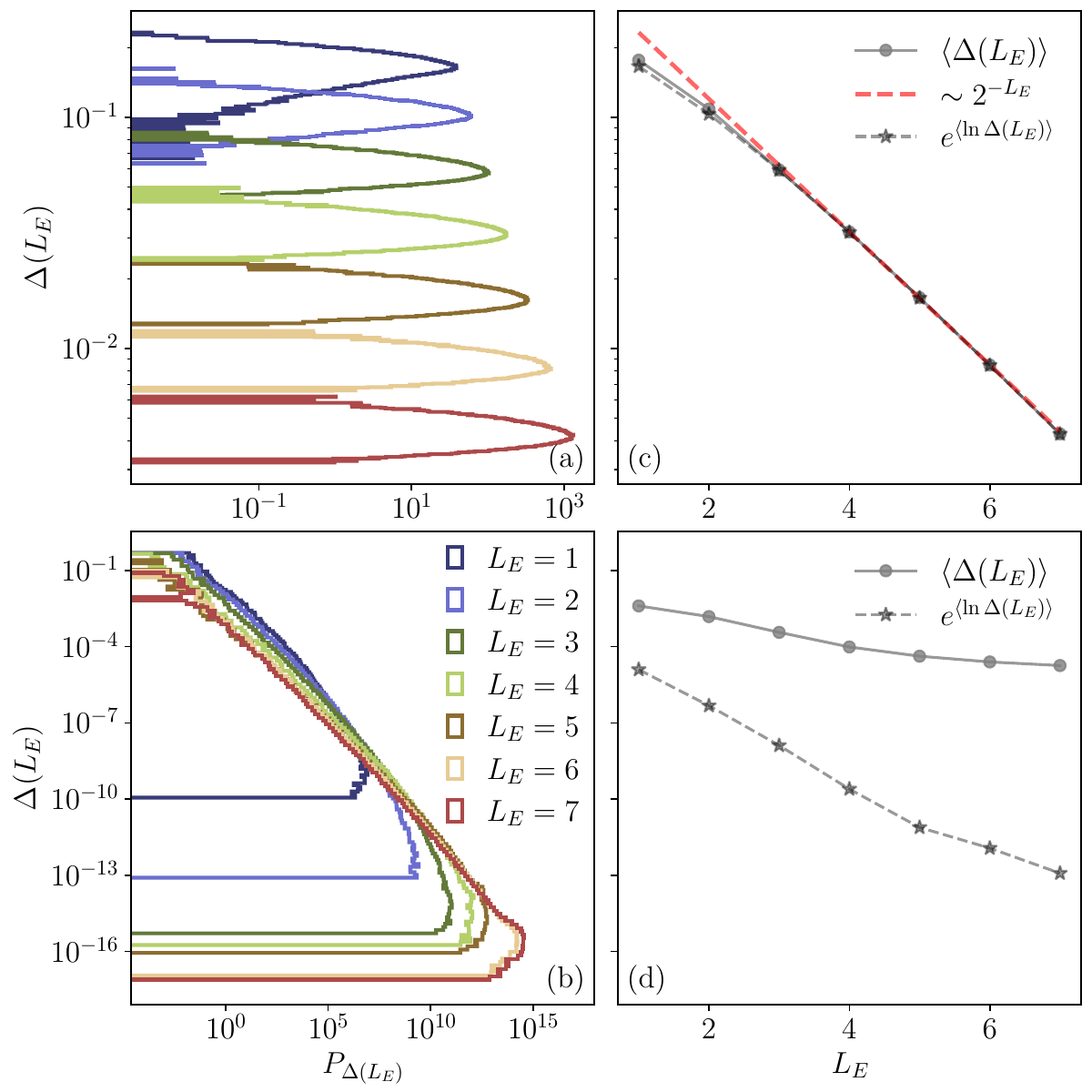}
\caption{The left columns show the distributions of $\Delta(L_E)$ over eigenstates in the (a) ergodic and (c) MBL phase respectively for different values of $L_E$ (different colours). While in the former, the distributions are quite sharp, in the latter, they have broad power law tails. Panels (c) and (d) show mean and typical $\Delta(L_E)$ in the two phases respectively as a function of $L_E$. In the ergodic phase [panel (c)] the data for both, the mean and typical, follows the $D_E^{-1}$ behaviour whereas in the MBL phase [panel (d)] the mean is significantly larger than the typical due to the tails in the aforementioned distribution. }
\label{fig:ppe-eig}
\end{figure}

}


\bibliography{refs}

\end{document}